\documentclass[12pt,a4paper]{article}
\usepackage[usenames, dvipsnames]{xcolor}
\usepackage[normalem]{ulem}
\usepackage{jcapmod}

\usepackage{tocloft,ulem}
\usepackage[]{todonotes}
\usepackage{verbatim}
\usepackage{mathrsfs}
\usepackage{cleveref}
\usepackage[shortlabels]{enumitem}
\usepackage{pgf}
\usepackage{tikz-cd}
\usetikzlibrary{shapes,arrows}
\usetikzlibrary{calc}
\usepackage{pgfplots}
\pgfplotsset{compat=1.12}
\usepackage{tikz-3dplot}

\usepackage{amsthm}

\usepackage{tikz}
\usepackage{setspace,caption}
\usepackage{lipsum}
\usetikzlibrary{matrix,arrows,calc}

\usepackage{amsmath}
\usepackage{amssymb}
\usepackage{amsfonts}
\usepackage{mathtools}

\newcommand{\I}{\mathrm{i}}

\newcommand{\oF}{\overline{F}}

\def\Re{{\rm Re\,}}
\def\Im{{\rm Im\,}}
\newcommand{\dd}{\mathrm{d}}

\usepackage{environ}
\usepackage{changepage}

\newcommand{\be}{\begin{equation}}
\newcommand{\ee}{\end{equation}}
\newcommand{\ol}{\overline}

\begin{document}
	\newcommand{\main}{.}
\begin{titlepage}

\setcounter{page}{1} \baselineskip=15.5pt \thispagestyle{empty}
\setcounter{tocdepth}{2}
\bigskip\

\vspace{.1cm}
\begin{center}
{\LARGE \bfseries Effective potentials, warping,\\[.4cm]
and implications for $\boldsymbol{F}$-term uplifting}
\end{center}

\vspace{0.55cm}

\begin{center}
\scalebox{0.95}[0.95]{{\fontsize{14}{30}\selectfont 
Arthur Hebecker,$^{a}$ Severin Lüst,$^{b}$ Andreas Schachner,$^{c,d}$ and Simon Schreyer$^{d}$\vspace{0.25cm}}}

\end{center}

\begin{center}

\vspace{0.15 cm}
{\fontsize{11}{30}
\textsl{$^{a}$Institute for Theoretical Physics, Heidelberg University, Philosophenweg 19, 69120 Heidelberg, Germany}\\
\textsl{$^{b}$Sorbonne Université, CNRS, Laboratoire de Physique Théorique et Hautes Énergies, LPTHE, F-75005 Paris, France}\\
\textsl{$^{c}$Department of Physics, Cornell University, Ithaca, NY 14853 USA}\\
\textsl{$^{d}$Arnold Sommerfeld Center for Theoretical Physics, Ludwig–Maximilian–University Munich, Theresienstraße 37, 80333 Munich, Germany}\\
}
\vspace{0.25cm}

\vspace{1.1cm}
\normalsize{\bf Abstract} \\[8mm]
\end{center}

\begin{center}
    \begin{minipage}[h]{15.0cm}
        
    We analyse warping corrections to the scalar potential in flux compactifications of Type~IIB string theory, focusing on their effect on $F$-term de Sitter uplifting in Calabi-Yau orientifold models.
    A systematic inverse-volume expansion allows us to derive the four-dimensional off-shell potential in the presence of warping and non-ISD $3$-form fluxes. 
    This corresponds to integrating out all massive Kaluza-Klein modes using the ten-dimensional equations of motion.
    We further propose a warped Kähler potential in four-dimensional $\mathcal{N}=1$ supergravity, and show that it is consistent with our ten-dimensional results.
    In the KKLT framework, we find that classical warping corrections, as well as mixed corrections involving non-ISD fluxes and quantum effects, are dominant, rendering the scenario effectively uncontrollable with current methods.
    By contrast, in LVS-like constructions these corrections are suppressed by inverse powers of the volume,
    specifically ${\cal V}^{1/2}$ or ${\cal V}^{1/6}$, depending on the concrete model.
    
    \end{minipage}

\end{center}

\vfill

\noindent December 2025

\end{titlepage}
\tableofcontents\newpage

\section{Introduction}

Finding string theory backgrounds compatible with an accelerated expansion of our universe, as suggested by cosmological observations, is a major challenge for string phenomenology. In the simplest case, such a universe is modelled by a four-dimensional de Sitter spacetime. However, it is still under debate whether de Sitter vacua can be realised in string theory.

Two of the leading and most studied proposals for meta-stable de Sitter vacua in string theory, the KKLT scenario \cite{Kachru:2003aw} and LVS \cite{Balasubramanian:2005zx}, commonly rely on the anti-D3-brane uplifting mechanism to generate a positive cosmological constant, see e.g.~\cite{Crino:2020qwk,McAllister:2024lnt} for explicit attempts.\footnote{We note that no fully explicit realisation of these scenarios has been found to date and that a possible conflict between KKLT and AdS/CFT holography has been identified \cite{Lust:2022lfc}, see also \cite{Bena:2024are, Apers:2025pon, Bedroya:2025fie}. Notably, this issue is already present at the supersymmetric level, prior to the uplift. For explicit model building progress in this direction see~\cite{Demirtas:2021ote}.}
However, since its conception, numerous potential problems related directly or indirectly to the anti-D3-brane uplift have been identified and discussed controversially \cite{Bena:2009xk,McGuirk:2009xx,Bena:2011hz,Dymarsky:2011pm,Bena:2011wh,Bena:2012bk,Gautason:2013zw,Dymarsky:2013tna,Blaback:2014tfa,Michel:2014lva,Polchinski:2015bea,Cohen-Maldonado:2015ssa,Bena:2018fqc,Armas:2018rsy,Carta:2019rhx,Blumenhagen:2019qcg, Blaback:2019ucp, Randall:2019ent,Bena:2019sxm,Dudas:2019pls,Gao:2020xqh,DallAgata:2022abm,Junghans:2022exo,Gao:2022fdi,Junghans:2022kxg,Hebecker:2022zme,Schreyer:2022len,Schreyer:2024pml,Moritz:2025bsi}, see e.g.~\cite{VanRiet:2023pnx} for a recent review.
In particular, to date the control problems 
identified in
\cite{Carta:2019rhx, Gao:2020xqh,Junghans:2022exo,Gao:2022fdi,Junghans:2022kxg,Hebecker:2022zme,Schreyer:2022len,Schreyer:2024pml,Moritz:2025bsi} remain unsolved. 
It is therefore worthwhile investigating alternative uplifting mechanisms (see \cite{Westphal:2006tn, Cremades:2007ig, Cicoli:2015ylx, Louis:2012nb} for some proposals and \cite{McAllister:2023vgy} for a recent review).

In this work, we focus on what we perceive as a particularly promising alternative to the anti-brane: $F$-term uplifts based on spontaneous supersymmetry breaking in the complex structure sector. This was first proposed in \cite{Saltman:2004sn, Denef:2004ze, Denef:2004cf} and later studied in \cite{Gallego:2017dvd,Honma:2019gzp,Hebecker:2020ejb,Krippendorf:2023idy, Lanza:2024uis}. Our main objective is to analyse the impact of backreaction effects on the classical $F$-term flux potential that arises when moving away from supersymmetric configurations.

Effective actions and potentials for Type~IIB flux compactifications in the presence of warping have been extensively studied, yet are not fully understood.
In the limit where warping corrections can be neglected, the effective potential and its $\mathcal{N}=1$ supergravity formulation were obtained in \cite{Gukov:1999ya, Giddings:2001yu} (see also \cite{Dasgupta:1999ss, Grana:2000jj, Grana:2001xn, Grimm:2004uq}). This was followed by general analyses of the effective actions of warped flux compactifications  \cite{DeWolfe:2002nn, Giddings:2005ff, Burgess:2006mn, Frey:2006wv, Douglas:2007tu, Koerber:2007xk, Shiu:2008ry, Douglas:2008jx, Frey:2008xw,Marchesano:2008rg,Martucci:2009sf,Douglas:2009zn,Underwood:2010pm,Marchesano:2010bs,Grimm:2012rg,Frey:2013bha,Martucci:2014ska,Grimm:2014efa,Grimm:2015mua,Martucci:2016pzt,Lust:2022xoq,Frey:2025rvf,Agarwal:2025rqd}. By now, the effect of warping is well understood for the case in which the complex structure moduli are frozen at an $F$-term minimum of the potential. However, despite partial progress, a complete description of the off-shell effective potential away from its $F$-term minima remains elusive.

\vspace{0.7em}

Both the KKLT scenario and LVS are based on compactification of Type~IIB string theory on a Calabi-Yau orientifold $X_6$. At the classical level, 3-form fluxes generate a non-trivial potential that takes the schematic form
\begin{equation}\label{eq:intr:Vflux}
    V_\mathrm{flux} \sim \frac{1}{\mathcal{V}^2} \! \int_{X_6} \left|G_- \right|^2 \,,
\end{equation}
where $G_-$ denotes the imaginary anti-self-dual (IASD) component of the three-form flux $G_3$, and $\mathcal{V}$ the volume of $X_6$. This potential is positive semi-definite and has Minkowski minima, $\left<V_\mathrm{flux}\right> = 0$, at the points in moduli space where $G_-=0$. This is a condition on the Type~IIB axio-dilaton and the complex structure moduli of the Calabi-Yau geometry, and therefore allows for their stabilisation.

In an effective four-dimensional supergravity formulation, the flux potential can be written as a no-scale $F$-term potential,
\begin{equation}\label{eq:intr:noscale}
    V_\mathrm{flux} \sim \mathrm{e}^K \left | D_I W \right|^2 \,,
\end{equation}
with $W = \int G_3 \wedge \Omega$ the GVW superpotential \cite{Gukov:1999ya}. Here the index $I$ collectively labels all complex structure moduli and the axio-dilaton. The condition $G_- = 0$ corresponds to the vanishing of the $F$-terms: $F_I = D_I W = 0$.

The K\"ahler moduli enter the classical potential $V_{\rm flux}$ only through the volume prefactor $\mathcal{V}^{-2}$, or, equivalently, the K\"ahler potential in $\mathrm{e}^K$.
Therefore, the K\"ahler moduli can only be stabilised by additional quantum or curvature corrections that break the no-scale structure in \eqref{eq:intr:noscale}.
The resulting corrected potential is expected to have a non-trivial AdS vacuum with negative cosmological constant,\footnote{An alternative proposal suggests that $\langle V_{\mathrm{corr}} \rangle > 0$ may suffice to yield a dS minimum~\cite{Balasubramanian:2004uy,Westphal:2005yz,Westphal:2006tn}, as revisited in~\cite{AbdusSalam:2025twp}. These minima arise from the interplay between $\alpha'$ corrections to the K\"ahler potential and non-perturbative effects in the superpotential $W$, but typically lie in regimes where perturbative control is difficult to ensure.}
\begin{equation}\label{eq:intr:combined}
    \left <V_\mathrm{flux} + V_\mathrm{corr} \right> \approx \left < V_\mathrm{corr} \right> < 0 \,,
\end{equation}
where it is assumed that the effect on the stabilisation of the complex structure moduli and the axio-dilaton is negligible, such that $\left <V_\mathrm{flux} \right>$ remains approximately zero.

Therefore, an additional energy source $V_\mathrm{uplift} > 0$ has to be introduced into the system such that a positive cosmological constant can be obtained. Traditionally, this is realised by placing an anti-D3-brane in a strongly warped region of the Calabi-Yau geometry. An alternative approach has been suggested in \cite{Saltman:2004sn}: Instead of stabilising the complex structure moduli at a point where $D_I W = 0$, one can search for non-trivial minima of the classical flux potential \eqref{eq:intr:Vflux} and \eqref{eq:intr:noscale} where
\begin{equation}
    \partial_I V_\mathrm{flux} = 0 \qquad\text{but}\qquad D_I W \neq 0 \,.
\end{equation}
This will result in a positive classical expectation value,
\begin{equation}
    \left <V_\mathrm{flux} \right> > 0 \,.
\end{equation}
If, for suitably chosen fluxes $G_3$, a minimum with an appropriate $\left <V_\mathrm{flux} \right>$ can be found, and if the stabilisation of the K\"ahler moduli goes through as before, the combined potential \eqref{eq:intr:combined} will have a positive expectation value, corresponding to a de Sitter minimum. Here, the uplifting energy comes from the non-vanishing $F$-terms $F_I = D_I W$, motivating the name $F$-term uplift.

\vspace{0.7em}

However, already at the classical level, this procedure has its problems. To appreciate this, let us first recall the success story of GKP in case of a supersymmetric complex structure sector \cite{Giddings:2001yu}: In such backgrounds, the three-form flux is imaginary self-dual (ISD). Its backreaction is limited to the presence of a warp factor and the five-form flux $F_5$. Apart from this, the geometry remains conformally Calabi-Yau, the axio-dilaton is constant, and both the warp-factor and $F_5$ are given in terms of the same function. Evaluating the ten-dimensional Type~IIB action on such configurations leads to a series of cancellations such that the potential is identically zero, cf.~\eqref{eq:intr:Vflux} with $G_-=0$. Equivalently, the $F$-term minima of the potential in \eqref{eq:intr:noscale} indeed solve the higher-dimensional equations of motion.

By contrast, when moving away from the supersymmetric minima of the flux potential, this reasoning no longer applies. A non-vanishing IASD component of the flux, $G_- \neq 0$, backreacts on the geometry in a more complicated way and thus destroys the aforementioned structure. Consequently, the cancellations in the Type~IIB action do not occur any more and additional terms in the potential can arise. It is then not clear whether the simple expressions \eqref{eq:intr:Vflux}, \eqref{eq:intr:noscale}, evaluated away from the $F$-term minima, still correctly describe the effective physics of the flux compactification. The primary objective of our paper is to analyse these additional classical correction terms to the potential and to scrutinise their effect on moduli stabilisation.

For this purpose, we need to derive a genuine off-shell potential that is valid at all points in moduli-space, and not only at its minima or critical points. This poses an intrinsic conceptual difficulty: Away from the critical points of the potential, no static solution to the higher-dimensional equations of motion exists. Consequently, it is not obvious which field configurations should be used when evaluating the higher-dimensional action in order to derive the potential.

The resolution of this problem is canonical: One performs a mode decomposition of the higher-dimensional fields so that the four-dimensional effective potential becomes a function of a finite number of light modes only. The heavy degrees of freedom, corresponding to the higher Kaluza-Klein (KK) modes, are eliminated from the four-dimensional EFT by integrating them out. This requires an analogous decomposition of the higher-dimensional equations of motion into a subset determining the light zero-modes, and a subset for the heavy KK modes. While the former only allow for static solutions at the critical points of the potential, the latter are always solvable and can be used to integrate out the heavy degrees of freedom. This fixes the higher-dimensional configurations as functions of the dynamical fields in the four-dimensional EFT.

We implement this procedure in a systematic expansion in the inverse overall volume $\mathcal{V}$. To make this explicit, let us denote our set of 10d fields by $\Psi$. A solution can then be given in terms of the expansion $\Psi = \sum_n \Psi^{(n)} c^{-n}$. Here the expansion parameter scales as $c\sim {\cal V}^{2/3}$, with a precise definition given below. At each order, we obtain a set of Poisson-type equations,
\begin{equation}\label{eq:intr:inversevolume}
    \Delta \Psi^{(n)} = f\!\left(\Psi^{(0)}, \dots, \Psi^{(n-1)} \right) \,,
\end{equation}
which allows for an iterative solution.\footnote{Equation~\eqref{eq:intr:inversevolume} is schematic. To be more precise, some of the fields from $\Psi^{(n)}$ also appear on the right hand side. However, the system is triangular at each order $n$, such that the overall iterative solvability is not inhibited as long as each Laplacian can be inverted.}
In these equations, $\Delta$ represents a suitable Laplacian operator on the internal space.

By their linearity, the equations \eqref{eq:intr:inversevolume} allow for a decomposition in eigenmodes of the Laplacian $\Delta$. In particular, whenever the source-term on their right-hand side contains a component proportional to a zero-mode of $\Delta$, the corresponding equation cannot be solved. We demonstrate that this is precisely the situation that arises away from the critical points of the potential, and identify the zero-modes of $\Delta$ with the light degrees of freedom in the 4d EFT. These light modes are left unfixed. One can then easily project the equations onto the orthogonal, higher eigenmodes by subtracting the zero-mode component from the right-hand side.\footnote{We invite the reader to consult Appendix~\ref{app:KK} for a detailed illustration of this procedure by means of a simple toy-model.}
This gives rise to a well-defined procedure of integrating out KK modes systematically at each order in $1/c$.

Subsequently, by inserting the inverse volume expansion back into the Type~IIB supergravity action, we obtain an analogous expansion for the effective potential. We find that at the leading order in $1/c$ the effective potential is still given by the expression \eqref{eq:intr:Vflux}. At this level, we provide an explicit, leading order matching between the ten-dimensional equations of motion and the 4d $\mathcal{N}=1$ supergravity scalar potential for non-vanishing $G_-$ flux. As one of the main results of this paper, we compute the first corrections to this expression given in \eqref{eq:V2} which arise at sub-leading order in $1/c$. We further show that, at all orders in $1/c$, the potential remains at least quadratic in $G_-$ so that $G_- = 0$ corresponds to a Minkowski minimum of the full potential. 

Since the scalar potential is usually discussed from the 4d $\mathcal{N}=1$ supergravity perspective, it would be desirable to find a K\"ahler potential reproducing the absence of terms linear terms in $G_-$ at the classical level, including warping. While we are not able find the full K\"ahler potential for warped compactifications, we make a proposal, based on \cite{Frey:2008xw,Martucci:2014ska} that ensures the absence of linear terms in $G_-$. 

In concrete moduli stabilisation scenarios, such as KKLT or LVS, the volume modulus $c$ gets stabilised at a finite value. This means that corrections to the potential cannot be made arbitrarily small, making them potentially relevant. In particular, the requirement of a controlled de Sitter uplift provides a non-trivial relationship between $c$ and the size of the $F$-terms $|F|\sim G_-$. 
This then fixes the parametric size of the warping corrections as well as of perturbative and non-perturbative quantum effects in the presence of $G_-$ flux. With all these corrections at hand, we compare them with the uplifting term and analyse the stability of the scalar potential at putative dS minima. In LVS with D7-branes on the volume 4-cycle, the leading corrections are suppressed by a factor of $g_s^{5/4}/\mathcal{V}^{1/6}$. Thus, while parametric control is possible, the requirements of the size of the volume may be stronger than expected. In cases where the volume 4-cycle is not wrapped by a D7-brane stack, the corrections are suppressed by $1/(g_s^{1/4}\mathcal{V}^{1/2})$, leading to weaker constraints on the size of the volume.

For KKLT, the situation is much more problematic. The underlying reason is the parametric smallness of the vacuum expectation value $W_0$ of the flux superpotential.
Realising a consistent $F$-term uplift in this regime necessitates the parametric relation $|F| \sim |W_0|$. By contrast, the overall volume --- and hence the degree of volume suppression governing both warping and quantum corrections --- is only moderately large, scaling typically as ${\cal V}\sim \ln(1/|W_0|)$. Taken together, these observations place KKLT in a delicate parametric regime for reasons which we now describe in detail.

It is well known that, already at the level of the leading order flux potential \eqref{eq:intr:Vflux}, obtaining a positive definite mass matrix in the regime $|F|\sim |W_0|\ll 1$ generically requires an additional fine tuning of particular contractions of the third covariant derivatives of the superpotential, $D^3W \equiv D_I D_J D_K W$ \cite{Denef:2004cf,Marsh:2011aa,Marsh:2014nla}.\footnote{A concise review of the argument is provided in Appendix~\ref{app:DD}.} 
This additional tuning leads to a light direction, with mass of order $|W_0|^2$, and thus renders the leading order vacuum, computed in the absence of warping effects and of mixing between quantum corrections and non-ISD flux $G_-$, highly sensitive to nominally subdominant contributions. Indeed, using our explicit expression for the warped effective potential, we find that warping induced terms are not automatically suppressed relative to the leading order contributions, even when the tuning of $D^3W$ is imposed. Furthermore, even if the warping effects could be sufficiently suppressed, mixing terms between non-ISD flux $G_-$ and quantum corrections, most notably loop effects, remain parametrically unsuppressed and continue to pose a serious obstruction to a controlled $F$-term uplift in KKLT.

Overall, we find that the resulting corrections are not parametrically suppressed and therefore cannot be controlled with present technology, which severely undermines the viability of $F$-term uplifting in the KKLT regime. More specifically, we will argue that parametric control would require the condition $W_0\,{\cal V}^{2/3}\gg 1$, which is incompatible with an exponentially small $W_0$. Consequently, any viable KKLT-like construction based on $F$-term uplifting would have to operate at moderately small values of $W_0$, accompanied by correspondingly large compactification volumes.

\medskip

The remainder of this work is organised as follows: In Section~\ref{sec:IIB}, we review the equations of motion for warped Type~IIB compactifications and demonstrate how they can be solved in a systematic inverse volume expansion. In Section~\ref{sec:10danalysis}, we derive the effective potential within this expansion, including the first non-trivial sub-leading correction terms. In Section~\ref{sec:4d}, we attempt to cast the corrected potential in four-dimensional supergravity form. In Section~\ref{sec:implications}, we discuss quantum corrections and the implications for the KKLT and LVS. We conclude in Section \ref{sec:conclusions}.

The appendices are structured as follows: In Appendix~\ref{app:EFE}, we derive the 10d Einstein equations. Appendix~\ref{app:KK} provides a simple toy model for integrating out the KK modes while keeping the zero modes unfixed. Appendix~\ref{app:identities} summarises some identities from special geometry. Equations of motion at second order in the inverse volume expansion are discussed in Appendix~\ref{app:g2mn}.
Further details on the four-dimensional $\mathcal{N}=1$ scalar potential are given in Appendix~\ref{app:4d}.
Appendix~\ref{app:DD} generalises results on the masses of the complex structure moduli for non-SUSY flux vacua for no-scale potentials.

\section{Ten-dimensional perspective on Type~IIB string theory}
\label{sec:IIB}

In this section we review the equations of motion of Type~IIB supergravity for warped flux-compactifications, and systematically expand them in inverse powers of the internal volume.
Throughout this paper, we consider Calabi-Yau orientifold compactifications of Type~IIB string theory with O3/O7 orientifold planes, D3/D7-branes, and $G_3$-flux, along the lines of \cite{Giddings:2001yu}.\footnote{Since it is not relevant for our results, we will usually not distinguish between the orientifold and its covering space.
For example, we refer to the relevant Hodge numbers simply as $h^{1,1}$, and $h^{2,1}$, without decomposing them into their even and odd components $h^{1,1}_\pm$ and $h^{2,1}_\pm$.}
We assume that the $G_3$-flux is chosen such that the D3-tadpole is cancelled, and that the D7 tadpole is cancelled locally, so that there is no backreaction on the axio-dilaton by the O7/D7 (corresponding to the IIB limit \cite{Sen:1996vd} of F-theory \cite{Vafa:1996xn}). Finally, we note that one could also choose as a starting point non-Calabi-Yau manifolds with Ricci-flat metric. SUSY would then be broken independently of the flux and the analysis would be more complicated. In particular, one would not be able to use all the topological features of Calabi-Yaus which will facilitate our analysis below. We will not pursue this option.

\subsection{Equations of motion in warped backgrounds}

We use a warped metric ansatz of the form \cite{Giddings:2001yu} 
\begin{equation}
    \dd s^2_{10} = \mathrm{e}^{2A(y)} \tilde g_{\mu\nu} \dd x^\mu \dd x^\nu +  \mathrm{e}^{-2A(y)} \tilde g_{mn}(y) \dd y^m \dd y^n \,.
    \label{eq:metricansatz}
\end{equation}
Here $\tilde g_{mn}$ is the metric on a compact, six-dimensional manifold $X_6$ with coordinates $y^m$, $m = 1, \dots, 6$, and $\tilde g_{\mu\nu}$ is the maximally symmetric 4d external metric. We also allow for a generically non-constant axio-dilaton $\tau = \tau(y)$, for a self-dual five-form flux $\tilde F_5$, and for 3-form fluxes $F_3$, $H_3$ which are conveniently given in complexified form:
\begin{align}
\label{eq:bianchiF5}
    \tilde F_5 &= (1 + \star) \, \dd \alpha \wedge \dd x^0 \wedge \dd x^1 \wedge \dd x^2 \wedge \dd x^3 \,,\\[0.3em]
    G_3 &= F_3 - \tau H_3 \,.
\end{align}
The integrals of $F_3$, $H_3$ over 3-cycles obey the standard quantisation conditions and the Bianchi identities $\dd F_3 = \dd H_3 =0$ are fulfilled.

Following \cite{Baumann:2008kq, Baumann:2010sx}, we introduce
\begin{equation}
    \label{eq:def}
    \Phi_\pm = \mathrm{e}^{4A} \pm \alpha \,, \qquad G_\pm = (\star_6 \pm \mathrm{i}) G_3\,,\qquad \Lambda = \Phi_+ G_- + \Phi_- G_+\,,
\end{equation}
where $\star_6$ denotes the six-dimensional Hodge-star operator with respect to the metric $\tilde g_{mn}$.
The equations of motion can then readily be derived in terms of these fields (see \cite{Giddings:2001yu,Baumann:2008kq,Baumann:2010sx,Gandhi:2011id,McGuirk:2012sb} and also App.~\ref{app:EFE} for details of the derivation). 
For $\Phi_\pm$ we find from the superposition of the trace of the 4d Einstein equations and the Bianchi identity of $\tilde F_5$
\begin{equation}
    \begin{split}
    \label{eq:eomPhi}
    \tilde\nabla^2\Phi_\pm & =   \frac{(\Phi_++\Phi_-)^2}{96\Im\tau} G_\pm\tilde\cdot\,\bar G_\pm+ \frac{2}{\Phi_++\Phi_-} |\tilde\partial\Phi_\pm|^2+ \tilde{\mathcal{R}}_4 \\ & \qquad+ \sqrt{2}\kappa_{10}^2(\Phi_++\Phi_-)^{1/2} \left(\frac{1}{4}(T^m_m-T^\mu_\mu)^\mathrm{loc}\pm T_3\rho_3^\mathrm{loc} \right)\,,
    \end{split}
\end{equation}
where $T^\mu_\mu = - T_p \delta(\Sigma)$ and $T^m_m=- T_p (\Pi^\Sigma)^m_m \delta(\Sigma)$. Here, $\delta(\Sigma)$ and $(\Pi^\Sigma)^m{}_n$ denote the delta distribution and projector of the cycle $\Sigma$ wrapped by the localised object. Crucially, $\delta(\Sigma)$ involves the metric and hence the warp factor. To make this explicit, let $\Sigma$ be $k$-dimensional and choose the coordinate system such that the cycle is locally defined as the $x=0$ hyperplane in $\mathbb{R}^6 = \mathbb{R}^k \times \mathbb{R}^{6-k}\ni y=(z,x)$. Then, locally,
\begin{equation}
    \delta(\Sigma)=\delta^{(6-k)}(x)/\sqrt{g_\perp}\,,
\end{equation}
with $g_\perp$ the determinant of the induced metric on the transverse hyperplane at fixed $z\in \mathbb{R}^k$. Further, we defined the localised D3 charge density $ \rho^{\mathrm{loc}}_3$. It receives contributions from O3/D3 and curved O7/D7 branes and is normalised such that a single D3 gives
\begin{equation}
    \rho^{\mathrm{loc}}_3 = \delta^{(6)}(y-y_i)/\sqrt{g_\perp}\,.
\end{equation}
Analogously to what was explained above for a general cycle, there is an explicit warp factor dependence due to the inverse perpendicular metric determinant. In this specific case, $\sqrt{g_\perp}=\sqrt{\tilde g_\perp}\exp((p-9)A)=\sqrt{\tilde g}\exp(-6A)$.

Moving on to the equations of motion for $G_3$ and $\tau$, we find
\begin{align}
    \label{eq:eomG3}
    0 & = \dd\Lambda +\frac{\mathrm{i}}{2\,\Im\tau }\,\dd\tau\wedge (\Lambda+\bar\Lambda)\,,\\
    \label{eq:eomtau}
    0 & = \tilde\nabla^2\tau +\frac{\mathrm{i}}{\Im\tau} (\tilde\partial\tau)^2 +\frac{\mathrm{i}}{48}(\Phi_++\Phi_-) \, G_+\tilde \cdot\, G_- - \frac{4 (\Im\tau)^2}{\sqrt{-g}} \frac{\delta S_\mathrm{D7}}{\delta\bar\tau}\,.
\end{align}
Finally, the internal Einstein equations read
\begin{equation}
    \begin{split}
    \label{eq:eomR}
    \hspace{-.3cm}\tilde R_{mn} & =  \frac{\partial_{(m}\tau\partial_{n)}\bar\tau}{2(\Im\tau)^2}  +\frac{2\partial_{(m}\Phi_+\partial_{n)}\Phi_-}{(\Phi_++\Phi_-)^2} - \frac{\Phi_++\Phi_-}{32\Im\tau} \left( G_{+(m}^{~~~\widetilde{pq}}\bar G_{-n)pq} + G_{-(m}^{~~~\widetilde{pq}}\bar G_{+n)pq}\right)
      \\
     & \,\, + \frac{\tilde g_{mn}}{4(\Phi_++\Phi_-)} \Biggl( -\tilde\nabla^2\Phi_+ + \frac{(\Phi_++\Phi_-)^2}{96\Im\tau}\, G_+\tilde\cdot\,\bar G_+ +\frac{2|\tilde\partial\Phi_+|^2}{\Phi_++\Phi_-} + \mathcal{T}^{\mathrm{loc}}_{\Phi_+} \\
    & \,\,-\tilde\nabla^2\Phi_- + \frac{(\Phi_++\Phi_-)^2}{96\Im\tau}G_-\,\tilde\cdot\,\bar G_- +\frac{2|\tilde\partial\Phi_-|^2}{\Phi_++\Phi_-} \Biggr)+ 
    \kappa_{10}^2\left( T^{\mathrm{loc}}_{mn} -\frac{g_{mn}}{4} ( T^p_p)^{\mathrm{loc}} \right)\, ,
    \end{split}
\end{equation}
where $\tilde R_{mn} = \tilde R^l_{~mln}$,
and $\tilde{\mathcal{R}}_4 = \tilde g^{\mu\nu} \tilde R_{\mu\nu}$. We also defined $\mathcal{T}^{\mathrm{loc}}_{\Phi_+}$ as the local term of the $\Phi_+$ equation of motion (see second line in \eqref{eq:eomPhi}) and the last term can again be rewritten in terms of the tilded metric. Note that in \eqref{eq:eomR} the terms in the large bracket proportional to $\tilde g_{mn}$  are exactly those of \eqref{eq:eomPhi}, such that they cancel on-shell up to the term $\tilde{\mathcal{R}}_4$.
In the set of equations above we used the notation
\begin{equation}
    X_{(p)} \tilde \cdot\, Y_{(p)} =  X_{m_1\cdots m_p} Y^{\widetilde{m_1\cdots m_p}}\,, \qquad\qquad |\tilde X_{(p)}|^2 = X_{(p)} \tilde \cdot \, \bar X_{(p)}\,.
    \label{eq:productconv}
\end{equation}

From the equations of motion it is readily verified that, for a GKP-type solution with
\begin{equation}
\label{eq:BPSconditions}
    \Phi_- = G_- = \tilde R_{mn} = \tilde{\mathcal{R}}_4 = \partial_m \tau = 0 \,,
\end{equation}
the equations of motion collapse to a Poisson-like equation for the inverse warp-factor \cite{Giddings:2001yu},
\begin{equation}
\label{eq:GKPwarping}
    - \tilde\nabla^2 \mathrm{e}^{-4A} = \frac{G_+\tilde\cdot\,\bar G_+}{48 \Im\tau} +2\kappa_{10}^2 T_3 \tilde \rho_3^{\mathrm{loc}}\,.
\end{equation}

The Ricci-flat metric $\tilde g_{mn}$ underlying \eqref{eq:BPSconditions} is that of a Calabi-Yau three-fold $X_6$.
The self-duality condition $G_-=0$, or equivalently
\begin{equation}
\star_6 G_3 = \mathrm{i}\, G_3 \,,
\end{equation}
imposes a constraint on the complex structure moduli of $X_6$, which is in general fulfilled only at particular points in complex structure moduli space.
This can be elegantly implemented as the $F$-term constraint for the GVW superpotential \cite{Giddings:2001yu,
Gukov:1999ya}. Its solutions are the Minkowski minima of the corresponding 4d effective potential on complex structure moduli space.

Let us note that the remarkable cancellations underlying this famous result persist in F-theory, where $\tilde{R}_{mn}$ and $\partial_m \tau$ are non-zero \cite{Giddings:2001yu}. It would be interesting to analyse the backreaction effects of a non-vanishing complex structure $F$-term also in this case, but we will not do so in the present paper.

\subsection{Perturbative solution in the inverse volume}

In what follows, we focus on models with D3/O3s and locally cancelled D7 tadpole. We include the contribution of curved D7/O7 to the tadpole in $\rho_3^{\mathrm{loc}}$ since this is a leading order effect. 
Due to the locally cancelled D7 tadpole, the energy momentum tensor of the D7/O7s as well as $\delta S_{\mathrm{D7}}/\delta\bar\tau$ in \eqref{eq:eomtau} drop out of the 10d equations of motion. 

We recall that, at the level of the GKP analysis \cite{Giddings:2001yu}, in the absence of supersymmetry breaking fluxes $G_-$,
\begin{equation}
\label{eq:GKPwarpfactorsolution}
   2\Phi_+^{-1} = \mathrm{e}^{-4A(y)}=c+\{\mbox{$y$-dependent}\} \,, \qquad \Phi_-=0 \,.
\end{equation}
The integration constant $c$ remains unfixed by the warp-factor equation \eqref{eq:GKPwarping}, and can be identified with the volume modulus \cite{Giddings:2005ff,Douglas:2007tu,Frey:2008xw}.
The large volume limit corresponds to $c\to \infty$. 

We are interested in solving the 10d equations of motion in the presence of IASD flux $G_- \neq 0$. We will do so employing a systematic $1/c$ expansion for all quantities appearing in the ten-dimensional equations of motion, 
\begin{equation}
\begin{aligned}
\label{eq:1overcexpansion}
    c \Phi_\pm &= \Phi^{(0)}_\pm + \frac{1}{c} \Phi^{(1)}_\pm + \cdots \,, &
    \tau &= \tau^{(0)} + \frac{1}{c} \tau^{(1)} + \cdots \,, \\
    \tilde g_{mn} &= \tilde g_{mn}^{(0)} + \frac{1}{c} \tilde g_{mn}^{(1)} + \cdots \,, &
    G_\pm &= G_\pm^{(0)} + \frac{1}{c} G_\pm^{(1)} + \cdots \,, 
\end{aligned}
\end{equation}
At lowest order in the $1/c$ expansion, the equations of motion are solved by an unwarped Calabi-Yau geometry with constant dilaton,\footnote{Note that the equations of motion only demand $\Phi_\pm^{(0)}  = \mathrm{const}$. The combination $\Phi^{(0)}_+ - \Phi^{(0)}_-$ drops out of the equations of motion. This allows us to set $\Phi^{(0)}_- = 0$. Furthermore, any rescaling of $\Phi^{(0)}_+$ can be absorbed into the definition of $c$, so we choose to set $\Phi^{(0)}_+=2$ in order to match \eqref{eq:GKPwarpfactorsolution}.}
\begin{equation}
    \Phi_+^{(0)} = 2 \,,\qquad \Phi_-^{(0)} = 0 \,,\qquad \tau^{(0)} = \mathrm{const.} \,, \qquad \tilde R_{mn} \left(\tilde g^{(0)}\right) = 0 \,,
\end{equation}
and vanishing four-dimensional cosmological constant.
The $1/c$ expansion of the internal metric $\tilde g_{mn}$ in \eqref{eq:1overcexpansion} also induces an analogous expansion of the Hodge-star operator, $\star_6 = \star^{(0)} + \frac1c \star^{(1)} + \cdots$. The leading-order fluxes $G^{(0)}_\pm$ are imaginary (anti-) self-dual with respect to the Ricci-flat Calabi-Yau metric $\tilde g^{(0)}$,
\begin{equation}
    \label{eq:AISDleading}
     \star^{(0)} G^{(0)}_\pm = \pm \mathrm{i} \, G^{(0)}_\pm \,,
\end{equation}
and satisfy the equation of motion
\begin{equation}
\label{eq:Gharmonicleading}
    \dd \star^{(0)} G^{(0)} = 0 \,.
\end{equation}
This means that at lowest order, the fluxes $G^{(0)}_\pm$ are harmonic with respect to the Calabi-Yau metric.

This leading order solution is independent of the presence of a supersymmetry breaking $G_-$ component of the three-form flux. At finite volume, or at sub-leading order in the $1/c$ expansion, this background receives corrections, both if supersymmetry is preserved as well as due to supersymmetry breaking $G_-$ fluxes. These corrections induce non-trivial warping, a dilaton profile, and a non-vanishing Ricci-curvature of the internal metric.

\subsection{The first non-trivial order } \label{sec:eomnontrivial}

To make this explicit, let us now proceed with the equations at order $\mathcal{O}(1/c)$. In general, the Poisson-type equations at each order in $1/c$ are not solvable because their right-hand sides have non-trivial projections onto the kernels of the Laplacians. We propose a solution for this problem in Section \ref{sec:10danalysis}. Here, we disregard this issue.

Starting with $\Phi_+^{(1)}$, we find (using $\Phi_+^{(0)}=2$ from now on)
\begin{equation}
\label{eq:eompp}
    \tilde \nabla^2\Phi_+^{(1)} = \frac{G_+^{(0)} \tilde\cdot \,\bar G_+^{(0)} }{24\Im\tau^{(0)}} + 4 \kappa_{10}^2 T_3  \,\tilde\rho^{\mathrm{loc}}_3 +\tilde{\mathcal{R}}_4\,.
\end{equation}
In the case $\Phi_- = 0$, corresponding to $\alpha = \exp(4A)$, this is related to the standard warp factor equation \eqref{eq:GKPwarping} and determines the varying part of the GKP warp factor due to the backreaction of the ISD-fluxes.

For all fields other than $\Phi_+$, the correction at order $1/c$ take us beyond GKP (for non-vanishing $G_-^{(0)}$). The corresponding equations of motion determining these corrections read (using additionally $\Phi_-^{(0)}=0$ from now on) 
\begin{align}
\label{eq:eompm}
    \tilde\nabla^2 \Phi^{(1)}_- & = \frac{G^{(0)}_-\,\tilde \cdot \,\bar G^{(0)}_-}{24\,\Im \tau^{(0)}} +\tilde{\mathcal{R}}_4 \,,\\
    \tilde \nabla^2\tau^{(1)} & = - \mathrm{i} \,\frac{G^{(0)}_+ \,\tilde\cdot \,G^{(0)}_- }{24}\,, \label{eq:eomtau1} \\
    \begin{split}
     \Delta \tilde g^{(1)}_{mn} & =  \frac{1}{8\Im\tau^{(0)}} \left(G_{+(m} ^{(0)~~\widetilde{pq}} \bar G_{-n)pq}^{(0)} +   G_{-(m}^{(0)~~\widetilde{pq}} \bar G_{+n)pq}^{(0)}\right) \\
      & \quad+ \frac{\tilde g^{(0)}_{mn}}{4} \left( \tilde \nabla^2\Phi_+^{(1)} - \frac{G_+^{(0)} \tilde\cdot \,\bar G_+^{(0)} }{24\Im\tau^{(0)}} - 4 \kappa_{10}^2 T_3  \,\tilde\rho^{\mathrm{loc}}_3 +\tilde\nabla^2 \Phi^{(1)}_- - \frac{G^{(0)}_-\,\tilde \cdot \,\bar G^{(0)}_-}{24\,\Im \tau^{(0)}} \right) .
     \label{eq:eomg1}
     \end{split}
\end{align}
Note that, if solutions of \eqref{eq:eompp} and \eqref{eq:eompm} exist, they can be inserted into the second line of \eqref{eq:eomg1}, thereby reducing the second line to $\frac12 \tilde g^{(0)}_{mn} \tilde{\mathcal{R}}_4$.
We have also defined the metric kinetic operator (which equals the Lichnerowicz Laplacian in de Donder gauge) in \eqref{eq:eomg1} as
\begin{equation}\label{eq:metriclaplacian}
    \Delta  \tilde g^{(1)}_{mn} = \tilde \nabla^2\tilde g^{(1)}_{mn} + \tilde \nabla_m \tilde\nabla_n\left((\tilde g^{(0)})^{pq}\,\tilde g^{(1)}_{pq}\right) - 2\tilde\nabla^p\tilde\nabla_{(m} \tilde g^{(1)}_{n)p} \,. 
\end{equation}
All the gradient operators are constructed using the Ricci flat CY metric $\tilde g^{(0)}_{mn}$. 
Furthermore, at order $1/c$, the self-duality condition for the three-form flux becomes
\begin{equation}
\label{eq:G31selfduality}
    \left(\star^{(0)}  \mp \mathrm{i} \right) G^{(1)}_{\pm} =  - \star^{(1)} G^{(0)}_{\pm} \,,
\end{equation}
while the equation of motion and the Bianchi identity for $G_3$ yield 
\begin{equation}
    \dd G_\pm^{(1)} = -  \frac12 \Bigl( \dd \Phi_+^{(1)} \wedge G_-^{(0)} + \dd \Phi_-^{(1)} \wedge G_+^{(0)} \Bigr) - \frac{\mathrm{i} }{\Im \tau^{(0)}} \, \dd \tau^{(1)} \wedge \Re G_\pm^{(0)} \,.
    \label{g1bia}
\end{equation}
This implies that the correction $G_3^{(1)}$ is in general no longer harmonic with respect to $\tilde g_{mn}^{(0)}$.
It is easy to see that we can build a Poisson equation for $G_3^{(1)}$ from the two linearly independent equations \eqref{g1bia} which guarantees uniqueness on compact manifolds up to harmonic forms. 
The Poisson equation is derived by using the definition $\Delta^{(0)} = -( \dd\star^{(0)} \dd\star^{(0)}+\star^{(0)}\dd\star^{(0)}\dd)$ together with \eqref{eq:G31selfduality}.

We note that the set of perturbed equations of motion is much simpler compared to the full set of equations as we are dealing with Poisson equations. In particular, the equations of motion at $\mathcal{O}(1/c)$ as displayed above have triangular form. By this we mean that, proceeding from top to bottom, the left hand side is defined in terms of the right hand side and the previously solved equations. While this feature has also been observed in \cite{Gandhi:2011id} in an expansion 
around an ISD background, we emphasise that it holds more generally in the inverse volume expansion. 

We reiterate the key issue that our equations are generically unsolvable due to the non-trivial kernel of the Laplacian. 
For example, \eqref{eq:eomtau1} has the constant $\tilde{\mathcal{R}}_4$ on the right-hand side. This term depends on the four-dimensional metric, which is an input parameter of the 4d effective action and can a priori take any value, making the zero-mode projection of the equation unsolvable. The resolution we propose in Section \ref{sec:10danalysis} is straightforward: The kernels of the Laplacians correspond to the moduli of the compact space. They are the dynamical fields of our desired 4d EFT and should hence {\it not} be determined at this stage of the analysis. Of course, at the minimum of the effective potential, the 4d fields are stabilised and the set of equations \eqref{eq:eompp} -- \eqref{eq:eomg1} becomes solvable.

The $1/c$-expansion above can be continued to any desired order, see App.~\ref{app:g2mn} for the equations at order $1/c^2$. For most of our purposes, however, the first non-trivial order is sufficient.

\section{The warped effective potential} \label{sec:10danalysis}

In this section we come to our main object of interest: The 4d off-shell effective potential.\footnote{
Our approach is somewhat different from that of \cite{Douglas:2009zn}, see the discussion in Sect.~\ref{sec:constraint}.} 
We assume scale separation between the KK scale of the compactification and the masses of `light fields'. Then, the low-energy dynamics will be characterised by a four-dimensional effective action which, ignoring fermions and vectors for simplicity, takes the leading-order form 
\begin{equation}
\label{eq:effaction}
    S = M_p^2 \int \dd^4x\sqrt{-\tilde g_{4}}\left( \frac{\tilde{R}_{4}}{2} + K_{M\overline N} (\phi,\bar{\phi}) \, \partial_\mu \phi^M \tilde\partial^\mu \bar \phi ^{\overline N} -M_p^2 V_{\mathrm{eff}}(\phi,\bar{\phi})\right)\,.
\end{equation}
Here, $M_p$ is the 4d Planck mass, $\phi^M$ are the light scalars, and $K_{M\overline N}$ is the metric on their field space. Note that $\tilde R_4$ and $\tilde{\mathcal{R}}_4$ are related by a Weyl rescaling: $\tilde R_4$ is given in 4d Einstein frame and $\tilde{\mathcal{R}}_4$ is in 10d Einstein frame since it appears in the 10d equations of motions. 

In general, due to the presence of the kinetic terms, the computation of the full effective action \eqref{eq:effaction} can be rather subtle, as discussed, for example, in \cite{Giddings:2005ff,Shiu:2008ry,Douglas:2008jx,Lust:2022xoq}. Promoting the deformation modes $\phi^M$ of the ten-dimensional solution to space-time dependent fields $\phi^M(x^\mu)$ requires a modification of the metric ansatz \eqref{eq:metricansatz} so that it includes off-diagonal terms. These terms can be understood as compensator fields, and can be eliminated by a suitable coordinate or gauge transformation. To correctly derive the scalar field space metric the compensator fields (or compensating gauge transformations) have to be determined by solving a set of constraint equations.

Here, however, we are not interested in the kinetic terms but only in the effective potential $V_{\mathrm{eff}}$, and can therefore ignore this issue.
If the light deformation modes of the ten-dimensional solution have been identified correctly, the constraint equations are solvable, and we can assume that the compensator fields have been gauged away.
The computation of the potential itself does not depend on a specific choice of coordinates on the compactification space, and is therefore gauge independent.
Consequently, it does not require solving the constraint equations that determine the compensator fields.

Furthermore, as we discuss below, the massless moduli of the unwarped solution, namely the axio-dilaton, as well as the K\"ahler and complex structure deformations of the Calabi-Yau metric, remain the correct light modes in our effective action \eqref{eq:effaction} in the large volume limit.

\subsection{Off-shell scalar potential in Type~IIB}\label{sec:offshellpotential}

The effective potential $V_{\mathrm{eff}}$ in \eqref{eq:effaction} is readily derived by inserting the metric ansatz \eqref{eq:metricansatz} into the action of Type~IIB string theory \eqref{eq:SIIB} (see e.g.~\cite{Giddings:2001yu,Giddings:2005ff}), and by assuming that all four-dimensional spacetime derivatives vanish (see also the discussion in Appendix~\ref{app:KK}),
\begin{equation}
\begin{split}
    V_{\mathrm{eff}} & =\frac{\kappa_{10}^2}{{\cal V}_{4, {\rm w}}^2} \int \dd^6y \sqrt{\tilde{g}_6} \Biggl(\frac{\mathrm{e}^{4A}}{24\Im\tau} G_3\,\tilde\cdot\, \bar{G}_3 + \frac{\mathrm{e}^{-8A}}{4} \bigl(\tilde\nabla\alpha\bigr)^2 + 4 \bigl(\tilde\nabla A\bigr)^2 +\frac{\bigl(\tilde\nabla\tau\bigr)^2}{4(\Im\tau)^2}\\
    & \qquad\qquad\qquad\qquad\qquad -\frac{\tilde{R}_6}{2} + \mathrm{e}^{-2A}\kappa_{10}^2 \mu(y)\Biggr) \,.
\end{split}
\label{eq:VeffD}
\end{equation}
Here $\mu(y) = \sum_i T^i_{p} \delta^{(9-p)}(y_\perp-y_\perp^i)/\sqrt{ g_\perp}$ can be understood as a mass density induced by localised sources. In our case with locally cancelled D7-tadpoles, $\mu(y) = T_3 \rho^\mathrm{loc}_3(y)$. Moreover, the warped volume ${\cal V}_{4, {\rm w}}$ which scales at leading order as ${\cal V}_{4, {\rm w}}\sim c$ is defined as
\begin{equation}
\label{eq:Vwarped}
    {\cal V}_{4, {\rm w}} =\int \dd^6y\sqrt{\tilde g_6}\exp(-4A) =\int \dd^6y\sqrt{\tilde g_6} \,\frac{2}{\Phi_++\Phi_-}\,.
\end{equation}
The prefactor $\kappa_{10}^2/{\cal V}_{4, {\rm w}}^2$ comes from the Weyl rescaling to 4d Einstein frame, i.e.,~$\tilde g_{\mu\nu}\to M_p^2\kappa_{10}^2/{\cal V}_{4, {\rm w}}\,\tilde g_{\mu\nu}$. From now on, we set $\kappa_{10}=1$.\footnote{If one instead works in the commonly used units where $2\pi\sqrt{\alpha'} = 1$, one finds $\kappa_{10}^2 = 1/(4\pi)$. To correctly account for this factor from the 4d supergravity perspective, the superpotential $W$ must then be rescaled by a factor of $1/\sqrt{4\pi}$.}

Using the Bianchi identity of $F_5$, the potential \eqref{eq:VeffD} can be simplified further.
The Bianchi identity reads
\begin{equation}
    \dd \tilde F_5 = H_3 \wedge F_3 + 2\kappa_{10}^2 T_3 \rho_3^{{\mathrm{loc}}},
    \label{eq:F5}
\end{equation}
and yields in its integrated form the tadpole cancellation condition
\begin{equation}\label{eq:tadpole}
     \int \frac{\mathrm{i}\,G_3 \wedge \bar G_3}{2\Im\tau} + 2\kappa_{10}^2 T_3 Q_3^{{\mathrm{loc}}}= 0 \,,
\end{equation}
where we used that $H_3\wedge F_3=\mathrm{i}G_3\wedge \bar G_3/(2\Im\tau)$.  Substituting 
\eqref{eq:bianchiF5} for $\tilde F_5$ in \eqref{eq:F5} and using \eqref{eq:metricansatz}, one finds\footnote{We work in GKP conventions where $\star_4 \left(\dd x^0\wedge\dd x^1\wedge \dd x^2\wedge\dd x^3\right) = 1$.}
\begin{equation}
    \tilde\nabla^2\alpha = \frac{\mathrm{i} \,\mathrm{e}^{8A}}{12\Im\tau} G_{mnp} (\star_6\,\bar G)^{\widetilde{mnp}} + 2 \mathrm{e}^{-4A} \left(\tilde\partial^m \mathrm{e}^{4A}\right) \partial_m \alpha + 2 \mathrm{e}^{2A} \kappa_{10}^2 T_3 \rho_3^{\mathrm{loc}}\,.
    \label{eq:bianchi}
\end{equation}
This equation is the difference of the two equations in \eqref{eq:eomPhi}.
The potential \eqref{eq:VeffD} can then be reformulated by adding and subtracting $\exp(-4A)$ times \eqref{eq:bianchi}. After partial integration one obtains
\begin{equation}
\label{eq:Vscaling}
    \begin{split}
        V_{\mathrm{eff}} & = \frac{1}{{\cal V}_{4, {\rm w}}^2} \int \dd^6y\sqrt{\tilde g_6}\left(\frac{\Phi_++\Phi_-}{ 96 \Im\tau} |\tilde G_-|^2
        +\frac{(\tilde \partial \Phi_-)^2 }{(\Phi_++\Phi_-)^2} + \frac{\partial_m\tau \,\partial^{\tilde m}\bar\tau}{4(\Im\tau)^2}- \frac{\tilde R_6}{2}\right).
    \end{split}
\end{equation}
The potential \eqref{eq:Vscaling} remains a rather formal expression as long as we do not specify precisely which metric and field profiles we use.  At the level of GKP, the potential is exactly zero, which follows by substituting the solutions \eqref{eq:BPSconditions}. One may interpret this by noting that the self-duality constraint $G_{-} = 0$ is a condition on the complex structure moduli $z^i$ and the axio-dilaton $\tau$, fixing them at a Minkowski minimum of the potential: $V_\mathrm{eff} = 0$. More generally, the potential is a function over moduli space and it is not clear a priori whether GKP-type loci with $V_{\rm eff}=0$ or other local minima exist. Our interest here is the systematic procedure for determining the potential. The study of its features for specific CY orientifolds is an independent challenge.

\subsection{10d equations of motion from the off-shell 4d perspective} \label{sec:lineom}

Our goal is now to analyse the potential \eqref{eq:Vscaling} for more general field configurations than the GKP solution by allowing for fluxes with an IASD component $G_- \neq 0$.
This means that we allow the complex structure moduli $z^i$ and the axio-dilaton $\tau$ to take values outside the loci where the corresponding $F$-terms are zero and the fluxes are ISD. As a result, the potential $V_{\mathrm{eff}}$ will generally be non-zero. Even worse, at generic points $z^i$ we will not even be at an extremum of the potential such that it is not possible in principle to satisfy all equations of motion unless we allow for some non-trivial four-dimensional dynamics. 

Still, we want to use the equations of motion derived in Section~\ref{sec:IIB} to integrate out all KK modes so that we can obtain a sensible effective potential for the light degrees of freedom.
For this purpose, we need to separate the equations of motion into a part corresponding to light degrees of freedom, and a part corresponding to the heavy KK modes.

We implement this procedure in the inverse volume expansion that was introduced in Section~\ref{sec:IIB}.
At the level of the linear equations that have been derived there, the projection onto the heavy KK modes can be implemented by subtracting the zero-mode(s) of the corresponding Laplacian operator from the right-hand side of the equations.
As illustrated by means of a simple toy model in Appendix~\ref{app:KK}, the zero modes correspond to the light degrees of freedom in the EFT.
Their equations of motion can only be solved at a critical point of the effective potential. 
The remaining, orthogonal equations, however, can always be solved, and allow us to formally integrate out the KK modes.

We now make this explicit at the first non-trivial order in $1/c$ which has been discussed in Section~\ref{sec:eomnontrivial}.
We start with equations \eqref{eq:eompp} and \eqref{eq:eompm} for the scalar quantities $\Phi^{(1)}_\pm$.
In the scalar case, the Laplacian has only one zero-mode, given by the constant function on the internal space $X_6$.
It follows from the tadpole cancellation condition \eqref{eq:tadpole} that\footnote{To avoid cluttering notation, we write $\sqrt{\tilde g_6^{(0)}}\equiv\sqrt{\tilde g^{(0)}}$ in what follows.}
\begin{equation}
    \int \dd^6y\sqrt{\tilde g^{(0)}} \left( \frac{|\tilde G_+^{(0)}|^2}{96\Im\tau^{(0)}} - \frac{|\tilde G_-^{(0)}|^2}{96\Im\tau^{(0)}} \right)  + \kappa_{10}^2T_3  Q_3^\mathrm{loc} = 0 \,.
\end{equation}
Therefore, \eqref{eq:eompp} and \eqref{eq:eompm} have the same zero-mode contributions, and their projection onto the higher-order modes reads
\begin{align}
    \tilde \nabla^2\Phi_+^{(1)} &= \frac{G_+^{(0)} \tilde\cdot \,\bar G_+^{(0)}}{24\Im\tau^{(0)}}  + 4\kappa_{10}^2 T_3  \,\tilde\rho^{\mathrm{loc}}_3 +\tilde{\mathcal{R}}_4 - \mathcal{C}_{\Phi}\,, \label{eq:eomphi+1} \\
    \tilde\nabla^2 \Phi^{(1)}_- &= \frac{G^{(0)}_-\,\tilde \cdot \,\bar G^{(0)}_-}{24\Im\tau^{(0)}} +\tilde{\mathcal{R}}_4-\mathcal{C}_{\Phi} \,.\label{eq:eomdeltaphiM}
\end{align}
Here $\mathcal{C}_\Phi$ is defined by
\begin{equation}\label{eq:cPhi}
    \mathcal{C}_{\Phi} - \tilde{\mathcal{R}}_4 =   \frac{1}{V_\mathrm{CY}} \int \dd^6y\sqrt{\tilde g^{(0)}}\,  \frac{|\tilde G_-^{(0)}|^2}{24\,\Im\tau^{(0)}}  \,,
\end{equation}
where we also chose to normalise the internal volume to unity: $V_\mathrm{CY} =\int\dd^6y\sqrt{\tilde g^{(0)}} \equiv 1$. Note that $\mathcal{C}_\Phi$ here is merely a technical tool making Eqs.~\eqref{eq:eomphi+1}, \eqref{eq:eomdeltaphiM} solvable, thereby defining the non-constant or KK-mode parts of $\Phi^{(1)}_\pm$. The zero-mode part of $\Phi^{(1)}_+ - \Phi^{(1)}_-$ corresponds to a constant shift of $\alpha$ and is pure gauge. By contrast, the zero-mode part of $\Phi^{(1)}_+ +\Phi^{(1)}_-$ corresponds to a constant  shift of $e^{-4A} $ and hence of the volume modulus or $c$. The latter is an, at this point arbitrary, argument of the effective potential we are calculating.

For $\tau^{(1)}$ the situation is very similar and we readily obtain the projection of \eqref{eq:eomtau1},  
\begin{equation}
    \label{eq:eomdeltatau}
    \tilde \nabla^2\tau^{(1)} = - \frac{\mathrm{i}\, G^{(0)}_+ \,\tilde\cdot \,G^{(0)}_-}{24} - \mathcal{C}_{\tau}\,,
\end{equation}
with
\begin{equation}\label{eq:tauconstant}
     \mathcal{C}_{\tau} = - \int \dd^6y\sqrt{\tilde g^{(0)}} \frac{\mathrm{i} \, G^{(0)}_+ \,\tilde\cdot \,G^{(0)}_-}{24} \,.
\end{equation}
We note that when using this varying correction $\tau^{(1)}$ to calculate the full dilaton profile $\tau$ (cf.~\eqref{eq:1overcexpansion}), one may in principle encounter strong-coupling regions even if $\tau^{(0)}$ was chosen in the perturbative regime. At large $c$, which is our main case of interest, such effects are generically avoided.

For the internal metric $\tilde g^{(1)}_{mn}$ the mode decomposition is slightly more interesting, as the Lichnerowicz Laplacian \eqref{eq:metriclaplacian} allows for multiple independent zero-modes $\left(\psi_M\right)_{mn} \in \ker \Delta$.
On a Calabi-Yau background, it is well-known that they can be decomposed into K\"ahler and complex structure deformations.
Splitting the index $M$ into $M = (A,i)$, with $A = 1, \dots, h^{1,1}$ and $i = 1, \dots, h^{2,1}$, the modes corresponding to K\"ahler deformations are given by
\begin{equation}
\label{eq:wA}
    \left(\psi_A\right)_{\mu\bar \nu} = \left(\omega_A\right)_{\mu\bar \nu} \qquad \mathrm{for} \qquad \omega_A\in H^{1,1} \,.
\end{equation}
The modes corresponding to complex structure deformations are given by
\begin{equation}\label{eq:cszeromode}
    \left(\psi_i\right)_{\bar\mu \bar\nu} = \mathrm{e}^{K_\mathrm{cs}} \left(\chi_i\right)_{(\bar \mu}{}^{\bar\kappa \bar\lambda} \, \bar \Omega_{\bar \nu) \bar \kappa \bar\lambda} \qquad \mathrm{for} \qquad \chi_i \in H^{2,1} \,,
\end{equation}
together with their complex conjugates, where $\bar \Omega$ is the anti-holomorphic $(0,3)$-form.

With this preparation, the projection of \eqref{eq:eomg1} onto the KK modes reads
\begin{equation}\begin{split}
\label{eq:eomdeltag}
     \hspace{-.7cm}\Delta \tilde g^{(1)}_{mn} =  \frac{1}{8\Im\tau^{(0)}} \left(G_{+(m} ^{(0)~\widetilde{pq}} \bar G_{-n)pq}^{(0)} +   G_{-(m}^{(0)~\widetilde{pq}} \bar G_{+n)pq}^{(0)}\right)  \\
      + \frac{\tilde g^{(0)}_{mn}}{2} \left( \tilde{\mathcal{R}}_4 - \mathcal{C}_\Phi \right)  - \mathcal{C}^M \left(\psi_M\right)_{mn} . 
\end{split}\end{equation}
Expanding the K\"ahler form in terms of the K\"ahler parameters $t^A$ as $J = \omega_A t^A$, $A=1,\ldots, h^{1,1}$, we find that the constants $\mathcal{C}^A$ corresponding to K\"ahler deformations are given by
\begin{equation}\label{eq:cA}
    \mathcal{C}^A =  \frac{t^A}{2} \left( \tilde{\mathcal{R}}_4-\mathcal{C}_\Phi  \right) = - t^A \int \dd^6y\sqrt{\tilde g^{(0)}}  \frac{|\tilde G_-^{(0)}|^2}{48\Im\tau^{(0)}} \,.
\end{equation}
With this choice, the last two terms in \eqref{eq:eomdeltag} cancel exactly. The constants corresponding to complex structure deformations are given by
\begin{equation}
    \mathcal{C}^i = \frac{K^{i \bar\jmath}}{N} \int\dd^6y \sqrt{\tilde g^{(0)}} \frac{1}{8\Im\tau^{(0)}} \left(G_{+(m} ^{(0)~~\widetilde{pq}} \bar G_{-n)pq}^{(0)} +   G_{-(m}^{(0)~~\widetilde{pq}} \bar G_{+n)pq}^{(0)}\right) \left(\bar \psi_{\bar \jmath}\right)^{mn} \,,
    \label{eq:defci}
\end{equation}
with $K^{i \bar\jmath}$ the inverse Weil-Petersson metric \eqref{eq:WPmetric} on the complex structure moduli space, and $N$ a numerical normalisation factor.

However, as we will show below in Section~\ref{sec:matching10dand4d}, the contractions of $G_+^{(0)}$ and $G_-^{(0)}$ that appear in \eqref{eq:eomdeltag} are themselves in the kernel of the Lichnerowicz Laplacian. They can be uniquely decomposed into a sum of the complex structure deformations $(\psi_i)_{\bar\mu\bar\nu}$ and their complex conjugates, and do not contain any higher KK modes. Consequently, the right-hand side of \eqref{eq:eomdeltag} vanishes identically, and our projected equation of motion simply reads
\begin{equation}\label{eq:g1CYcondition}
    \Delta \tilde g^{(1)}_{mn}  = 0 \,.
\end{equation}
Therefore, even at the first sub-leading order $1/c$ in the large volume expansion the background remains Calabi-Yau. Without losing generality, we can absorb any non-trivial $\tilde g^{(1)}_{mn}$ into a redefinition of the leading order $\tilde g^{(0)}_{mn}$, and therefore set $\tilde g^{(1)}_{mn} = 0$.

In combination with \eqref{eq:G31selfduality}, this result also implies that the fluxes $G^{(1)}_{\pm}$ remain self-dual with respect to the Calabi-Yau metric $\tilde g^{(0)}_{mn}$ at order $1/c$. However, contrary to $G^{(0)}_{\pm}$ the corrections $G^{(1)}_{\pm}$ are generally not harmonic any more due to the non-vanishing derivatives of $\Phi^{(1)}_\pm$ and $\tau^{(1)}$ on the right hand side of \eqref{g1bia}.

\subsection{The four-dimensional curvature term and the Hamiltonian constraint}\label{sec:constraint}

Having established that the higher KK modes decouple at order $1/c$ and that the background remains Calabi–Yau at this level, we are now in a position to address a subtlety that plays an important role in connecting the ten- and four-dimensional descriptions for the light degrees of freedom: The appearance of the four-dimensional curvature term $\tilde{\mathcal{R}}_4$ and the associated Hamiltonian constraint as discussed, for example, in \cite{Giddings:2005ff,Douglas:2009zn,DeLuca:2021pej,Lust:2022xoq,Lust:2025vyz}.

As we just saw above, $\tilde{\mathcal{R}}_4$ contributes only to the zero-mode components that we have projected out.
We therefore do not need to specify a concrete value for $\tilde{\mathcal{R}}_4$ and can keep it undetermined at the level of our analysis.
Any change in $\tilde{\mathcal{R}}_4$ can readily be absorbed into a redefinition of the constant $\mathcal{C}_\Phi$, and does not change the equations for the higher KK modes.

We expect that if there is a value for the moduli fields at which their effective potential has a critical point, there is also a specific value for $\tilde{\mathcal{R}}_4$ so that all zero-mode components $\mathcal{C}_\Phi$, $\mathcal{C}_\tau$, $\mathcal{C}^i$, and $\mathcal{C}^A$ of the above equations vanish.
In this case, there exists a static solution of the full ten-dimensional equations of motion, including their zero-mode components.
Consistency of the four-dimensional EFT requires then that the critical value for $\tilde{\mathcal{R}}_4$, corresponding to the four-dimensional cosmological constant, agrees with the value of the effective potential at its critical point (up to a factor of ${\cal V}_{4, {\rm w}}$ that is needed to translate between ten and four-dimensional Planck units).

It was argued in the literature \cite{Giddings:2005ff, Douglas:2009zn} that a similar relation between $\tilde{\mathcal{R}}_4$ and the effective potential has to be imposed as a constraint on the off-shell configuration space.
This constraint equation is given by a particular linear combination of \eqref{eq:eomPhi} and the trace of \eqref{eq:eomR}, and corresponds to the Hamiltonian constraint in the Hamiltonian formulation of general relativity.
Since in our procedure we demand the KK components of all equations of motion to be solved, this applies in particular also to the non-trivial KK part of said constraint equation.%
\footnote{In \cite{Douglas:2009zn} it was argued that the Hamiltonian constraint is required to guarantee that the effective potential is bounded from below.
This is already achieved by imposing its KK component.}

If we would choose to also solve the zero-mode part of the constraint, we would find an additional relation between the constant $\mathcal{C}_\Phi$ and $\tilde{\mathcal{R}}_4$, that, at order $1/c$, is solved by\footnote{The corresponding equation can be obtained from the sum of the two equations in \eqref{eq:eomPhi} and $\frac14 \mathrm{e}^{4A}$ times the trace of \eqref{eq:eomR}.}
\begin{equation}
    \mathcal{C}_\Phi  = \frac32 \tilde{\mathcal{R}}_4  =  \int \dd^6y\sqrt{\tilde g^{(0)}}  \,\frac{|\tilde G_-^{(0)}|^2}{8\Im\tau^{(0)}}  \,.
\end{equation}
However, we would like to stress that, as explained above, imposing this additional relation is not required in our analysis.
Instead, $\tilde{\mathcal{R}}_4$, being the curvature of the four-dimensional metric, should rather be treated as a dynamical field in the four-dimensional EFT, along the same lines as the other light degrees of freedom that are not fixed by the ten-dimensional equations of motion.

We also note that an over-counting of the zero-mode components is avoided by imposing the normalisation condition $V_\mathrm{CY} = \frac16 \kappa_{ABC} t^A t^B t^C = 1$ on the internal Calabi-Yau volume.
Therefore, \eqref{eq:cA} reduces the number of independent zero-mode components by one, so that their total number matches precisely the number of independent light fields in the EFT, namely the Type~IIB axio-dilaton, the K\"ahler moduli, and the complex structure moduli of the internal Calabi-Yau geometry.
We elaborate further on this matching in the next subsection.

\subsection{Matching minimum conditions in 10d and 4d}
\label{sec:matching10dand4d}

In Section \ref{sec:lineom} we showed how to project out the zero-modes from the ten-dimensional equations of motion. The resulting equations can, in principle, be used to integrate out the massive KK modes, and to obtain a sensible lower-dimensional effective potential. We now turn to the remaining zero mode components of the equations of motion. As we will see, these cannot be solved fully unless the four-dimensional fields take values at a critical point of their effective potential. We demonstrate this in detail below, which allows us to identify the leading-order effective potential, and to verify that it correctly captures the low-energy effective physics.

It follows from \eqref{eq:Gharmonicleading} that the fluxes $G^{(0)}_\pm$ are harmonic three-forms with respect to the Calabi-Yau metric $g^{(0)}_{mn}$ on $X_6$.
Taking into account their self-duality, we can therefore expand them in a basis of harmonic forms as
\begin{equation}\label{eq:Gexpansion}
    G_+^{(0)} = A^i \chi_i + \bar B \bar\Omega \,, \qquad G_-^{(0)} = A\Omega +\bar B^{\bar \imath} \bar\chi_{\bar \imath} \,,
\end{equation}
with $A$, $A^i$, $\bar B^{\bar \imath}$, and $\bar B$ constant over $X_6$.

With the help of the identities given in Appendix~\ref{app:identities}, we can work out the contractions of $G^{(0)}_\pm$ that appear in the different equations of motion at order $1/c$.
Starting with \eqref{eq:eomdeltaphiM} for $\Phi^{(1)}_-$ we find
\begin{equation}
    G^{(0)}_- \tilde \cdot\, \bar G^{(0)}_- = 3! \| \Omega \|^2 \left( \left| A \right|^2 +   B^i \bar B_i \right) + F_{i\bar\jmath} \bar B^i B^{\bar \jmath} \,,
    \label{eq:G-G-}
\end{equation}
where $\|\Omega\|^2=\Omega_{\mu\nu\rho}\bar\Omega^{\mu\nu\rho}/3!$\,.
The first term is constant over $X_6$, and thus contains only the zero mode.
On the other hand, $F_{i\bar\jmath}$ integrates to zero over $X_6$.
It is hence orthogonal to the zero mode and contains only higher-order KK modes.
Therefore, \eqref{eq:G-G-} in combination with \eqref{eq:cPhi} yields
\begin{equation}\label{eq:cPhiresult}
    \mathcal{C}_{\Phi} - \tilde{\mathcal{R}}_4 
    = \frac{\mathrm{e}^{-K_\mathrm{cs}}}{4\, \Im \tau^{(0)}} \left( \left| A \right|^2 +   B^i \bar B_i \right) \,,
\end{equation}
and a similar result for the constants $\mathcal{C}^A$ in \eqref{eq:cA}.

Similarly, in \eqref{eq:eomdeltatau} for $\tau^{(1)}$ we have
\begin{equation}
    G_+^{(0)} \,\tilde\cdot\, G_-^{(0)} = 3! \| \Omega \|^2 \left( A \bar B + A^i \bar B_i \right) + F_{i\bar\jmath} A^i\bar B^{\bar\jmath} \,.
    \label{eq:G+G-}
\end{equation}
Again, the first term contributes only to the zero mode, and the second term only to the higher-order KK modes.
Inserting this back into \eqref{eq:tauconstant} and integrating over $X_6$ gives
\begin{equation}
    \mathcal{C}_{\tau} = - \frac{\mathrm{i}}{4} \mathrm{e}^{-K_\mathrm{cs}} \left( A \bar B + A^i \bar B_i \right) \,.
\end{equation}

Eventually, for the contractions of $G_\pm$ with open indices that appears in the equation of motion \eqref{eq:eomdeltag} for $\tilde g^{(1)}_{mn}$ we find
\begin{equation}
\begin{aligned}
& G_{+(\mu} ^{(0)~\widetilde{pq}} \bar G_{-\sigma)pq}^{(0)} +   G_{-(\mu}^{(0)~\widetilde{pq}} \bar G_{+\sigma)pq}^{(0)}
   = (\bar\chi_{\bar\imath})_{\bar\nu\bar\rho(\mu} \Omega_{\sigma)}{}^{\bar\nu\bar\rho} \left(B \bar B ^{\bar\imath} + A\bar A^{\bar\imath} -\mathrm{i} \mathrm{e}^{K_\mathrm{cs}} A^jB^k \kappa_{jk}{}^{\bar\imath}\right) .
    \label{eq:G+G-openindices}
\end{aligned}
\end{equation}
Importantly, the contraction of $\bar \chi_{\bar\imath}$ and $\Omega$ that appears on the right-hand side of this equation is the same as in \eqref{eq:cszeromode}.
It is a zero mode of the Laplacian operator \eqref{eq:metriclaplacian}.
Therefore, contrary to \eqref{eq:G-G-} and \eqref{eq:G+G-}, \eqref{eq:G+G-openindices} does not contain any higher-order KK modes.
This also allows us to read off the zero mode constants $\mathcal{C}^i$ directly from \eqref{eq:eomdeltag}, and we find
\begin{equation}
    \mathcal{C}^{\bar \imath} = \frac{\mathrm{e}^{-K_\mathrm{cs}}}{8\Im\tau^{(0)}} \left(B \bar B ^{\bar\imath} + A\bar A^{\bar\imath} -\mathrm{i} \mathrm{e}^{K_\mathrm{cs}} A^jB^k \kappa_{jk}{}^{\bar\imath}\right) \,.
\end{equation}

We now argue that the zero mode components of these equations vanish at the critical points of the four-dimensional effective potential.
This first requires identifying a suitable potential.
It is well known that the expansion coefficients in \eqref{eq:Gexpansion} can be expressed in terms of the GVW superpotential $W_\mathrm{GVW}(z^i, \tau)$ and its derivatives.
In terms of the three-form flux $G_3$, it is given by \cite{Gukov:1999ya}
\begin{equation}\label{eq:GVW}
    W_\mathrm{GVW} = \int G_3 \wedge \Omega\,.
\end{equation}
We have seen above that warping effects are negligible at leading order in $1/c$.
It is hence reasonable to expect that the leading order potential can be obtained from the standard K\"ahler potential without warping corrections.
Following \cite{Giddings:2001yu}, we have
\begin{equation}\label{eq:unwarpedkaehler}
    K = K_\mathrm{cs}(z^i, \bar z^{\bar \imath}) + K_\mathrm{ad}(\tau, \bar \tau) + K_\mathrm{K\ddot{a}hler}(T^A, \bar T^{\bar A})  \,,
\end{equation}
with
\begin{equation}
    K_\mathrm{cs} = -\ln\left( \mathrm{i} \int \Omega\wedge\bar\Omega \right) \,,\quad
    K_\mathrm{ad} = - \ln\left( -\mathrm{i}(\tau^{(0)}-\bar\tau^{(0)})\right) \,,
\end{equation}
and 
\begin{equation}
    K_\mathrm{K\ddot{a}hler} = - 3 \ln(c) - 2 \ln \left(\frac16 \int J \wedge J \wedge J \right) \,.
\end{equation}
Here $J$ and $\Omega$ denote the K\"ahler form and holomorphic 3-form with respect to the leading order Calabi-Yau metric $\tilde g^{(0)}_{mn}$. As before, we assume the volume of the internal Calabi-Yau metric to be normalised to one, corresponding to $\frac16\int J^3 = 1$. Therefore, to obtain the correct volume dependence in the potential, we include an additional $c$-dependent term in $K_\mathrm{K\ddot{a}hler}$, see for instance \cite{Martucci:2014ska}. 

This K\"ahler potential satisfies the no-scale condition $K_A K^{A\bar B} K_{\bar B} =3$ in the K\"ahler sector. 
In turn, the resulting $\mathcal{N}=1$ supergravity potential takes the familiar form 
\begin{equation}\label{eq:noscalepotential}
    V_\mathrm{flux} = \mathrm{e}^K K^{I \bar J} F_I \bar F_{\bar J}  \,,
\end{equation}
where $F_I=D_IW_\mathrm{GVW}$.
Here, the index $I=(i, \tau)$ runs over all complex structure moduli $z^i$ ($i=1, \dots, h^{2,1}$) as well as the axio-dilaton $\tau$. 

The derivative of $V_\mathrm{flux}$ with respect to the K\"ahler parameters\footnote{At the level of the classical theory, there is no potential generated for the $C_4$ axions and thus it suffices to take derivatives with respect to the K\"ahler parameters $t^A$.} $t^A$ is readily obtained, 
\begin{equation}\label{eq:kaehlerderivative}
    \partial_{t^{A}} V_\mathrm{flux} = (\partial_{t^{A}}K) V_\mathrm{flux} \,,
\end{equation}
or, for the universal volume modulus, $\partial_c V = -3 V/c$. As anticipated, a critical point with respect to the volume exists only if $V=0$, otherwise there is a run-away to $\mathcal{V} \rightarrow \infty$. On the other hand, the derivative of the potential \eqref{eq:noscalepotential} with respect to the complex structure and axio-dilaton directions can be compactly written as \cite{Denef:2004cf}
\begin{equation}
    \partial_I V_\mathrm{flux} = \mathrm{e}^K \left( Z_{IJ} \bar F^J + \overline W_\mathrm{GVW} F_I \right) \,,
    \label{eq:4dminimumcond}
\end{equation}
where $Z_{IJ}=D_IF_J$, and may allow for non-trivial critical points for certain values of $z^i$ and $\tau.$ 

To compare the derivatives \eqref{eq:kaehlerderivative} and \eqref{eq:4dminimumcond} of the potential with the zero-mode components of the ten-dimensional equations of motion, we insert the expansion \eqref{eq:Gexpansion} into the superpotential \eqref{eq:GVW} and find
\begin{equation}\begin{gathered}\label{eq:Wcoefficients}
    W_\mathrm{GVW} = \frac{1}{2} \mathrm{e}^{-K_\mathrm{cs}} \bar B \,, \qquad 
    F_i = \frac{1}{2} \mathrm{e}^{-K_\mathrm{cs}} K_{i\bar \jmath} \bar B^{\bar \jmath} \,, \qquad
    F_\tau =  \frac{ \mathrm{e}^{-K_\mathrm{cs}}}{ 2 \left(\bar \tau - \tau\right)} \bar A \,, \\
    Z_{ij} =  -\frac{\mathrm{i}}{2} \kappa_{ijk} A^k \,, \qquad
    Z_{\tau i} =  \frac{\mathrm{e}^{-K_\mathrm{cs}}}{ 2 \left(\bar \tau - \tau\right)} K_{i \bar \jmath}\bar A^{\bar \jmath} \,,\qquad
    Z_{\tau\tau} = 0 \,, \\
\end{gathered}\end{equation}
where the Yukawa couplings $\kappa_{ijk}$ are defined as
\begin{equation}
\label{eq:kappaijk}
    \kappa_{ijk} = -  \int \Omega \wedge D_i D_j D_k \Omega \,.
\end{equation}
Inserting this back into the potential \eqref{eq:noscalepotential}, we find
\begin{equation}
    V_\mathrm{flux}
    = \frac{\mathrm{e}^{- K_\mathrm{cs}}}{8 c^3 \, \Im \tau} \left( \left| A \right|^2 +   B^i \bar B_i \right) \,.
\end{equation}
Comparison with \eqref{eq:cPhiresult} and \eqref{eq:cPhi} shows that this potential agrees with
\begin{equation}\label{eq:G-potential}
    V_\mathrm{flux} = \frac1{c^3} \int \dd^6y\sqrt{\tilde g^{(0)}}  \,\frac{|\tilde G_-^{(0)}|^2}{48\Im\tau^{(0)}} \,.
\end{equation}
Again, by inspecting \eqref{eq:cPhiresult}, we can relate the volume derivative of the potential to the zero-mode of the equation of motion for $\Phi^{(1)}_\pm$,
\begin{equation}
    \partial_c V_\mathrm{flux} = -\frac{3}{2c^4} \left(\mathcal{C}_{\Phi} - \tilde{\mathcal{R}}_4 \right) \,.
\end{equation}
We can also relate the derivative of the potential with respect to an arbitrary K\"ahler direction to the zero-modes $\mathcal{C}^A$ in the equation of motion for $\tilde g^{(1)}_{mn}$,
\begin{equation}\label{eq:minkaehler}
    \partial_{t^{A}} V_\mathrm{flux} = \frac1{c^3} (\partial_{t^{A}}\partial_{t^{B}}K)  \mathcal{C}^B.
\end{equation}

Similarly, by inserting \eqref{eq:Wcoefficients} into \eqref{eq:4dminimumcond} we find that the $\tau$-component of the derivative of the potential is given by
\begin{equation}
\label{eq:mintau}
    \partial_\tau V_\mathrm{flux} = \frac{\mathrm{e}^{- K_\mathrm{cs}}}{8 \mathcal{V}^2 \, \Im \tau} \left( \bar A B + \bar A_i  B^i \right) = \frac{\mathrm{i}}{2 c^3 \, \Im \tau}
    \,\mathcal{C}_\tau \,,\\
\end{equation}
whereas the $i$-derivative becomes
\begin{equation}
    \partial_i V_\mathrm{flux} = \frac{\mathrm{e}^{- K_\mathrm{cs}}}{8 \mathcal{V}^2 \, \Im \tau} \left( A\bar A_{ i} + B\bar B_i - \mathrm{i}\mathrm{e}^{K_\mathrm{cs}} \kappa_{ijk} B^j A^k \right) = \frac1{c^3}\, \mathcal{C}_i \,,
    \label{eq:mincs}
\end{equation}
where indices are raised or lowered with the Weil-Petersson metric \eqref{eq:WPmetric}.

We have successfully demonstrated that critical points of the potential \eqref{eq:G-potential} are in one-to-one correspondence with solutions of the ten-dimensional equations of motion at order $1/c$.
The zero-mode components of these equations that obstruct their solvability are directly proportional to the first derivatives of the potential.
Static solutions therefore exist only if the derivatives of the potential vanish.

This correspondence is of course also reflected in the aforementioned agreement between the zero modes of the ten-dimensional fields and the light scalar fields in the effective, four-dimensional description.
The former comprise the constant part $\tau^{(0)}$ of the axio-dilaton in \eqref{eq:mintau}, the K\"ahler and complex structure deformations of the internal Calabi-Yau metric in \eqref{eq:minkaehler} and \eqref{eq:mincs}, as well as the zero-mode of the warp factor $\mathrm{e}^{-4A}$, given by $c$.
An over-counting between the K\"ahler moduli $t^A$ and the universal volume modulus $c$ is avoided by the normalisation constraint $V_\mathrm{CY} = 1$.
In \eqref{eq:cA} we found an analogous relation between the zero-mode in the warp-factor equation and K\"ahler modes in the equation of motion for the internal metric.

The potential suggested in \eqref{eq:G-potential} is, of course, very similar to the first term in the effective potential \eqref{eq:Vscaling}.
In the following, we are going to demonstrate that the remaining terms in \eqref{eq:Vscaling} are indeed sub-leading in our $1/c$ expansion, so that \eqref{eq:G-potential} manifests the correct effective potential at leading order in $1/c$.

\subsection{Inverse volume expansion of the scalar potential} \label{sec:scaling}

We are now in the position to insert our $1/c$ expansion \eqref{eq:1overcexpansion} of the ten-dimensional fields back  into \eqref{eq:Vscaling} to obtain an analogous inverse volume expansion of the effective potential,
\begin{equation}
    V_\mathrm{eff} = \frac{1}{{\cal V}_{4, {\rm w}}^2} \sum_{n=0}^\infty \frac{V^{(n)}}{c^n} \,.
\end{equation}
Here, the prefactor that we have decided to pull out of the expansion contributes a universal volume factor of
\begin{equation}
    \frac{1}{{\cal V}_{4, {\rm w}}^2} = \frac{1}{c^2} + \mathcal{O}\left(\frac{1}c\right)\,.
\end{equation}
From our expansion ansatz we immediately find
\begin{equation}
    V^{(0)} = 0 \,,
\end{equation}
implying that there is no potential in the $c \rightarrow \infty$ limit even before Weyl-rescaling the four-dimensional metric to the Einstein frame.
At the first non-trivial order we recover the familiar expression
\begin{equation}\label{eq:V1}
    V^{(1)} = \int \dd^6y\sqrt{\tilde g^{(0)}}\,\frac{|\tilde G_-^{(0)}|^2}{48\Im\tau^{(0)}} \, ,
\end{equation}
which is related to $V_{\mathrm{flux}}$ in Eq.~\eqref{eq:G-potential} via $V_{\mathrm{flux}} = V^{(1)}/c^3$.
This is of course consistent with our analysis in Section~\ref{sec:matching10dand4d}, where we determined that critical points of the same potential, given in \eqref{eq:G-potential}, are in one-to-one correspondence with solutions to the 10d equations of motion at the first sub-leading order in $1/c$.

Using our previous results, we can also give the next sub-leading correction to the potential,
\begin{equation}\begin{split}\label{eq:V2}
    V^{(2)} = \! \int \! \dd^6y \sqrt{\tilde g^{(0)}} \Biggl\{ & \! \frac{1}{96\Im\tau^{(0)}} \! \Biggl[\! \left( \!\Phi_+^{(1)}+\Phi_-^{(1)}-\frac{2 \Im\tau^{(1)}}{\Im\tau^{(0)}}\right) \! |\tilde G^{(0)}_-|^2 +4 \Re G^{(0)}_-\tilde\cdot\bar G_-^{(1)} \! \Biggr] \\
    & + \frac14 (\tilde\partial \Phi^{(1)}_-)^2 + \frac{|\tilde\partial\tau^{(1)}|^2}{4 \left(\Im\tau^{(0)}\right)^2}
    \! \Biggr\} \,,
\end{split}\end{equation}
where $\Phi_\pm^{(1)}$ and $\tau^{(1)}$ can be obtained from solving \eqref{eq:eomphi+1}, \eqref{eq:eomdeltaphiM} and \eqref{eq:eomdeltatau}, and $\tilde G^{(1)}_-$ is given as a solution of \eqref{g1bia}.

We note that only the first three terms in \eqref{eq:Vscaling} contribute to $V^{(2)}$.
The curvature term contributes only at order $V^{(4)}$.
This can be seen as follows:
First, we recall from \eqref{eq:g1CYcondition} that the first correction to the internal metric $\tilde g^{(1)}_{mn}$ remains Ricci-flat.
Therefore, the first non-vanishing contribution to the curvature term can only come from the second correction $\tilde g^{(2)}_{mn}$.
However, any first order variation of the Ricci-tensor of a Ricci-flat metric is just given by the Lichnerowicz Laplacian \eqref{eq:metriclaplacian}.
Its trace is a total derivative and does not contribute to the integral.%
\footnote{The same can also be seen by recalling that the first variation of the Einstein-Hilbert term is--up to a total derivative--given by the Einstein tensor which vanishes on a Ricci-flat background.}
Consequently, the first non-vanishing contribution to the curvature term must be quadratic in $\tilde g^{(2)}_{mn}$, and therefore scales like $c^{-4}$.

In the following, we would also like to determine the parametric scaling of $\Phi^{(1)}_\pm$, $\tau^{(1)}$, and $G^{(1)}_\pm$ in order to estimate the scaling of $V^{(2)}$.
The IASD component $G^{(0)}_-$ takes the role of a supersymmetry breaking parameter, that we, in the context of $F$-term uplifting, require to be small.
Therefore, we denote its parametric scaling by
\begin{equation}
    G_-^{(0)} \sim \varepsilon \,.
\end{equation}
On the other hand, the ISD component $G^{(0)}_+$ can be comparably large, and we write
\begin{equation}
    G_+^{(0)}\sim \sqrt{N} \,,
\end{equation}
so that, for $\varepsilon^2 \ll N$, the induced D3 charge of $G_3$ scales like $g_s N$.
Moreover, from above we have $\Im\tau^{(0)}=1/g_s$, $\Phi_-^{(0)}=0$, and $\Phi_+^{(0)}=2$. 

Using this notation, we readily obtain the following scaling for the leading order potential given in \eqref{eq:V1},
\begin{equation}
    V^{(1)} \sim g_s \varepsilon^2 \,.
\end{equation}
To establish the scaling behaviour of the first correction to the potential in \eqref{eq:V2}, we first need to determine the scaling of the first-order corrections to the fields.
From their equations of motion \eqref{eq:eomphi+1} -- \eqref{eq:eomdeltatau} we find that they scale like
\begin{equation}
    \Phi_+^{(1)} \sim g_s N\,, \qquad
   \Phi_-^{(1)} \sim g_s\,\varepsilon^2\,,\qquad \tau^{(1)} \sim \sqrt{N}\,\varepsilon \,,
    \label{eq:scalings}
\end{equation}
Inserting this into \eqref{g1bia}, we also obtain the scaling of the first correction to the fluxes,
\begin{equation}
    G_\pm^{(1)} \sim g_s N\,\varepsilon \,.
    \label{eq:scalingG1}
\end{equation}
With this preparation, we can employ an $\varepsilon$-expansion for $V^{(2)}$, and find to leading order
\begin{equation}
\begin{aligned}
\label{eq:V2eps}
    V^{(2)} = \! \int \! \dd^6y \frac{\sqrt{\tilde g^{(0)}}}{4\Im\tau^{(0)}} \! \left( \! \frac{\Phi_+^{(1)}}{24}  |\tilde G^{(0)}_-|^2 + \frac16 \Re G^{(0)}_-\tilde\cdot \,\bar G_-^{(1)} + \frac{|\tilde\partial\tau^{(1)}|^2}{\Im\tau^{(0)}} \right) + \mathcal{O}(\varepsilon^3) 
     \,,
\end{aligned}
\end{equation}
and therefore
\begin{equation}
    V^{(2)}\sim g_s^2 N\,\varepsilon^2 \, .
\end{equation}
We see that both $V^{(1)}$ and $V^{(2)}$ scale to leading order quadratically in $\varepsilon$.
In fact, one can easily argue that all corrections $V^{(n)}$ will scale at least quadratically in $\varepsilon$, 
and that there cannot be any linear terms in $\varepsilon$ in the potential.
This is trivially true for the first term in \eqref{eq:Vscaling}, since by definition $|G_-|^2 \sim \varepsilon^2$.
Moreover, any higher-order correction to $\Phi_-$, $\tau$, or $\tilde g_{mn}$ vanishes in the case $G_- = 0$, and must therefore be at least linear in $\varepsilon$.
However, the last three terms in \eqref{eq:Vscaling} are at least quadratic in the corrections.
This is immediately clear for the second and third terms, and was demonstrated to hold for the last term in the discussion below \eqref{eq:V2}.
Therefore, any correction originating from these terms will likewise scale at least quadratically in $\varepsilon$.
This, of course, reflects the fact that the full potential has a Minkowksi minimum at $G_- \sim \varepsilon = 0$.

Let us briefly summarise these results:
At leading order in $1/c$, the potential is given by (recall Eq.~\eqref{eq:G-potential})
\begin{equation}\label{eq:VGKP}
    V_\mathrm{flux} = \frac{1}{c^3} \int \dd^6y\sqrt{\tilde g^{(0)}}\,\frac{|\tilde G_-^{(0)}|^2}{48\Im\tau^{(0)}} \sim \frac{g_s \varepsilon^2}{c^3} \,,
\end{equation}
where, for completeness, we have displayed again the leading scaling behaviour with the fluxes.
We also determined the first sub-leading correction to the potential which it is given by
\begin{equation}
\label{eq:deltaVeff_summary}
    \delta V_\mathrm{warp} = \frac{1}{c^4} \left(V^{(2)} -2  V^{(1)} \delta {\cal V}_{4, {\rm w}} \right) \sim \frac{g_s^2 N \varepsilon^2}{c^4} \,,
\end{equation}
with $V^{(1)}$ and $V^{(2)}$ given in \eqref{eq:V1} and \eqref{eq:V2eps}, and 
where $\delta {\cal V}_{4, {\rm w}}$ denotes the first correction to the warped volume \eqref{eq:Vwarped}.%
\footnote{By expanding ${\cal V}_{4, {\rm w}}$ in $1/c$ by writing ${\cal V}_{4, {\rm w}}=c+\delta {\cal V}_{4, {\rm w}}$, one finds 
\begin{equation}
    \delta {\cal V}_{4, {\rm w}} = -\frac12\int d^6y \sqrt{\tilde g^{(0)}} \left(\Phi_+^{(1)}+\Phi_-^{(1)} \right) \sim g_s N \,.
\end{equation}
Thus, both terms in \eqref{eq:deltaVeff_summary} have the same leading scaling behaviour in $g_s N$.
}
In particular, we see that the first correction to the potential is suppressed as
\begin{equation}\label{eq:correctionssuppression}
    \frac{\delta V_\mathrm{warp}}{V_\mathrm{flux}} \sim \frac{g_s N}{c}
\end{equation}
compared to the leading order term.
The correction is hence negligible in the dilute flux regime where $g_s N \ll c$.
However, in regions with strong warping, such as at the tip of a Klebanov-Strassler throat, it may become relevant.

The leading order potential $V_\mathrm{flux}$ in \eqref{eq:VGKP} is of course the familiar expression for the flux potential in the unwarped case \cite{Giddings:2001yu}.
It is well-known that this potential can be recast into a manifestly four-dimensional $\mathcal{N}=1$ supergravity formulation in terms of the superpotential \eqref{eq:GVW} and the K\"ahler potential \eqref{eq:unwarpedkaehler}.
In the next section, we will discuss how to implement warping-corrections to the potential in the $\mathcal{N}=1$ supergravity formalism.
Moreover, the implications of these corrections on KKLT and LVS will be discussed in detail in Section \ref{sec:implications}. 

\section{Four-dimensional perspective}\label{sec:4d}

In addition to the 10d analysis, it is obviously desirable to consider the 4d $\mathcal{N}=1$ supergravity perspective. The use of a supersymmetric EFT needs justification since SUSY is broken both by ISD and non-ISD fluxes. It is well-known, see e.g.~footnote 1 in~\cite{Burgess:2005jx}, that a SUSY EFT description is valid if the SUSY-breaking mass-splittings are small compared to the mass gap characterizing the heavy states which are integrated out, i.e.~the KK modes. In our case, this is guaranteed since all such mass splittings are induced by 3-form-flux, which dilutes as $1/R^3$, with $R\sim {\cal V}^{1/6}$. By contrast, the KK masses are induced by gradient terms in the Calabi-Yau, which scale only as $1/R$.

After these preliminary remarks, we start with the simplest non-trivial setting:

\subsection{A simple model with two moduli}\label{sec:4dsinglemod}

Consider first the case of a single K\"ahler and a single complex structure modulus,
\begin{equation}
    K^{(0)}=K^{(0)}_{k}(T,\ol{T}) + K_{\mathrm{cs}}(z,\ol{z})\, ,
\end{equation}
with 
\begin{equation}\label{bkpo}
    K^{(0)}_{k}=-3\ln(T+\ol{T}) \quad,\qquad K_{\mathrm{cs}}=-\ln\biggl (\mathrm{i}\int \Omega\wedge \ol{\Omega}\biggl ) \,.
\end{equation}
For simplicity, we neglect the axio-dilaton in this toy example, so that the main content of our idea is more transparent.
It was noted in \cite{Frey:2008xw} and argued in more detail in \cite{Martucci:2014ska}
that in this simple case warping corrections are encoded in a $z$-dependent additive shift of $T+\ol{T}$, such that the warped model is characterised by
\begin{equation}
    K=K_{k}(T,\ol{T},z,\ol{z}) + K_{\mathrm{cs}}(z,\ol{z})
    \label{kkomb}
\end{equation}
with
\begin{equation}
    K_k(T,\ol{T},z,\ol{z})=-3\ln\big(T+\ol{T} + f(z,\ol{z})\big)\,.
    \label{kkfu}
\end{equation}
The arguments of \cite{Frey:2008xw,Martucci:2014ska} do not involve complex structure dynamics, so it has not been demonstrated that \eqref{kkomb} and \eqref{kkfu} fully characterise a model where both K\"ahler and complex structure moduli are dynamical. For now, we make this assumption and work out the resulting scalar potential 
\begin{equation}\label{eq:FtermPotentialGVW}
    V = \mathrm{e}^K \left( K^{M\overline N} (D_M W_\mathrm{GVW})  (D_{\overline N} \overline W_\mathrm{GVW}) -3|W_\mathrm{GVW}|^2 \right)\,.
\end{equation}
Here $M,N$ run over all moduli, for the moment just $T$ and $z$, and $W_\mathrm{GVW}$ is the Gukov-Vafa-Witten superpotential \cite{Gukov:1999ya}
\begin{equation}\label{eq:ansatzW}
    W_\mathrm{GVW}(z)=\int G_3\wedge \Omega\, .
\end{equation}
We proceed by expressing the K\"ahler metric
\begin{equation}\label{eq:KMN_propf}
    K_{M\overline{N}} = \dfrac{3}{\mathcal{T}^2}
    \left (
        \begin{array}{cc}
            \; 1 & f_{\bar{z}} \\[0.5em]
            \;  f_{z} & \quad |f_z|^2 + \tfrac{1}{3}\mathcal{T}\mathcal{A}
        \end{array} 
    \right )\quad\,\,\mbox{with}\qquad \mathcal{A} = \mathcal{T}\partial_{z}\partial_{\bar{z}}K_{\mathrm{cs}}-3f_{z\bar{z}}\, ,
\end{equation}
in terms of
\begin{equation}
    \mathcal{T} = T+\bar{T}+f(z,\bar{z}) \,\,\, ,\qquad f_z=\partial_z f\,\,\, ,\qquad f_{\bar{z}}=\partial_{\bar{z}} f\,\,\, ,\qquad f_{z\bar{z}}=\partial_z\partial_{\bar{z}} f\, .
\end{equation}
Its inverse reads
\begin{equation}\label{eq:InvKM1}
    K^{M\overline{N}} = \dfrac{\mathcal{T}}{3\mathcal{A}}\, \left (
        \begin{array}{cc}
          \mathcal{A}\mathcal{T}+3|f_z|^2  & \,\,\,\,-3f_z\\[0.6em]
           -3f_{\bar{z}} &  3
        \end{array} 
    \right )\,.
\end{equation}
Note that, though not apparent in the present form, in the limit Re$T\to \infty$ this metric approximates a direct product of K\"ahler and complex structure moduli spaces.

When working out the contractions in 
\begin{equation}
    K^{M\overline{N}}\, K_M\, K_{\overline{N}} = 3+\dfrac{\mathcal{T}}{\mathcal{A}}\, (\partial_z K_{\mathrm{cs}})(\partial_{\bar{z}} K_{\mathrm{cs}})\, ,
    \label{contr}
\end{equation}
various cancellations occur such that the result remains simple.\footnote{Similar observations were made previously in \cite{Blumenhagen:2019qcg}, although in a linearised treatment and with a specific form for $f(z,\bar{z})$ that arises in the conifold case.}
Using this formula in the standard expression \eqref{eq:FtermPotentialGVW} for the full potential, one arrives at the final result
\begin{equation}\label{eq:potentialCS}
    V =\mathrm{e}^{K}K^{z\bar{z}} \, \bigl |(D_zW_\mathrm{GVW})^{(0)}\bigl |^2\,.
\end{equation}
Here $K$ is the full K\"ahler potential and $K^{z\ol{z}}$
(cf.~\eqref{eq:InvKM1}) is the complex structure part of the full inverse K\"ahler metric
\begin{equation}\label{eq:KZZInv}
    K^{z\bar{z}} = \dfrac{\mathcal{T}}{\mathcal{A}} = \dfrac{T+\ol{T}+f}{(T+\ol{T}+f)\partial_{z}\partial_{\bar{z}}K_{\mathrm{cs}}-3f_{z\bar{z}}}\,.
\end{equation}
However, crucially, $(D_zW_\mathrm{GVW})^{(0)}$ is the {\it uncorrected} complex structure $F$-term,
\begin{equation}
    (D_zW_\mathrm{GVW})^{(0)} = \partial_z W_\mathrm{GVW}+(\partial_z K_{\mathrm{cs}})W_\mathrm{GVW}\, .
\end{equation}
As expected, the term $\sim -3|W_\mathrm{GVW}|^2$ drops out, cf.~\eqref{contr}.
We emphasise that the global minima of the potential are still determined by the solutions of $(D_zW_\mathrm{GVW})^{(0)}=0$. However, warping corrections affect the loci of minima of $V$ with non-vanishing $F$-terms $(D_zW_\mathrm{GVW})^{(0)}\neq 0$.
By expanding \eqref{eq:KZZInv} and $\mathrm{e}^{K}$ to linear order in $1/c=2/(T+\overline T)$, we find
\begin{equation}\label{eq:potentialCS1}
    V =\mathrm{e}^{K^{(0)}}K_{\mathrm{cs}}^{z\bar{z}}\biggl (1-\dfrac{3(f-K_{\mathrm{cs}}^{z\bar{z}}f_{z\bar{z}})}{T+\ol{T}}+\ldots \biggl ) \, \bigl |(D_zW_\mathrm{GVW})^{(0)}\bigl |^2\,,
\end{equation}
where $K_{\mathrm{cs}}^{z\bar{z}}=(\partial_{z}\partial_{\bar{z}}K_{\mathrm{cs}})^{-1}$ is the leading-order inverse metric.
The two terms scale parametrically like
\begin{equation}\label{eq:potentialCS2}
    V \sim \;\dfrac{g_s\varepsilon^2}{c^3}\; +\;\dfrac{g_s\varepsilon^2 (f-K_{\mathrm{cs}}^{z\bar{z}}f_{z\bar{z}})}{c^4}\;  + \;\ldots \,.
\end{equation}
Here, the second term should match \eqref{eq:deltaVeff_summary} which suggests that $f,f_{z\bar{z}}$ should scale like $f,f_{z\bar{z}} \sim g_sN$. This is in fact in agreement with \cite{Martucci:2014ska} where it was argued that schematically
\begin{equation}
    f \sim  \int \frac{\mathrm{i}\,G_3 \wedge \bar G_3}{2\Im\tau}+\ldots \sim g_s N+g_s \varepsilon^2+\ldots\, .
\end{equation}

\subsection{Towards the general case} \label{sec:K4dgeneral}

In retrospect, the simple result just found is not too surprising since we knew from GKP that warping respects the no-scale structure. Even more: The 10d analysis of GKP shows that the SUSY locus in complex structure moduli space, i.e.,~the locus where complex structure $F$-terms and hence the scalar potential vanish, is not modified by warping.

One may then hope that the discussion of the previous subsection generalises as follows: Consider the case with multiple K\"ahler and complex structure moduli. Assume that warping corrections manifest themselves only in a complex structure-dependent shift of 
real 4-cycle volumes, such as in the in transition from $K_k^{(0)}$ in \eqref{bkpo} to $K_k$ in \eqref{kkfu}. In other words, let us consider a supergravity model with shift symmetric (K\"ahler) moduli $T^A$ and (complex structure) moduli $z^i$ in which the K\"ahler potential takes the form
\begin{equation}\label{eq:Kk}
    K[T^A,\overline{T}^{A},\phi^I,\bar{\phi}^I] = K_k[T^A+\overline{T}^{A}+f^A(\phi^I,\bar{\phi}^I)] + K_{\mathrm{cs}}[z^i,\bar{z}^i]+K_{\mathrm{ad}}[\tau,\bar{\tau}]\,.
\end{equation}
Here $\phi^I=(z^i,\tau)$ denotes the complex structure moduli $z^i$ and the axio-dilaton $\tau$, and $K_k$ takes the standard no-scale form. By this we mean that $K_k=-2\ln{\cal V}$ with ${\cal V}$ being homogeneous of degree 3/2 in its arguments ${\cal T}^A\equiv T^A+\overline{T}^{A}+f^A(\phi^I,\bar{\phi}^I)$. This last point is, of course, not the critical part of the assumption -- this is just the standard K\"ahler moduli dependence of Calabi-Yau 3-folds. The critical part is that, motivated by the findings of \cite{Martucci:2014ska, Martucci:2016pzt}, we introduced corrections which are parametrised by the $f^A$ and represent complex structure dependent additive shifts of the 4-cycle volumes $T^A+\ol{T}^{\ol A}$. We conjecture that this encodes the complete effect of warping. 

Once this assumption is made or, if one is instead interested in a model of this type purely from the 4d supergravity perspective, one may work out the scalar potential \eqref{eq:FtermPotentialGVW}. The indices run over all moduli, i.e.,~$M=\{A,I\}$. The calculation, which we report in some detail in App.~\ref{app:many}, is similar to analyses discussed in \cite{Burgess:2020qsc} (see also \cite{Grimm:2004uq}), where closely related structures have been discussed. We find that the result takes precisely the form observed in the simpler case in the previous subsection:
\begin{equation}
    V = \mathrm{e}^K K^{I\bar J} (D_I^{(0)} W_\mathrm{GVW})\, \overline{(D_J^{(0)} W_\mathrm{GVW})}\, .
    \label{fullp}
\end{equation}
Crucially, as before, $K$ is the full K\"ahler potential and $K^{I\bar J}$ is the complex structure and axio-dilaton part of the full inverse K\"ahler metric $K^{M\ol{N}}$. By contrast,
\begin{equation}
    D_I^{(0)}W_\mathrm{GVW}=\partial_I W_\mathrm{GVW} + (K_{\mathrm{cs}}+K_{\mathrm{ad}})_I W_\mathrm{GVW}
\end{equation}
are the {\it uncorrected} complex structure and axio-dilaton $F$-terms, without any involvement of the functions $f^A$ which we would like to think of as warping corrections.

We emphasise again that the above is strictly speaking merely an intriguing observation in the context 4d supergravity models. Claiming that this is the general form of the warping-corrected scalar potential for Calabi-Yau orientifolds would require establishing our assumptions. While we are unable to do so at the moment, we should quote the results of \cite{Martucci:2014ska, Martucci:2016pzt} (see also the brief summary given in \cite{Gao:2022uop}) in support of this proposal. There, it has been shown that if the complex structure moduli are treated as fixed, e.g.,~because they have been integrated out, then the warping corrections to the K\"ahler moduli K\"ahler potential do indeed take precisely the form just discussed: The four-cycle variables are shifted by an explicitly calculable and complex structure-dependent geometrical quantity. This is consistent with our proposal that this form is valid even before complex structure stabilisation.

Our proposal is furthermore in agreement with the results found in Section \ref{sec:10danalysis}: When expanding the potential \eqref{fullp} in $1/c$, the leading correction is suppressed by $g_sN/c$ and is obtained when expanding the inverse metric.

\section{Implications for moduli stabilisation} \label{sec:implications}

\subsection{Including (non-)perturbative corrections to the scalar potential} \label{sec:qcorrections}

So far, we have systematically analysed all contributions to the flux-induced scalar potential following from the leading-order 10d Type~IIB effective action including warping effects. As a key result, all terms that we found are at least quadratic in the IASD flux $G_-$. In other words, no term linear in $G_-$ has emerged.

At this level of the analysis, the K\"ahler moduli are either flat (for $G_-=0$) or runaway directions. To achieve full moduli stabilisation, perturbative and non-perturbative corrections in $\alpha'$ and $g_s$ have to be taken into account, as done e.g.~in the KKLT~\cite{Kachru:2003aw} and LVS~\cite{Balasubramanian:2005zx} proposals. Once such corrections are included, terms linear in $G_-$ will arise. It is the goal of this subsection to identify the dominant contributions to the scalar potential that are of this type. We explicitly do {\it not} discuss all the standard correction terms used in KKLT and LVS as long as they are independent of $G_-$.

Since quantum corrections are incorporated most easily in the 4d $\mathcal{N}=1$ supergravity approach, we will use the corresponding standard formula for the scalar potential in the following:
\begin{equation}
\label{eq:FtermPotential}
    V = \mathrm{e}^K \left( 
    K^{M\overline N} D_M W \bar D_{\overline{N}} \overline W -3|W|^2 \right)\,.
\end{equation}
Here $M,\overline N$ run over all moduli. We write the K\"ahler potential and superpotential as
\begin{equation}
    K = K_\mathrm{cl} + \delta K\qquad\mbox{and}\qquad W = W_\mathrm{GVW} +\delta W\,,
    \label{cordef}
\end{equation}
where $K_\mathrm{cl}$ and $W_\mathrm{GVW}$ encode the classical contributions (in the case of $K_\mathrm{cl}$, including warping) whereas $\delta K$ and $\delta W$ are induced by quantum corrections.  In particular, $\delta W$ arises due to gaugino condensation or Euclidean D3-brane (ED3) instantons:
\begin{equation}
    \delta W = \sum\limits_B A_B(z^i,\tau) \, \exp(-2\pi a_B\,T^B)\,.
    \label{npef}
\end{equation}
Here $A_B(z^i,\tau)$ is the Pfaffian, depending in general on the complex structure moduli and the axio-dilaton.\footnote{
A $\tau$ dependence arises e.g.~from fluxed instantons. Since we do not expand in $g_s$, in our analysis this effect is not sub-leading.
}
For an ED3 instanton effect or gaugino condensation on an $SO(8)$ stack of D7-branes one has $a_B=1$ and $a_B=1/6$, respectively.
Note that the full expression for $K_\mathrm{cl}$ for generic warped compactifications is difficult to derive, as we have explained in Section \ref{sec:4d}.

Let us pause to comment on the validity of the non-renormalization theorem for the superpotential $W$ in our setting as studied in \cite{Burgess:2005jx}. As the authors of \cite{Burgess:2005jx} emphasise, their theorem is in general not valid for varying axio-dilaton. However, while such a variation does indeed arise in our context, this happens only at sub-leading order in $c$ (cf.~\eqref{eq:1overcexpansion}). In other words, the variation of $\tau$ is suppressed by the volume and hence does not appear at leading order in $\alpha'$. As a result, the non-renormalization proof of \cite{Burgess:2005jx} remains applicable at this order. Then the rest of the proof also goes through, the key insight being that Kähler moduli cannot appear in $W$ and hence effects of higher order in $\alpha'$ are absent. The only exception are non-perturbative effects, as in \eqref{npef}, where Kähler moduli can appear in the exponent, consistently with their gauged discrete shift symmetry. While the prefactor $A$ can involve $\tau$, a sensitivity to its variation is again excluded since this would come with a power-like dependence on Kähler moduli.

The contributions to $V$ that we are interested in are the `mixed' terms, involving a quantum correction and a factor linear in the IASD flux $G_-$. The latter is proportional to $D_I W_{\rm GVW}$, with the index $I=(i,\tau)$ representing a complex structure modulus $z^i$ or the axio-dilaton $\tau$. Thus, these contributions take the form 
\begin{equation}
    \label{eq:deltaVmix}
    \delta V_\mathrm{mix} = \left[   
    \mathrm{e}^K K^{M\bar I}\left(\partial_M \delta W +\delta K_M W_\mathrm{GVW}\right)
    +\delta\left(\mathrm{e}^K K^{A\ol I}\right) D_A W_{\rm GVW}
    \right] \bar D_{\bar I} \overline{W}_{\mathrm{GVW}} + \mathrm{c.c.},
\end{equation}
where $\delta K_M \equiv \partial_M \delta K$, and $\delta(\exp(K)K^{A\bar I})$ denotes the leading correction to $\exp(K)K^{A\bar I}$ induced by quantum corrections. As before, $M=(A,I)$ runs over all moduli, $I$ labels the complex structure moduli and the axio-dilaton, and $A$ the K\"ahler moduli.

Let us analyse \eqref{eq:deltaVmix} term by term, starting with the first term where we first choose $M$ to be a complex structure modulus or $\tau$. With $K^{\tau \bar\imath}=0$ at leading order, we estimate parametrically\footnote{Here, on the right hand side, the index structure is merely symbolic. In other words, there are exponential terms in $T^A$, for different $A$, and there are terms involving $\delta K_{i,\tau}$, for different complex moduli $z^i$ or the axio-dilaton $\tau$. \label{istr}
}
\begin{align}
\label{eq:deltaVtau}
    \mathrm{e}^K K^{i\bar \jmath}_{\rm cl}\left(\partial_i \delta W +\delta K_i W_\mathrm{GVW}\right) \bar D_{\bar \jmath} \overline{W}_{\mathrm{GVW}} & \sim  \frac{g_s\,\varepsilon}{\mathcal{V}^2}\left( \mathrm{e}^{-2\pi a_A T^A}+\delta K_i W_\mathrm{GVW}\right) \,,\\
    \label{eq:deltaVi}
    \mathrm{e}^K K^{\tau\bar \tau}_{\rm cl}\left(\partial_\tau \delta W +\delta K_\tau W_\mathrm{GVW}\right) \bar D_{\bar \tau} \overline{W}_{\mathrm{GVW}} & \sim \frac{g_s\,\varepsilon}{\mathcal{V}^2} \left(\mathrm{e}^{-2\pi a_AT^A}+\delta K_\tau W_\mathrm{GVW}\right) \,.
\end{align}
To arrive at this result, we used
$D_i W_\mathrm{GWV}\sim G_-^{(0)}$ and 
$D_\tau W_\mathrm{GWV}\sim g_s G_-^{(0)}$, which is justified since we are only interested in the dominant effect and can hence employ the unwarped Kähler metric in the covariant derivatives. We also used 
$|G_-^{(0)}|\sim\varepsilon$, $K^{\tau\bar\tau}_{\rm cl}\sim 1/g_s^2$, $K^{i\bar\jmath}_{\rm cl}\sim \mathcal{O}(1)$, $\partial_i(\delta W,\delta K)\sim (\delta W,\delta K)$, and $\partial_\tau (\delta W,\delta K) \sim g_s (\delta W,\delta K)$.

Next, we choose $M$ to be a K\"ahler modulus in the first term in \eqref{eq:deltaVmix}. Parametrically, the corresponding contribution reads
\begin{equation}
    \mathrm{e}^K K^{A\bar I}_{\rm cl} \left(\partial_A \delta W  +   \delta K_A W_\mathrm{GVW}\right) \bar D_{\bar I} \overline W_\mathrm{GVW}  \sim K^{A\bar I}_{\rm cl}\,\frac{g_s\,\varepsilon}{\mathcal{V}^2} \left(\mathrm{e}^{-2\pi a_A T^A}+\delta K_A W_\mathrm{GVW}\right) ,
\end{equation}
where we assumed $\partial_A\delta W\sim \delta W$. One may naively think that this is sub-leading compared to \eqref{eq:deltaVtau} and \eqref{eq:deltaVi} since $K^{A\bar I}_{\rm cl}=0$ without warping.
However, this is not the case. Indeed, due to warping effects, the complex structure and K\"ahler sectors get mixed and hence $K^{A\bar \imath}_{\rm cl}\neq0$, as discussed in Section~\ref{sec:4d}. Since the precise form of the warped K\"ahler potential is not known, the best we can do is to estimate $K^{A\bar \imath}_{\rm cl}$ using the ansatz \eqref{eq:Kk} for the K\"ahler potential from Section~\ref{sec:K4dgeneral}.
The results \eqref{eq:KAj} and \eqref{eq:KiB} then imply
\begin{equation}
\label{eq:mixedinversemetric}
    K^{A\bar\jmath}_{\rm cl} \sim K^{i\bar B}_{\rm cl} \sim \mathcal{O}(1)\,.
\end{equation}

Finally, we consider the second term in \eqref{eq:deltaVmix}. We start by rewriting it as\footnote{
Note 
that, in the SUSY vacuum underlying KKLT, this term vanishes at leading order since the K\"ahler moduli are stabilised by $D_A W\approx 0$. However, here our approach is to consider corrections on the basis of GKP. From this perspective, $D_A W \simeq (\partial_A K_{\rm cl})W_{\rm GVW}$
is a leading-order, non-zero expression.
}
\begin{equation}
   \left(\mathrm{e}^{K_{\rm cl}} \delta K^{A\bar I} +\delta \left( \mathrm{e}^K \right) K^{A\bar I}_{\rm cl}\right) D_AW_{\rm GVW} \bar D_{\bar I} \overline{W}_{\rm GVW} +\mathrm{c.c.}
   \label{eq:dVmixlast}
\end{equation}
Here $\delta K^{A\bar I}$ is the leading quantum correction to the inverse metric, which may be written as
\begin{equation}
\label{eq:Kinverseleadingcorr}
    \delta K^{A\bar I} = -  K_\mathrm{cl}^{A\bar M} \delta K_{\bar M N} K_\mathrm{cl}^{N\bar I} +\mathcal{O}\left((\delta K)^2\right)\,.
\end{equation}
The first term in \eqref{eq:dVmixlast}, which is also the leading contribution, may then be estimated as
\begin{equation}
    \mathrm{e}^K K_\mathrm{cl}^{A\bar B} \delta K_{\bar B j} K_\mathrm{cl}^{j\bar \imath} (\partial_A K_\mathrm{cl})W_{\rm GVW}\, G_-^{(0)} \lesssim \frac{g_s \varepsilon\,\Re(\bar T^{\bar B})}{\mathcal{V}^{2}} W_{\rm GVW}\,\delta K_{\bar B j}\,,
    \label{dkcd}
\end{equation}
where we used that $K^{j\bar\imath}_{\rm cl}\sim \mathcal{O}(1)$. 
Furthermore, we used the relation (as derived e.g.~in~\cite{Cicoli:2007xp}) $K^{A\bar B}_{\rm cl}\partial_A K_\mathrm{cl}=-2\Re(\bar T^{\bar B})$, where $T^{B}$ is the complexified 4-cycle K\"ahler modulus. 

The scaling of the second term in \eqref{eq:dVmixlast} may be estimated as
\begin{equation}
    \mathrm{e}^{K_\mathrm{cl}} \delta K K^{A\bar\imath}_{\rm cl} \partial_AK_\mathrm{cl} W_\mathrm{GVW} \,G_-^{(0)} \lesssim \frac{g_s\varepsilon}{\mathcal{V}^2}\delta K\,W_\mathrm{GVW}\,.
    \label{dkc}
\end{equation}
Here we used  \eqref{eq:mixedinversemetric} and the fact that $\partial_A K_\mathrm{cl}< {\cal O}(1)$ in the geometric regime. Using also that $\delta K< {\cal O}(1)$, we see that \eqref{dkc} is parametrically smaller or at most comparable with respect to the second term in \eqref{eq:deltaVtau} and \eqref{eq:deltaVi}. It may hence be disregarded.

Summarising the results of this section, we have found that the leading quantum corrections to $V$ which are linear in $G_-$ scale as 
\begin{equation}
\label{eq:deltaVmix_summary}
    \delta V_\mathrm{mix} \lesssim \frac{g_s\,\varepsilon}{\mathcal{V}^2} \left( \mathrm{e}^{-2\pi a_A T^A} + \delta K_MW_\mathrm{GVW} + \Re(\bar T^{\bar B})W_\mathrm{GVW}\,\delta K_{\bar B j}\right)\,.
\end{equation}
Here, as explained in footnote~\ref{istr}, the index structure is symbolic.

To identify the dominant term in \eqref{eq:deltaVmix_summary}, we need to determine the leading contributions to $\delta K_M$ and $\delta K_{\bar B j}$.
We repeat that $\delta K$ does not include warping effects which are already taken into account in Section~\ref{sec:10danalysis}.
It is known from the literature \cite{vonGersdorff:2005bf,Berg:2005ja,Berg:2005yu,Berg:2007wt,Cicoli:2007xp,Cicoli:2021rub,Gao:2022uop} that the leading terms in $\delta K$ are homogeneous functions of degree $-1$ in 4-cycle K\"ahler moduli. They are of the form $\delta K_\mathrm{loc} \sim g_s \mathcal{I}(t^A,z^j,\bar z^{\bar\jmath})/\mathcal{V}$, where $\mathcal{I}(t^A,z^j,\bar z^{\bar\jmath})$ is a homogeneous function of degree $1$ in (real) 2-cycle K\"ahler moduli $t_A$ with a complicated complex structure moduli dependence.
From the EFT perspective, such terms arise due to $g_s$ suppressed higher-order $\alpha'$ corrections on D7-branes/O7-planes \cite{Gao:2022uop} and are therefore called local $\alpha'$ corrections in \cite{Gao:2022uop}.\footnote{Note that in the literature they are also called KK-type loop corrections as suggested in \cite{Berg:2007wt}.} Even though at the time of writing no example is known where $\mathcal{I}(t^A,z^j,\bar z^{\bar\jmath})$ contains higher order ratios of 2-cycles it can also not be excluded. 
Now, we want to estimate the largest contributions of the last two terms in \eqref{eq:deltaVmix_summary}. For the term $\sim \delta K_M$, the largest contribution is obtained when choosing $M=i$. One then finds the scaling
\begin{equation}
    \label{eq:diKKK}
    \partial_i\delta K_\mathrm{loc} \sim\partial_i\frac{g_s\mathcal{I}(t^A,z^j,\bar z^{\bar\jmath})}{\mathcal{V}} \,, \qquad\qquad\partial_{\bar B}\partial_j \delta K_\mathrm{loc} \sim \partial_{\bar B}\partial_j\frac{g_s\mathcal{I}(t^A,z^j,\bar z^{\bar\jmath})}{\mathcal{V}}\, .
\end{equation}

In cases of fibered geometries, loop corrections are particularly dangerous as also noted in \cite{Gao:2022uop}. The reason is that they do not necessarily feature a suppression by the overall volume but instead by some power of some smaller cycle. Take for instance the case of a K3 fibration, with volume 4-cycle $\tau_f$, over a $\mathbb{CP}^1$ with volume 2-cycle $t_b$. The volume is then given by $\mathcal{V}\sim \sqrt{\tau_f}\tau_2$, where $\tau_2\sim t_b\sqrt{\tau_f}$. Having two brane stacks intersecting on the 2-cycle volume of the base leads to an Einstein-Hilbert term induced at 1-loop order \cite{Epple:2004ra,Haack:2015pbv}. This in turn implies a corrections of the form (neglecting complex structure moduli dependence) $\delta K_\mathrm{loc}\sim g_s t_b/\mathcal{V}\sim g_s/\tau_f$. In cases of a small fiber this correction is large and, a priori, dangerous for any moduli stabilisation scenario independently of the uplift. 

In the following, we specify to swiss-cheese (sc) type geometries which are commonly used in LVS. In this case, one finds 
\begin{equation}
\label{eq:diKKK_swiss}
    \partial_i\delta K_\mathrm{loc}^\mathrm{(sc)} \lesssim \frac{g_s}{\mathcal{V}^{2/3}} \,, \qquad\qquad\partial_{\bar B}\partial_j \delta K_\mathrm{loc}^\mathrm{(sc)} \sim g_s \partial_{\bar B}\frac{t_A}{\mathcal{V}}\,.
\end{equation}
Here in the second term the K\"ahler moduli $A$ and $\bar B$ can be either the `small cycle' or the `volume cycle' and the scaling depends on which case is considered.

Note that for cases without 7-branes, $\delta K_\mathrm{loc}=0$ and the leading contributions to $\delta K$ would be either the BBHL term \cite{Becker:2002nn} or
genuine loop corrections \cite{vonGersdorff:2005bf,Berg:2005ja,Berg:2005yu,Berg:2007wt,Cicoli:2007xp,Cicoli:2021rub,Gao:2022uop}.
The BBHL term induces a correction to the K\"ahler potential of the form
\begin{equation}
\label{eq:dKBBHL}
    \delta K^\mathrm{(BBHL)} \sim \frac{1}{\mathcal{V}\,g_s^{3/2}}\,,
\end{equation}
which only depends on two moduli -- the volume $\mathcal{V}$ and the dilaton $\Im\tau \sim 1/g_s$.

The genuine loop corrections are homogeneous functions of degree $-2$ in 4-cycle K\"ahler moduli which generically occur whenever the tower of KK modes that induces the loop corrections displays an $\mathcal{N}=1$ instead of an $\mathcal{N}=2$ spectrum \cite{Gao:2022uop}.\footnote{In the literature, such loop corrections are also called winding-type loop corrections \cite{Berg:2007wt}. They have the same scaling but differ is their interpretation: The winding type loop corrections do only occur if 7-branes intersect. In this sense, the genuine loop corrections are more general because they also occur in setups without 7-branes.}
In the case of LVS, which we analyse below, one such genuine loop correction will become important: The correction related to the blow up modulus $\tau_s$ \cite{Gao:2022uop}. Suppressing the unknown dependence on the complex structure moduli, it reads
\begin{equation}
\label{eq:genuineLoop}
    \delta K_\mathrm{gen} \sim \frac{1}{\mathcal{V}\sqrt{\tau_s}} \,, \qquad\Rightarrow\qquad \partial_{\tau_s}\partial_j \delta K_\mathrm{gen}\sim\partial_{\tau_s} \delta K_\mathrm{gen} \sim \frac{1}{\mathcal{V}\,\tau_s^{3/2}}+\cdots\,.
\end{equation}

\subsection{Stability analysis of the scalar potential} \label{sec:stability}

So far, we have only considered the size of the contributions to the scalar potential. In this section we discuss the stability of the scalar potential (or in other words the positive-definiteness of the masses of the scalar fields) including the corrections discussed above. Such a stability analysis for SUSY-breaking vacua has been performed in \cite{Denef:2004cf,Marsh:2011aa,Kallosh:2014oja,Marsh:2014nla,Gallego:2017dvd}, albeit not with all the corrections derived here.

Let us start with a preliminary, general discussion, motivated by the supergravity context but not directly dependent on it: Consider a scalar potential on a real $d$-dimensional Riemannian manifold, parametrised by coordinates $x^a$, which takes the form
\begin{equation}
    V(x)=g_{ab}(x)v^a(x)v^b(x)\, ,
\end{equation}
with $v(x)$ a vector field. We are interested in situations where $V$ has a minimum at small value, which we take to be at $x=0$ without loss of generality. Thus, we define $|v(0)|=\epsilon\ll 1$, such that $V(0)\sim \epsilon^2$. The extremum condition $\partial_aV=0$ implies
\begin{equation}\label{eq:rightEV}
    v_b\,(D_a v^b)=0\,,
\end{equation}
at $x=0$, with $D$ the covariant derivative. Thus the $d\times d$ matrix $D_a v^b$ has an eigenvector with zero eigenvalue. Then its determinant is zero, the determinant of its transposed matrix is also zero, and hence the transposed matrix also has a zero-eigenvector:
\begin{equation}
    \hat{e}^a_+\,(D_a v^b)=0\,.
\end{equation}
We have chosen this eigenvector to have unit length, $|\hat{e}_+|=1$\,. We see that the vector field $v$ has exactly vanishing covariant derivative along $\hat e_+$ at $x=0$. From this, we conclude that our potential has not only a minimum at small value, $V\sim \epsilon^2$, but that it necessarily also has an almost flat direction:
\begin{equation}\label{ddv}
    (\hat{e}^c_+ D_c)^2 V = 2 g_{ab} v^a (\hat{e}_+^c D_c)^2 v^b \,\sim \, \epsilon\,.
\end{equation}
Here we have assumed that the second derivative of our vector field, $(\hat{e}^c_+ D_c)^2 v^b$ is not parametrically small. Our result is also clear at an intuitive level: If we vary $x$ around $x=0$, the vector $v$ varies. By the minimum condition for $V$, this variation must vanish at linear level in the direction parallel to the vector. The corresponding direction in $x$-space defines $\hat{e}_+$. Then, the potential is necessarily particularly flat along this direction because it is defined as the square of our vector field, which has a small value at $x=0$. This is the content of \eqref{ddv}.

Next, we will make this explicit for the no-scale $F$-term potential considered in this paper. While the almost flat direction of the potential has in this case already been discussed in \cite{Denef:2004cf} (see also \cite{Marsh:2014nla}), we will in the following need the additional fact that the $F$-term itself, corresponding to the vector $v$ above, has vanishing covariant derivative along this direction.

Introducing a real index $a = (I, \bar I)$,
the matrix of K\"ahler covariant derivatives of the $F$-terms $F_a = \left(F_I, \bar F_{\bar I}\right)$ reads
\begin{equation}\label{eq:DFmatrix}
    \left(D_a F_b\right)
    = \begin{pmatrix}
        D_I F_J & D_I \bar F_{\bar J} \\
        \bar D_{\bar I} F_J & \bar D_{\bar I} \bar F_{\bar J}
    \end{pmatrix} 
    = \begin{pmatrix}
        Z_{IJ} & K_{I \bar J} \bar W \\
        K_{\bar I J} W & \bar Z_{\bar I \bar J} \\
    \end{pmatrix} \,,
\end{equation}
where we have used that $F_I = D_I W$ and $Z_{IJ} = D_I F_J$, as introduced in Section~\ref{sec:matching10dand4d}.

At critical points of the potential, where $\partial_a V = 0$, $\bar F^a = \left(\bar F^I, F^{\bar I}\right)$ is a zero-eigenvector of this matrix,
\begin{equation}
   \left(D_a F_b\right) \bar F^b = 0 \,,
\end{equation}
as can be seen from the explicit expression for $\partial_I V$ given in \eqref{eq:4dminimumcond}.
This is the K\"ahler covariant version of \eqref{eq:rightEV}.
As discussed above, this implies that also the transpose of $\left(D_a F_b\right)$ has a zero-eigenvector,
\begin{equation}\label{eq:EHatEV}
    \hat{e}^a_+ \left(D_a F_b\right) = 0 \,.
\end{equation}
It can be verified by direct calculation that this vector is given by
\begin{equation}
\label{eq:light_direction}
   \hat e^a_+ = \frac{1}{\|W F\|} \left(W \bar F^I,\overline W F^{\bar I}\right) \,,
\end{equation}
where we included a normalisation factor. 

By means of a similar argument as in \eqref{ddv}, it was explicitly demonstrated in \cite{Denef:2004cf} that $\hat{e}^a_+$ represents an almost flat direction of the potential,
\begin{equation}
    (\hat{e}^a_+D_a)^2 V  \sim \mathcal{O}(\varepsilon) \,,
\end{equation}
suppressed in the size of the $F$-term $|F| \sim \varepsilon \ll 1$. 

As a key novel point, we want to argue that this property persist for all classical backreaction corrections to the potential. To do so, we need to understand how the IASD flux $G_-$ varies in the direction $\hat{e}^a_+$.
According to \eqref{eq:Gexpansion} and \eqref{eq:Wcoefficients}, it can be written as 
\begin{equation}
\label{eq:G-exp}
    G_-^{(0)} = F_i \bar \alpha^i + \bar F_{\bar \tau} \alpha^{\bar \tau}  \,,
\end{equation}
where
\begin{equation}
    \bar \alpha^i = 2\mathrm{e}^{K_\mathrm{cs}}\, \bar \chi^i \,,\qquad \alpha^{\bar \tau} = 2\mathrm{i} \mathrm{e}^{K_\mathrm{cs}+K_\mathrm{ad}} \, K^{\bar\tau\tau}\, \Omega \,.
\end{equation}
Therefore
\begin{equation}\label{eq:DGMinus}
    \partial_a G_-^{(0)} =  \left( D_a F_i \right) \bar \alpha^i + F_i \left(D_a \bar \alpha^i\right) + \left( D_a \bar F_{\bar \tau} \right) \alpha^{\bar \tau}  + \bar F_{\bar \tau} \left(D_a \alpha^{\bar \tau} \right) \,,
\end{equation}
and at critical points of the potential
\begin{equation}
\label{eq:eG-}
   \hat{e}^a_+ \partial_a G_-^{(0)} = \hat{e}^a_+ \Bigl[ F_i \left(D_a \bar \alpha^i\right) + \bar F_{\bar \tau} \left(D_a \alpha^{\bar \tau} \right) \Bigr] \sim \mathcal{O}(\varepsilon) \,.
\end{equation}

Let us summarise the main ingredients underlying our subsequent analysis. A key result of \cite{Denef:2004cf} (see also \cite{Marsh:2011aa,Marsh:2014nla} and Appendix \ref{app:DD} for a derivation) is that, at a non-supersymmetric critical point of the $F$-term potential, there exists an almost flat direction in field space, spanned by the vector $\hat e^{a}_{+}$ defined in \eqref{eq:light_direction}.
The mass of this modulus and its corresponding complex partner, to which we associate the direction $\hat e^a_-$ in field space (to be defined in \eqref{eq:light_direction2}),
are given by \cite{Denef:2004cf} 
\begin{equation}\label{eq:DD:m1pmain}
    (m_F^{+})^2 = \frac{2}{|F|^2} \left(\Re ( \mathrm{e}^{2\I\theta} U_{IJK} \, \oF^I \, \oF^J \, \oF^K ) + R_{I\bar{J}K\bar{L}} \, \oF^I F^{\bar{J}} \oF^K F^{\bar{L}} \right)\,,
\end{equation}
and
\begin{equation}\label{eq:DD:m1mmain}
    (m_F^{-})^2 = 4|W|^2+\frac{2}{|F|^2} \left(-\Re ( \mathrm{e}^{2\I\theta} U_{IJK} \, \oF^I\, \oF^J \, \oF^K ) + R_{I\bar{J}K\bar{L}} \, \oF^I F^{\bar{J}} \oF^K F^{\bar{L}} \right)\,.
\end{equation}
In what follows, with the exception of the KKLT analysis in Section~\ref{sec:KKLT}, we assume that $|W|$ is large compared to $|F_I|$. In this regime the mass $(m^-_F)^2\sim |W|^2$ is parametrically heavier than $(m^+_F)^2\sim |F_I|\sim  \varepsilon$. Consequently, in 4d Planck units there is a single light complex structure modulus whose mass is given by
\begin{equation}
\label{eq:mf}
    (m_F^+)^2 \sim \frac{g_s F}{\mathcal{V}^2} \sim \frac{g_sG_-^{(0)}}{\mathcal{V}^2} \sim \frac{g_s\varepsilon}{\mathcal{V}^2}\,,
\end{equation}
whereas all other complex structure moduli masses are of the order (assuming $|W|\sim\mathcal{O}(1)$)
\begin{equation}\label{eq:mtyp}
    m_\mathrm{typ}^2 \sim \dfrac{g_s }{\mathcal{V}^2}\,.
\end{equation}
In addition, we have observed that, at critical points of this potential, also $G_-^{(0)}$ varies very slowly along the direction $\hat e^a_+$. Since the complete classical potential \eqref{eq:Vscaling}, including warping corrections, is sourced by $G_-^{(0)}$, this insight will be crucial when we now turn to estimating the size of derivatives of the potential and to its stability.

A light complex structure modulus with mass \eqref{eq:mf} which is suppressed by $\varepsilon$ runs the risk of being destabilised by some correction $\delta V$ to the scalar potential, as for instance examined above. In order to avoid this destabilisation, we have to require that
\begin{equation}\label{eq:mfcorrection}
    (m_F^+)^2 \gtrsim \left(\hat e^a_+ D_a\right)^2 \delta V\,,
\end{equation}
For the leading correction \eqref{eq:deltaVeff_summary} of Section \ref{sec:scaling}, one finds 
\begin{equation}
\label{eq:stabdVeff}
    \left(\hat e^a_+ D_a\right)^2 \delta V_{\mathrm{warp}} =\frac{1}{c^4} \left(\hat e^a_+ D_a\right)^2 \left( V^{(2)} -2 V^{(1)}\delta {\cal V}_{4, {\rm w}}\right)
    \sim \frac{g_s^2N\,\varepsilon}{\mathcal{V}^{8/3}}+\mathcal{O}\left(\frac{\varepsilon^2}{\mathcal{V}^{8/3}}\right)\,,
\end{equation}
where we used \eqref{eq:eG-}, and the fact that the second derivative of $G_-^{(0)}$ is generically not small to obtain the leading order scaling. One then finds 
\begin{equation}    
    \frac{\left(\hat e^a_+ D_a\right)^2 \delta V_{\mathrm{warp}}}{(m_F^+)^2} \sim \frac{g_sN}{ \mathcal{V}^{2/3}}\,,
    \label{eq:ddVcl}
\end{equation}
which is small at sufficiently large volumes. Hence the light modulus will not be destabilised by the classical corrections provided $|W|$ is large compared to $|F_I|$.

Next, we perform the same analysis for the corrections of Section \ref{sec:qcorrections}. The main difference to the classical corrections is that $G_-^{(0)}$ occurs only linearly. Therefore, when both derivatives act on $G_-^{(0)}$, these corrections are no longer suppressed by $\varepsilon$.
Applying this reasoning to \eqref{eq:deltaVmix_summary}, we find for the leading terms
\begin{equation}
    \left(\hat e^a_+ D_a\right)^2 \delta V_\mathrm{mix} 
    \sim \frac{g_s}{\mathcal{V}^2}  \left( \mathrm{e}^{-2\pi a_AT^A}+\delta K_M \,W_\mathrm{GVW}+ \Re(\bar T^{\bar B})\, \delta K_{\bar B j}\,W_\mathrm{GVW}\right).
    \label{eq:ddVqc}
\end{equation}
Thus,
\begin{equation}
    \frac{\left(\hat e^a_+ D_a\right)^2 \delta V_\mathrm{mix}}{(m_F^+)^2} \sim \frac{1}{\varepsilon } \left( \mathrm{e}^{-2\pi a_AT^A}+\delta K_M\,W_\mathrm{GVW}+\Re(\bar T^{\bar B})\,\delta K_{\bar B j}\,W_\mathrm{GVW}\right)\,.
    \label{eq:ddVqcmf}
\end{equation}

\subsection{Application to \texorpdfstring{$F$}{F}-term uplifting in concrete models}\label{sec:applications}

In this section we assess how the backreaction quantified in Section~\ref{sec:scaling}, together with the quantum corrections involving $G_{-}$ flux analysed in Section~\ref{sec:qcorrections}, affect the classical flux potential. This has implications for the stability and consistency of possible de Sitter vacua from $F$-term uplifting \cite{Saltman:2004sn}.

The full potential for the complex structure and Kähler moduli can be written as
\begin{equation}\label{eq:Vfull}
    V_\mathrm{tot} =  V_\mathrm{flux} + V_\mathrm{AdS}+ \delta V_{\text{warp}}+ \delta V_{\text{mix}}+\ldots \,.
\end{equation}
Here $V_\mathrm{flux}$ is the leading-order, flux-induced potential from \eqref{eq:VGKP}:
\begin{equation}\label{eq:VupGen}
    V_\mathrm{flux} = \frac{g_s|G^{(0)}_-|^2}{\mathcal{V}^2}\sim \frac{g_s D_I W \bar D^I\bar W}{\mathcal{V}^2} \sim \frac{g_s\varepsilon^2}{\mathcal{V}^2} \, .
\end{equation}
Solving $\partial_{I} V_{\mathrm{flux}} = 0$ for the complex structure moduli, with $\partial_{I} V_{\mathrm{flux}}$ evaluated as in \eqref{eq:4dminimumcond}, determines their values at the minimum. Since $V_\mathrm{flux}$ is by assumption positive at the minimum, it also acts as an uplifting term. Next, $V_\mathrm{AdS}$ is the sum of the Kähler moduli $F$-term potential and the supergravity contribution $-3\,e^K\,|W|^2$. By assumption, this stabilises the Kähler moduli in an AdS minimum. Finally, the corrections $\delta V_\mathrm{warp}$ and $\delta V_\mathrm{mix}$ are given by \eqref{eq:deltaVeff_summary} and \eqref{eq:deltaVmix_summary}, respectively.

The goal of this section is to estimate for KKLT and LVS to which extent $V_\mathrm{AdS}$, $\delta V_\mathrm{warp}$, and $\delta V_\mathrm{mix}$ affect the stabilisation of the complex structure moduli and the $F$-term uplifting in general. In other words, we ensure that we have correctly identified the leading order contributions for stabilising the complex structure and K\"ahler moduli.
We do so by comparing the size of the corrections against the size of the dS minimum, by analysing the stability of the potential including all the corrections, and by estimating the shift of the vacuum expectations values of the complex structure moduli induced by all corrections.

We start with $\delta V_\mathrm{warp}$ since, for this correction, the analysis is independent of the stabilisation scheme. Repeating for convenience \eqref{eq:correctionssuppression} and \eqref{eq:ddVcl}, we have
\begin{equation}
    \frac{\delta V_{\mathrm{warp}}}{V_\mathrm{flux}} \sim \frac{g_sN}{\mathcal{V}^{2/3}}\,,\qquad\qquad 
    \frac{\left(\hat e^a_+ D_a\right)^2 \delta V_{\mathrm{warp}}}{(m_F^+)^2} \sim \frac{g_sN}{ \mathcal{V}^{2/3}}\,.
    \label{eq:dVeffvsVup}
\end{equation}
Estimating the stability of the potential against $\delta V_\mathrm{warp}$ along a direction $\hat f^a$ corresponding to a heavy modulus with mass \eqref{eq:mtyp}, one finds
\begin{equation}
    \frac{(\hat f^a D_a)^2 \delta V_{\mathrm{warp}}}{m_\mathrm{typ}^2} \sim \frac{g_sN}{ \mathcal{V}^{2/3}}\,,
\end{equation}
where we used that $(\hat f^a D_a)^2 G_-^{(0)} \sim\mathcal{O}(1)$ based on our analysis in Section \ref{sec:stability}.

Finally, we check by how much the vacuum expectations values of the complex structure moduli are shifted when $\delta V_\mathrm{warp}$ is incorporated. One finds for the light direction 
\begin{equation}
\label{eq:dVwarpshift}
    \delta z^a \sim \frac{\hat e^a_+\partial_a \delta V_\mathrm{warp}}{(m_F^+)^2} \sim \frac{g_s N \varepsilon}{\mathcal{V}^{2/3}} \ll 1\,,
\end{equation}
where we used $\varepsilon\ll 1$. 
The result for the heavy directions is similar.
To summarise, corrections from $\delta V_\mathrm{warp}$ do not endanger the $F$-term uplifting provided $|W|$ is large compared to $|F_I|$.

The same analysis for $V_\mathrm{AdS}$ and $\delta V_\mathrm{mix}$ requires characteristic scaling relations of the supersymmetry-breaking parameter~$\varepsilon$ with the volume modulus $\mathcal{V}$, the vacuum expectations value of the superpotential $W_0$, and the string coupling~$g_s$. Since the precise relation depends on the stabilisation scheme, we will first focus on LVS in Section \ref{sec:LVS} and then on KKLT in Section \ref{sec:KKLT}.

\subsection{LVS} \label{sec:LVS}

For simplicity, we focus on the simplest LVS setting with only two K\"ahler moduli: The volume modulus $\mathcal{V}$ and the blow-up modulus $\tau_s$. In doing so we avoid the dangerous loop effects that can occur in fibered geometries (cf.~the discussion at the end of Section \ref{sec:qcorrections}).

In LVS with $F$-term uplift, the leading order scalar potential is given by $V_\mathrm{flux}+V_\mathrm{AdS}$ where $V_\mathrm{flux}$ is given by \eqref{eq:VupGen} and $V_\mathrm{AdS}$ reads schematically
\begin{equation}
\label{eq:VAdSaroundmin}
    V_\mathrm{AdS}\sim  \frac{|A_s|^2g_s\sqrt{\tau_s}\text{e}^{-4\pi a_s\tau_s}}{\mathcal{V}} - \frac{|A_s| g_s \tau_s |W_0| \text{e}^{-2\pi a_s \tau_s}}{\mathcal{V}^2} + \frac{ |W_0|^2}{\sqrt{g_s}\mathcal{V}^3}\,,
\end{equation}
where $A_s$ is the complex structure and dilaton dependent Pfaffian prefactor of the non-perturbative correction related to the blow up modulus $\tau_s$.
In the minimum, $V_\mathrm{AdS}$, the volume, and the blow up modulus are stabilised at  
\begin{equation}
    \label{eq:LVSvev}
    V_\mathrm{AdS,min}\sim  - \frac{ g_s \sqrt{\tau_s} \,W_0^2}{\mathcal{V}^3}\,,\qquad
    \mathcal{V}\sim W_0\sqrt{\tau_s}\mathrm{e}^{2\pi a_s\tau_s}\,\,\,,\qquad
    \tau_s \sim \frac{1}{g_s} +\mathcal{O}(1)\,.
\end{equation}
One obtains a small uplift to dS if $|V_\mathrm{AdS,min}|\approx V_\mathrm{flux}$, which enforces the IASD flux $G_-$ to be of the order \cite{Gallego:2017dvd}
\begin{equation}\label{eq:epsilonscalingLVS}
    \varepsilon \sim \frac{W_0}{\mathcal{V}^{1/2}\,g_s^{1/4}} \,,
\end{equation}
and therefore to be parametrically small. For arbitrary values of $\varepsilon$ one would find a runaway potential since then $V_\mathrm{flux}\gg |V_\mathrm{AdS,min}|$. 

First, we check that the terms in $V_\mathrm{AdS}$ do not affect the stabilisation of the complex structure moduli despite the flat direction that was analysed in Section \ref{sec:stability}.\footnote{We note that the procedure described above, in which the complex structure moduli are integrated out first, is strictly speaking not correct in LVS. The reason is that the blow up modulus is heavier than the light complex structure modulus with mass $m_F^+$. This subtlety does not change our parametric estimates below and we therefore keep treating the blow up as a dynamical field also from the perspective of the light complex structure modulus. Also the inverse effect of a dynamical light complex structure modulus on the stabilisation of the blow up modulus can be checked to be small.} To recap, the typical mass of a complex structure modulus is given by \eqref{eq:mtyp} and the light modulus has mass
\begin{equation}
    (m_F^+)^2 \sim \frac{g_s \varepsilon}{\mathcal{V}^2} \sim \frac{g_s^{3/4} W_0}{\mathcal{V}^{5/2}}\,.
\end{equation}
This has to be compared to the second derivative of \eqref{eq:VAdSaroundmin} with respect to the complex structure moduli and the axio-dilaton. Close to the minimum of the full potential, one finds 
\begin{equation}
    \frac{\left(\hat e^a_+ D_a\right)^2  V_{\mathrm{AdS}}}{(m_F^+)^2} \sim \frac{W_0}{g_s^{5/4} \mathcal{V}^{1/2}}\left(1+g_s^2\right)\,, \qquad \qquad \frac{(\hat f^a D_a)^2 V_{\mathrm{AdS}}}{m_\mathrm{typ}^2} \sim \frac{W_0}{g_s^{5/4} \mathcal{V}^{1/2}}\,,
    \label{eq:stabilityVads}
\end{equation} 
To obtain \eqref{eq:stabilityVads}, we used $D_IW\sim \varepsilon$, $(\hat e^a_+ D_a)^2 W_0 \sim \varepsilon$, and $(\hat f ^a D_a)^2 W_0 \sim \mathcal{O}(1)$ (as discussed in Sec.~\ref{sec:stability}). Furthermore, we worked with the conservative estimates $(\hat e ^a_+ D_a)^2 A_s \sim (\hat f ^a D_a)^2 A_s \sim \mathcal{O}(1)$. The leading contribution to the first ratio in \eqref{eq:stabilityVads} comes from the complex structure moduli dependence of the Pfaffian in the first two terms of \eqref{eq:VAdSaroundmin}. The $g_s^2$ suppressed piece comes from the third term in \eqref{eq:VAdSaroundmin}. For the second ratio in \eqref{eq:stabilityVads}, characterising the heavy directions, all terms in \eqref{eq:VAdSaroundmin} give contributions with the same leading order scaling. Hence this result remains unchanged if the Pfaffian does not depend on complex structure moduli.

To summarise, $V_\mathrm{AdS}$ does not destabilise complex structure moduli since both ratios in \eqref{eq:stabilityVads} are small at large volume. 

Second, we note that $V_{\rm AdS}$ generically includes a term linear in the complex structure moduli at the location of the minimum of $V_{\rm flux}$. We should check that the resulting shift of the complex structure moduli is small. Along the flat direction, this shift is
\begin{equation}
\label{eq:linearshiftLVS}
\delta z^a \sim \frac{\hat{e}^a_+ \partial_a V_{\rm AdS}}{(m_F^+)^2}\sim \frac{W_0}{g_s^{5/4}\mathcal{V}^{1/2}}\left(1+ g_s +\mathcal{O}(\varepsilon)\right)\,,
\end{equation}
where the leading order term comes from the moduli dependence of the Pfaffian, the $g_s$ suppressed piece comes from the last term in \eqref{eq:VAdSaroundmin}, and the contribution of $\mathcal{O}(\varepsilon)$ from derivatives of $W_0$. 
All in all, the ratio in \eqref{eq:linearshiftLVS} is small at large enough volumes, implying a small $V_{\rm AdS}$-induced shift of the light complex structure modulus. The result for the heavy direction is similar.

Now, we move on to $\delta V_{\rm mix}$
from Section \ref{sec:qcorrections}, as summarised in \eqref{eq:deltaVmix_summary}, and perform the same analysis as for $V_\mathrm{AdS}$ above. In the simplest LVS setting, the leading corrections are given by
\begin{equation}
\label{eq:VmixLVS}
    \delta V_\mathrm{mix} \sim \frac{g_s\varepsilon}{\mathcal{V}^2} \left(\mathrm{e}^{-2\pi a_s\tau_s}+ \delta K_MW_0+ \Re(\bar T^{\bar B})\,\delta K_{\bar B j} W_0\right)\,.
\end{equation}
Including the leading loop effect in swiss-cheese type geometries \eqref{eq:diKKK_swiss} and \eqref{eq:genuineLoop}, and the BBHL correction \eqref{eq:dKBBHL} one finds 
\begin{equation}
    \frac{\delta V_\mathrm{mix}}{V_\mathrm{flux}} \sim \underbrace{\frac{1}{\mathcal{V}^{1/2}g_s^{1/4}}}_{\text{from } \delta W}+ \underbrace{\frac{g_s^{5/4}}{\mathcal{V}^{1/6}}}_{\text{from }\delta K_\mathrm{loc}^{\mathrm{(sc)}}} + \underbrace{\frac{1}{\mathcal{V}^{1/2} g_s^{1/4}}}_{\text{from } \delta K^\mathrm{(BBHL)}}+ \underbrace{\frac{g_s^{5/4}}{\mathcal{V}^{1/2}}}_{\text{from } \delta K_\mathrm{gen}},
    \label{eq:qcscaling}
\end{equation}
where we displayed the leading effect induced by each different correction to $\delta W$ and $\delta K$.
Hence, in the case when $\delta K^\mathrm{(sc)}_\mathrm{loc}\neq 0$ (which requires non-trivial 7-brane configurations), and where $\delta K^\mathrm{(sc)}_\mathrm{loc}\sim g_s/\mathcal{V}^{2/3}$ (which requires 7-branes wrapping the volume 4-cycle),
the corrections $\delta V_\mathrm{mix}$ are suppressed by a factor of $g_s^{5/4}/\mathcal{V}^{1/6}$ compared to the uplifting term which generically is the dominant term in \eqref{eq:qcscaling}. The reason is that $\mathcal{V}\sim \exp(1/g_s)$ is exponentially large. 
In cases when $\delta K_\mathrm{loc}=0$, the leading correction in suppressed by $1/(\mathcal{V}^{1/2}g_s^{1/4})$. We note that the condition that the first term in \eqref{eq:qcscaling} is small can be rewritten as the condition $\delta W \ll |D_IW|$ as was also found in \cite{Marsh:2014nla}.

Next, we turn to the stability analysis of the $\delta V_\mathrm{mix}$ corrections. From \eqref{eq:ddVqcmf} we find, after a very similar calculation as in \eqref{eq:qcscaling},
\begin{equation}
    \frac{\left(\hat e^a_+ D_a\right)^2 \delta V_\mathrm{mix}}{(m_F^+)^2} \sim \frac{1}{g_s^{1/4}\mathcal{V}^{1/2}}+ \frac{g_s^{5/4}}{\mathcal{V}^{1/6}} \,.
    \label{eq:dVmixstab}
\end{equation}
Here, we have not listed the leading contribution from each correction but focussed on the overall leading corrections. 
As above, the corrections are parametrically suppressed either by $g_s^{5/4}/\mathcal{V}^{1/6}$ in cases where $\delta K^\mathrm{(sc)}_\mathrm{loc}\neq 0$ due to 7-branes on the volume 4-cycle, or by $1/(\mathcal{V}^{1/2}g_s^{1/4})$ when $\delta K_\mathrm{loc}=0$. 
As a final check for $\delta V_\mathrm{mix}$, we have to convince ourselves that the shift of the complex structure moduli induced by $\delta V_\mathrm{mix}$ is small (see the analogous calculation for $V_\mathrm{AdS}$ in \eqref{eq:linearshiftLVS}). 
A short calculation reveals that the shift follows the same scaling relations as in \eqref{eq:qcscaling} and \eqref{eq:dVmixstab} and is therefore small.

We conclude that in LVS on swiss-cheese type geometries, the $F$-term uplifting and the stability of the potential are parametrically controlled. The largest corrections are suppressed by $g_s^{5/4}/\mathcal{V}^{1/6}$ and only occur when 7-branes are wrapped on the volume 4-cycle.

Before closing this section, let us compare our results to the literature. A related control analysis has been performed in \cite{Marsh:2014nla,Gallego:2017dvd} where the implications of non-perturbative corrections $\delta W$ in particular to the mass matrix have been analysed. Our analysis here goes beyond this in two ways.
First, we also take into account perturbative corrections to the K\"ahler potential which generically are dominant compared to the corrections induced by $\delta W$. Second, besides checking the stability of the potential, we also carefully analyse the size of the corrections compared to the leading term and their influence on the stabilisation of the heavier moduli.

\subsection{KKLT}\label{sec:KKLT}

Let us now perform the analogous study for KKLT. For the purposes of the following parametric discussion, the detailed Calabi-Yau geometry (in particular the value of $h^{1,1}$ and the intersection numbers) is not important. Before stabilising the K\"ahler moduli, $V_\mathrm{AdS}$ is given by 
\begin{equation}
    V_\mathrm{AdS} = \mathrm{e}^K \left(K^{A\bar B} D_A W\bar D_{\bar B} \overline{W} - 3|W|^2\right)\,.
    \label{adsf}
\end{equation}
At the minimum, it reduces to
\begin{equation}\label{eq:AdSKKLT}
    V_\mathrm{AdS,min} = -3 \mathrm{e}^K |W|^2 \sim - \frac{g_s\, W_0^2}{\mathcal{V}^2} \,.
\end{equation}
Comparing this to the uplifting potential \eqref{eq:VupGen} and imposing $|V_\mathrm{AdS,min}|\approx V_\mathrm{flux}$, one finds the relation 
\begin{equation}\label{eq:VupliftKKLT}
    \varepsilon\simeq W_0\,.
\end{equation} 
This relation is problematic for the $F$-term uplift because positivity of the spectrum can no longer be ensured and tachyonic directions may appear. 

More specifically, as emphasised already in \cite{Denef:2004cf}, one of the masses in \eqref{eq:DD:m1pmain} and \eqref{eq:DD:m1mmain} is generically negative when $\varepsilon\simeq W_0$, rendering the critical point unstable. This can be understood as follows. For small $\varepsilon$, the first term in \eqref{eq:DD:m1pmain} scales linearly with $\varepsilon$ and dominates. If its sign is positive, it provides a positive contribution to $(m_F^+)^2$. However, in the regime $W_0\simeq |F_I|$, the corresponding second term in \eqref{eq:DD:m1mmain} then dominates and is negative. This implies an unstable direction unless additional structure is imposed. As argued in \cite{Denef:2004cf}, this instability can only be avoided by a fine tuning which ensures that
\begin{equation}\label{eq:finetuning}
    \frac{2}{|F|^2} |U_{IJK} \, \oF^I\, \oF^J \, \oF^K| < \mathcal{O}(\varepsilon^2)\,.
\end{equation}
The explicit form of $U_{IJK}$ for the Gukov-Vafa-Witten superpotential $W_{\mathrm{GVW}}$ is derived in \eqref{eq:DD:UGVW}, where it is shown that the components $U_{ijk}$ along the complex structure moduli directions generically contain an unsuppressed contribution.
Under the condition \eqref{eq:finetuning}, both masses can be made positive, at least in principle, and are parametrically of order
\begin{equation}\label{eq:lightcsKKLT}
    (m_F^\pm)^2\sim \frac{g_s\varepsilon^2}{\mathcal{V}^2} \, ,
\end{equation}
in 4d Planck units. The light complex structure moduli therefore have parametrically the same mass as typical K\"ahler moduli, with  $m_\text{K\"ahler}^2 \sim g_s W_0^2/\mathcal{V}^2$. The analysis of how $V_\mathrm{AdS}$ affects the light complex structure moduli is therefore not applicable since they have to be stabilised together with the K\"ahler moduli. 

Let us further note that the situation may in fact be more challenging than the analysis of 
\cite{Denef:2004cf} suggests: As argued in \cite{Marsh:2011aa}, the expressions \eqref{eq:DD:m1pmain} and \eqref{eq:DD:m1mmain} in general do not constitute reliable approximations to the true eigenvalues of the Hessian. The reason is that, after tuning both these expressions to be as small as $\sim |F|^2$, off-diagonal terms in the complete Hessian matrix become competitive. More precisely, terms suppressed by $|F|$ which mix the light and the typical complex structure directions enter the expressions for the lowest eigenvalues. To guarantee their positivity, more complicated expressions then have to be studied. The statistical analysis of \cite{Marsh:2011aa} suggests that the tuning becomes more severe. 

However, this is not our concern. We assume that the required tuning is possible and follow the same steps as in the LVS analysis of Section~\ref{sec:LVS}, highlighting the features that are specific to KKLT. 

We begin by examining how warping effects, encoded in $\delta V_{\mathrm{warp}}$, modify the standard $F$-term uplift in KKLT. Comparing $\delta V_{\mathrm{warp}}$ with $V_{\mathrm{flux}}$, one finds the same parametric behaviour as in the first equation of \eqref{eq:dVeffvsVup}, so that the warping correction remains small provided the overall volume is sufficiently large. Moreover, one can verify that the shifts in the vacuum expectation values of the complex structure moduli induced by $\delta V_{\mathrm{warp}}$ are likewise small, despite the fact that their masses scale as $\mathcal{O}(\varepsilon^2)$ rather than $\mathcal{O}(\varepsilon)$ as in the LVS case.

The reason for this is that $D_I G_-$ varies slowly along an entire complex direction in field space, rather than only along a single real direction as in LVS. 
The complex direction in field space is spanned by $\hat e_+$, as given in \eqref{eq:light_direction}, and by $\hat e_-$, defined as \cite{Denef:2004cf}
\begin{equation}
    \label{eq:light_direction2}
   \hat e^a_- = \frac{1}{\|W F\|} \left(-W \bar F^I,\overline W F^{\bar I}\right) \,.
\end{equation}
From \eqref{eq:eG-}, we already know that the $F$-term, and therefore $G_-$ varies slowly along $\hat e_+$. In KKLT, where $W\sim\varepsilon$, the same applies to the direction $\hat e_-$ since 
\begin{equation}   
    \hat e^a_- D_a F_b = \frac{-2|W| ^2}{\|WF\|} \left( F_J , - \bar F_{\bar J} \right) \sim \varepsilon\,.
\end{equation}
Making use of the fact that $\delta V_\mathrm{warp}$ is sourced only by $G_-^{(0)}$, one finds $(\hat e^a_\pm D_a)\delta V_\mathrm{warp} \sim \varepsilon^2/\mathcal{V}^{8/3}$ which proves that the shift of the vacuum expectation values of the light complex structure moduli induced by $\delta V_\mathrm{warp}$ is small.

We now turn to the stability analysis of the warped potential, which requires evaluating the second derivatives of $\delta V_{\mathrm{warp}}$ in \eqref{eq:deltaVeff_summary} while consistently imposing the fine-tuning condition \eqref{eq:finetuning}. A key observation is that \eqref{eq:deltaVeff_summary} contains several contractions of the background IASD flux $G_-^{(0)}$, most notably the second and third terms in \eqref{eq:V2}, which are not of the schematic form $|G_-^{(0)}|^2$ familiar from the leading-order potential \eqref{eq:VGKP}. 
For such contractions, there is no a priori reason for cancellations analogous to those enforced by \eqref{eq:finetuning} to occur. As a result, we generically expect the second derivatives of these terms along the (almost) flat directions $\hat e^a_\pm$ to scale as $(\hat e^a_\pm D_a)^2 \delta V_{\mathrm{warp}}\sim g_s^2 N \,\varepsilon/\mathcal{V}^{8/3}$. 
Establishing this behaviour explicitly is technically demanding, as it would require solving the full set of first-order equations of motion \eqref{eq:eomtau1} and \eqref{g1bia} to obtain the corrections $\tau^{(1)}$ and $G_-^{(1)}$, respectively.
Nonetheless, the parametric estimates above are sufficient to argue that the tuning \eqref{eq:finetuning} does not generically suppress all warping-induced contributions to the mass matrix, rendering stability non-generic and dependent on additional, highly contrived cancellations.

To summarise, we find that even after imposing the additional fine tuning \eqref{eq:finetuning}, the outcome mirrors that of \eqref{eq:stabdVeff}. Thus, we obtain
\begin{equation}\label{eq:dVwarpKKLT}
    \frac{\left(\hat e^a_\pm D_a\right)^2 \delta V_{\mathrm{warp}}}{(m_F^\pm)^2} \sim \frac{g_sN}{\varepsilon\, \mathcal{V}^{2/3}}\,.
\end{equation}
Since $\varepsilon\sim W_0$, this is clearly in strong conflict with an exponentially small $W_0$ in standard KKLT which usually requires $\varepsilon\ll 1/\mathcal{V}^{2/3}$.

Having discussed these issues at a general level, we now return to the model discussed in Section~\ref{sec:4d}, in which the effects of warping are incorporated through shifts in the Kähler coordinates $T^A$. Under this assumption, we computed the scalar potential \eqref{fullp}, including the proposed leading warping corrections. One might then ask whether, in this (arguably simplified) parametrisation of warping effects, the additional tuning imposed in \eqref{eq:finetuning} could nevertheless be sufficient to alleviate the potential instability. As demonstrated by the explicit analysis carried out in Appendix~\ref{app:DD:4d:stability}, this is not the case. In particular, the warping induced correction to the inverse Kähler metric on complex structure moduli space modifies the Hessian in such a way that the tuning condition \eqref{eq:finetuning} alone does not suffice to guarantee the stability of the scalar potential.

\medskip

Finally, let us come back to the general situation and study the implications of $\delta V_\mathrm{mix}$ for KKLT. 
The leading quantum corrections as summarised in \eqref{eq:deltaVmix_summary} are, in the case of KKLT, given by 
\begin{equation}
    \delta V_\mathrm{mix} \sim \frac{g_s\varepsilon}{\mathcal{V}^2}\left(\mathrm{e}^{-2\pi a_A T^A}+ \delta K_MW_0 + \Re(\bar T^{\bar B})\, \delta K_{\bar B j} W_0\right) \,, 
    \label{eq:dVmixKKLT}
\end{equation}
where, as explained in Footnote \ref{istr}, the index structure is symbolic and hence no sum over $A$ is implied. Thus, we get 
\begin{equation}
    \frac{\delta V_\mathrm{mix}}{V_\mathrm{flux}} \sim  \left(\mathrm{e}^{-2\pi a_AT^A}/\varepsilon+  \delta K_M+ \Re(\bar T^{\bar B})\,\delta K_{\bar B j} \right) \,.
     \label{eq:KKLTleadingqc}
\end{equation}
By computing $D_{T^A}W \approx 0$ explicitly, the leading order solution for the vacuum expectations values of the K\"ahler moduli yields
\begin{equation}
\label{eq:relVW0}
    \mathrm{e}^{-2\pi a_AT^A} \sim W_0\,\dfrac{t_A}{\mathcal{V}}\,.
\end{equation}
With $\delta K_M\lesssim1/\mathcal{V}^{2/3}$ and $\delta K_{\bar B j}\lesssim 1/\mathcal{V}^{2/3}$ from above, the first term in \eqref{eq:dVmixKKLT} is always parametrically bigger than or equal to the second independently of the leading contribution to $\delta K$. To find the leading term in $\delta V_\text{mix}$ it is therefore sufficient to consider the first term.
Focussing on the K\"ahler modulus with the largest contribution $\delta W\sim \mathrm{e}^{-2\pi a_A T^A}$ in \eqref{eq:KKLTleadingqc}, we find
\begin{equation}\label{eq:KKLTdVmixVflux}
    \frac{\delta V_\mathrm{mix}}{V_\mathrm{flux}}  \sim \dfrac{t_A}{\mathcal{V}}\ll 1\,.
\end{equation}

Moving on to the stability analysis for $\delta V_\mathrm{mix}$, we repeat that all terms in $\delta V _\mathrm{mix}$ are linear in $G_-^{(0)}$ and therefore the fine tuning can not affect the second derivatives of $\delta V_\mathrm{mix}$ as it applied to terms of the form $\sim|\tilde G_-^{(0)}|^2$. Consequently, the leading term in the second derivative of $\delta V_\mathrm{mix}$ is obtained when both derivatives act on $G_-^{(0)}$ and is therefore independent of $\varepsilon$. We get 
\begin{equation}\label{eq:KKLTstab}
    \frac{\left(\hat e^a_\pm D_a\right)^2 \delta V_\mathrm{mix}}{(m_F^\pm)^2} \sim  
    \frac{1}{\varepsilon}\biggl (\dfrac{t_A}{\mathcal{V}} \,+\,\delta K_M + \mathcal{V}^{2/3} \delta K_{\bar B j} \biggl )\;\lesssim \dfrac{1}{\varepsilon\, \mathcal{V}^{2/3}}\,.
\end{equation}
This result is independent of the uncertainties that we faced regarding the scaling of the second derivatives of $\delta V_\mathrm{warp}$ when taking into account the
tuning constraint.
We also note in passing that the shift of the vacuum expectations values of the complex structure moduli induced by $\delta V _\mathrm{mix}$ is small. 

To conclude, we have observed that the $F$-term uplift in standard KKLT with exponentially small $W_0$ cannot be realised in a controlled way.
The stability of the potential is strongly affected by the corrections $\delta V_\mathrm{warp}$ and $\delta V_\mathrm{mix}$.
This necessitates the explicit computation not only of classical warping corrections but also of presently unknown loop corrections.
Achieving a controlled $F$-term uplift in KKLT-like scenarios, with a parametric suppression of these corrections, would require finding examples satisfying
\begin{equation}
    \varepsilon\,\mathcal{V}^{2/3}\sim W_0 \mathcal{V}^{2/3}\gg 1\,,
\end{equation} 
that is, cases in which $W_0$ is not too small and the volume is comparably large, similar to the examples of \cite{McAllister:2024lnt}.
There, the volume is larger than naively expected due to the large number of K\"ahler moduli $h^{1,1}$.

\section{Conclusions}\label{sec:conclusions}

The primary goal of this work has been to investigate how warping affects $F$-term uplifting based on non-ISD three-form fluxes in Calabi-Yau orientifold compactifications of Type~IIB string theory. Our central strategy has been to define a four-dimensional effective off-shell potential for the light fields, obtained by systematically integrating out the heavy fields corresponding to the KK modes. This can be realized using the ten-dimensional equations of motion and an expansion in the inverse volume $1/c$. At each order in $1/c$, this procedure yields a set of Poisson-like equations in triangular form, which can be solved iteratively once the zero-mode contributions have been appropriately subtracted. These zero modes correspond precisely to the dynamical fields of the four-dimensional off-shell theory that remain unfixed.

We then employed this procedure to estimate the magnitude of the leading warping corrections to the off-shell scalar potential arising from non-trivial IASD three-form flux $G_-$. We showed that the leading corrections are suppressed by the volume four-cycle $c$, as expected for warping effects. More interestingly, every correction is at least quadratic in $G_-$ at any order in $1/c$. From the four-dimensional $\mathcal{N}=1$ supergravity perspective, this result is somewhat unexpected, as it implies that, at the classical level including warping, no term mixing complex structure and Kähler moduli $F$-terms, $D_zW\,D_TW$, can occur. Understanding this behaviour requires knowledge of the warped K\"ahler potential. In Section \ref{sec:4d}, we put forward a proposal for this Kähler potential, based on arguments from \cite{Frey:2008xw, Martucci:2014ska}, that ensures the absence of linear terms in $G_-$. As a consequence of this proposal, flux vacua with ISD flux remain unaffected by warping effects, whereas non-supersymmetric critical points are modified through a warping-induced correction to the Kähler metric on moduli space.

Using our general expression for the warped effective potential, we then examined the level of control in $F$-term uplifting in the context of KKLT and LVS in Section~\ref{sec:implications}. Already at the level of the leading order flux potential \eqref{eq:VGKP}, the tuning required for $F$-term uplifting, $|F|\sim \varepsilon \ll 1$, implies the presence of a parametrically light complex structure modulus \cite{Denef:2004cf}, which potentially compromises the stability of the uplifted vacuum. We identified two principal sources of dangerous corrections. The first source are warping effects, encoded in the sub-leading contributions $\delta V_{\mathrm{warp}}$ to the effective potential, as given in \eqref{eq:deltaVeff_summary}. The second source are mixing terms between the IASD flux $G_-\neq 0$ and quantum effects, such as non-perturbative corrections to the superpotential or loop corrections to the K\"ahler potential. This leads to further corrections $\delta V_{\mathrm{mix}}$ given in \eqref{eq:deltaVmix_summary} which involve quantum corrections and a factor linear in $G_-\sim \varepsilon$.

In LVS with D7-branes wrapping the large four-cycle, the leading corrections are suppressed by a factor of $g_s^{5/4}/\mathcal{V}^{1/6}$. While this suppression is sufficient to maintain parametric control, it implies a comparatively strong lower bound on the required size of the compactification volume. By contrast, if the volume four-cycle is not wrapped by a D7-brane stack, the leading corrections are instead suppressed by $1/(g_s^{1/4}\mathcal{V}^{1/2})$, resulting in a weaker constraint on $\mathcal{V}$. We conclude that $F$-term uplifting in LVS can be achieved in a parametrically controlled regime.

The situation is qualitatively different in the KKLT scenario. In this case, a consistent $F$-term uplift requires not only $|F|\sim \varepsilon \ll 1$, but also the additional parametric relation $|W_0|\sim |F|$. In this regime, the mass matrix obtained from the leading order flux potential \eqref{eq:VGKP} is no longer guaranteed to be positive definite \cite{Denef:2004cf,Marsh:2014nla}, as reviewed and generalized in Appendix~\ref{app:DD}. Already at leading order, stability requires an additional fine tuning involving the cubic derivative $D^3W$ of the flux superpotential as given in \eqref{eq:finetuning}. However, once this tuning is imposed, the resulting mass spectrum of the potential \eqref{eq:VGKP} contains two parametrically light mass eigenstates with $(m_F^{\pm})^2 \sim \varepsilon^2/\mathcal{V}^2$. These light modes are therefore significantly more vulnerable to destabilisation by sub-leading effects than in LVS, where one instead finds the generic hierarchy $(m_F^{+})^2 \sim \varepsilon/\mathcal{V}^2$ and $(m_F^{-})^2 \sim 1/\mathcal{V}^2$.

Our analysis shows that mass-squared corrections for the light eigenmodes arising from $\delta V_{\mathrm{warp}}$ and $\delta V_{\mathrm{mix}}$ generically scale as $\varepsilon/\mathcal{V}^{8/3}$. Consequently, the relative size of corrections to the light eigenvalues $(m_F^{\pm})^2$ is proportional to $1/(\varepsilon\,\mathcal{V}^{2/3})$. Achieving parametric control therefore requires the hierarchy $W_0\,\mathcal{V}^{2/3}\gg 1$, which is manifestly incompatible with an exponentially small value of $W_0$.

In principle, one might contemplate the (in our view unlikely) possibility that the aforementioned tuning \eqref{eq:finetuning} of $D^3W$ could also suppress the classical mass corrections induced by warping, such that $\delta V_{\mathrm{warp}}$ contributes only at order $\varepsilon^2/\mathcal{V}^{8/3}$ to the light mass spectrum.
By using the general 10d expression \eqref{eq:deltaVeff_summary},
we found indeed clear evidence that such a suppression is unlikely to arise due to the structure of source terms appearing in the 10d equations of motion. 
This conclusion is further supported by the explicit analysis in Appendix~\ref{app:DD:4d:stability}, which employs the 4d potential \eqref{fullp} obtained from our proposal for the warped K\"ahler potential inspired by \cite{Frey:2008xw,Martucci:2014ska}. Even setting this issue aside, and assuming for the sake of argument that such a suppression of $\delta V_{\mathrm{warp}}$ could be achieved, the conclusion does not change. The contributions from $\delta V_{\mathrm{mix}}$, which are linear in the non-ISD flux $G_-\sim \varepsilon$, remain unsuppressed by the tuning \eqref{eq:finetuning} and continue to generate potentially destabilising effects.
These, in particular, include loop effects which are notoriously difficult to calculate.
We therefore conclude that the condition $W_0\,\mathcal{V}^{2/3}\gg 1$ is a necessary requirement for parametric control. As a result, any KKLT-like scenario that aims to realise $F$-term uplifting in a controlled manner must operate at moderately small values of $W_0$, together with a volume much exceeding the standard parametric estimate ${\cal V}\sim \ln(1/|W_0|)$. 

\medskip

Our analysis raises a number of interesting directions for future research. A natural next step is to complement our analytic study of the inverse volume expansion with numerical analyses of warped Calabi-Yau backgrounds. Recent progress in the construction of numerical Calabi-Yau metrics opens the possibility of computing warping corrections directly in explicit compact models. In close analogy to the numerical strategy of \cite{DeLuca:2021pej}, one may attempt to solve the full set of warped Type~IIB equations on selected geometries, using both the recent advances in determining Ricci-flat metrics like \cite{Anderson:2020hux,Douglas:2020hpv,Jejjala:2020wcc,Larfors:2022nep} and the forthcoming numerical study of the warp factor~\cite{SeverinFabianSimon}. This would allow for a direct numerical evaluation of the warped effective potential, thus providing a valuable cross-check of the analytic large-volume expansion.

Beyond improving control over warping effects, an equally important open direction concerns genuinely $\mathcal{N}=1$ corrections to the 4d EFT. In the present work, we have restricted our analysis of $\mathcal{N}=1$ corrections to known loop effects in the K\"ahler potential \cite{vonGersdorff:2005bf,Berg:2005ja,Berg:2005yu,Cicoli:2007xp,Gao:2022uop}. Comparatively little is known about more general $\mathcal{N}=1$ quantum corrections, although some partial progress has been made in this direction recently, see e.g. \cite{Minasian:2015bxa,Klaewer:2020lfg,Kim:2023sfs,Kim:2023eut,Cho:2023mhw,Cvetic:2024wsj}. 
A systematic treatment of such $\mathcal{N}=1$ corrections, and of their interplay with non-ISD fluxes, is left for future work.
Potential obstructions to moduli stabilisation scenarios based on non-perturbative superpotential effects \cite{Sethi:2017phn,Lust:2022lfc,Bena:2024are,Apers:2025pon,Bedroya:2025fie} may also be relevant in the context of $F$-term uplifting; however, it remains unclear to what extent they intersect with or exacerbate the specific issues identified in the present analysis.

A complementary and equally promising direction is the construction of explicit examples of dS vacua from $F$-term uplifting, for which our analysis provides a number of concrete control criteria. Building on the idea of winding uplifts \cite{Hebecker:2020ejb,Carta:2021sms} (see also \cite{Hebecker:2017lxm,Hebecker:2018fln}) and employing the numerical techniques developed in \cite{Krippendorf:2023idy,Cole:2019enn,Dubey:2023dvu,Chauhan:2025rdj}, one may attempt to realise models in which supersymmetry-breaking fluxes generate a parametrically small uplift compatible with full moduli stabilisation. We note, however, that identifying flux choices which yield an appropriately small supersymmetry-breaking parameter $\varepsilon$ may be subtle and could face additional constraints, potentially including those related to the D3-tadpole \cite{Bena:2020xrh}. If examples could nevertheless be obtained, they would offer valuable benchmarks for assessing the viability of $F$-term uplifting as a genuine alternative to the anti-D3-brane scenario. 

\subsection*{Acknowledgments}
We thank Thibaut Coudarchet, Manki Kim, Liam McAllister, Fernando Quevedo, Muthusamy Rajaguru, and Max Wiesner for fruitful discussions. The research of AS is supported by NSF grant PHY-2309456. AH was supported by Deutsche Forschungsgemeinschaft (DFG, German Research Foundation) under Germany’s Excellence Strategy EXC 2181/1 - 390900948 (the Heidelberg STRUCTURES Excellence Cluster).

\appendix

\section{Derivation of the Einstein equations}\label{app:EFE}

In this Appendix we provide some more details on the derivation of the 10d Einstein equations. This is not new and the results can also be found in \cite{Giddings:2001yu,Gandhi:2011id,McGuirk:2012sb} but here, we carefully take into account all contributions together including local terms, and non-vanishing 4d curvature.
The general strategy is to first derive the 10d trace reversed Einstein equations and then specify to the internal and external components. 

The bosonic part of the Type~IIB action in Einstein frame is given by 
\begin{equation}\label{eq:SIIB}
    S_{\mathrm{IIB}}=  \frac{1}{2\kappa_{10}^2}\int\dd^{10}x \sqrt{-G} \left[ R_{10} -\frac{(\partial_M\tau)(\partial^M\overline{\tau})}{2\left(\Im\tau\right)^2} -\frac{G_{3}\cdot\overline{G}_{3}}{12\Im\tau} - \frac{\tilde{F}_{5}^2}{4\cdot5!}\right] + S_{\mathrm{CS}}\,.
\end{equation}
We want to derive the trace reversed Einstein equations 
\begin{equation}
    R_{MN} = \kappa_{10}^2 \left( T_{MN} -\frac{1}{D-2} G_{MN} T\right) \,,
\end{equation}
where $T_{MN} = T_{MN}^\mathrm{sugra}+T_{MN}^{\mathrm{loc}}$, and
\begin{equation}
G_{MN} =
    \begin{pmatrix}
        g_{\mu\nu}(x,y) & 0 \\
        0  & g_{mn} (y)
    \end{pmatrix}
    = 
    \begin{pmatrix}
        \mathrm{e}^{2A(y)} \tilde g _{\mu\nu}(x) & 0 \\ 0 & \mathrm{e}^{-2A(y)} \tilde g_{mn}(y)
    \end{pmatrix}\,,
\end{equation}
where $M,N=0,\dots,9$. The energy momentum tensor is defined as 
\begin{equation}
    T_{MN} = - \frac{2}{\sqrt{-G}}\, \frac{\delta S_\mathrm{IIB}}{\delta G^{MN}}\,,
\end{equation}
and in our case given by
\begin{equation}
\begin{split}
     T_{MN}^\mathrm{sugra} = & \frac{1}{\kappa_{10}^2} \Biggl[\frac{\partial_{(M}\tau\partial_{N)}\bar \tau}{2(\Im\tau)^2} + \frac{G_{(M}^{~PQ}\bar G_{N)PQ}}{4\Im\tau} + \frac{1}{4\cdot 4!} F_{(M}^{~PQRS}F_{N)PQRS} \\
     & - G_{MN} \left(\frac{\partial_{P}\tau\partial^{P}\bar \tau}{4(\Im\tau)^2} +\frac{|G_3|^2}{24\Im\tau} + \frac{\tilde F_5^2}{8\cdot 5!}\right)\Biggr]\,,
\end{split}
\end{equation}
and therefore 
\begin{equation}
    T^\mathrm{sugra} = (T^{~M}_M)^\mathrm{sugra} =- \frac{1}{\kappa_{10}^2} \left(2 \frac{\partial_{M}\tau\partial^{M}\bar \tau}{(\Im\tau)^2} + \frac{|G_3|^2}{6\Im\tau}\right)\,.
\end{equation}
In addition, one finds the local contribution
\begin{equation}
    T^{\mathrm{loc}}_{\mu\nu} = -T_p g_{\mu\nu} \,\delta(\Sigma)\qquad T^{\mathrm{loc}}_{mn} = -T_p \,(\Pi^\Sigma)_{mn}\,\delta(\Sigma)\,,
\end{equation}
where $\delta(\Sigma)$ and $(\Pi^\Sigma)_{mn}$ denote the delta distribution and projector on the cycle $\Sigma$ wrapped by the localised object, as also defined below \eqref{eq:eomPhi}. Note that for D3/O3, $T_{mn}^{\mathrm{loc}}=0$.  
Before writing down the internal Einstein equations, we note that 
\begin{align}
\label{eq:F52}
    \frac{\tilde F_5^2}{8\cdot 5!} & =  0 \,,\\
    \label{eq:F5external}
    \frac{ F_{(\mu}^{~PQRS}F_{\nu)PQRS}}{4\cdot 4!} & =  -\frac{\mathrm{e}^{-8A}g_{\mu\nu}}{4} (\partial\alpha)^2 = -\frac{\mathrm{e}^{-4A} \tilde g_{\mu\nu}}{4}\,(\tilde\partial\alpha)^2\,,\\
    \frac{F_{(m}^{~PQRS}F_{n)PQRS}}{4\cdot 4!} & =\mathrm{e}^{-8A}\left( -  \frac{\partial_{(m}\alpha \partial_{n)}\alpha}{2} + \frac{g_{mn}}{4} (\partial\alpha)^2\right)\,,\label{eq:F5internal}\\
    R_{mn} & =  \tilde R_{mn} + \tilde g_{mn} \tilde\nabla^2 A -8\tilde\nabla_mA\tilde\nabla_n A \label{eq:R6conformal} \\
    &= \tilde R_{mn} + \frac{\mathrm{e}^{-4A}\tilde g_{mn}}{4}\left( \tilde\nabla^2\mathrm{e}^{4A} - \mathrm{e}^{-4A} \left(\tilde\nabla \mathrm{e}^{4A}\right)^2\right) - \frac{\mathrm{e}^{-8A}}{2} \tilde\nabla_m \mathrm{e}^{4A} \tilde\nabla_n\mathrm{e}^{4A} \nonumber\,,
\end{align}
where $ R_{mn} = R^P_{~mPn}$ are the internal components of the 10d Ricci tensor, $\tilde R_{mn} = \tilde R^q_{~mqn}$ is the 6d Ricci tensor with respect to $\tilde g_{mn}$, and \eqref{eq:F52} is due to the self-duality property of $F_5$. In \eqref{eq:F5external} we used that the only non-vanishing components of $F_5$ are $F_{m\mu\nu\rho\sigma} = \epsilon_{\mu\nu\rho\sigma} \partial_m \alpha$ and $F_{mpqrs} = \epsilon_{mpqrst} \partial^t\alpha$. In the same way, one also finds \eqref{eq:F5internal}.
Equ.~\eqref{eq:R6conformal} follows from the conformal transformation of the Ricci tensor as can for instance be checked using \cite{Carroll:2004st} (see App.~G, equations (G.15), (G.17) and (G.18)).

Now we can put everything together and obtain the internal Einstein equations. They read
\begin{equation}
\label{eq:R6internalNew}
    \begin{split}
        \tilde R_{mn} & = \frac{\partial_{(m}\tau\partial_{n)}\bar\tau}{2(\Im\tau)^2} + \frac{\mathrm{e}^{4A}}{4\Im\tau} G_{(m}^{~~\widetilde{pq}}\bar G_{n)pq} + \frac{\mathrm{e}^{-8A}}{2} \left(\partial_{(m}\mathrm{e}^{4A} \partial_{n)}\mathrm{e}^{4A} - \partial_{(m}\alpha\partial_{n)}\alpha\right) \\
        & \qquad- \frac{\mathrm{e}^{-4A}\tilde g_{mn}}{4} \left( \frac{\mathrm{e}^{8A}|\tilde G_3|^2}{12\Im\tau } - \mathrm{e}^{-4A}\left((\tilde\partial\alpha)^2 + \left(\tilde\partial\mathrm{e}^{4A}\right)^2 \right)+ \tilde\nabla^2\mathrm{e}^{4A}\right)\\
        & \qquad + \kappa_{10}^2 \mathrm{e}^{-2A}\left(\tilde T^{\mathrm{loc}}_{mn} -\frac{\tilde g_{mn}}{8} T^{\mathrm{loc}} \right) \,.
    \end{split}
\end{equation}
It can be checked that $\tilde R_6=\tilde R^m_{~m}=0$ for GKP solutions with $G_-=0$ upon using the warp factor equation. 

Let us compare \eqref{eq:R6internalNew} with the internal Einstein equations obtained in the literature given by 
\begin{equation}
\label{eq:R6app}
    \begin{split}
    \tilde R_{mn} & = \frac{\partial_{(m}\tau\partial_{n)}\bar\tau}{2(\Im\tau)^2}  +\frac{2}{(\Phi_++\Phi_-)^2} \partial_{(m}\Phi_+\partial_{n)}\Phi_-
     - \tilde g_{mn}\, \frac{\tilde{\mathcal{R}}_4}{2(\Phi_++\Phi_-)} \\
     & \qquad - \frac{\Phi_++\Phi_-}{32\Im\tau} \left(\tilde G_{+(m}^{~~pq}\bar G_{-n)pq} + \tilde G_{-(m}^{~~pq}\bar G_{+n)pq}\right)  \,,
    \end{split}
\end{equation}
where $\tilde{\mathcal{R}}_4$ denotes the 4d Ricci scalar, i.e., $\tilde{\mathcal{R}}_4=\tilde g^{\mu\nu}\tilde R^\rho_{~\mu\rho\nu}$. It can be determined by the trace of the 4d Einstein equations:
\begin{equation}
\label{eq:warpeq}
    - \tilde{\mathcal{R}}_4 = - \tilde\nabla^2\mathrm{e}^{4A} + \frac{\mathrm{e}^{8A} |\tilde G_3|^2}{12 \Im\tau} + \mathrm{e}^{-4A}\left( \left(\tilde\partial\mathrm{e}^{4A}\right)^2 + (\tilde\partial\alpha)^2\right)+\frac{\kappa_{10}^2}{2} \mathrm{e}^{2A} (\tilde T^m_m-\tilde T^\mu_\mu )^{\mathrm{loc}}\,.
\end{equation}
Thus, using \eqref{eq:warpeq}, \eqref{eq:R6app} takes the form 
\begin{equation}
\begin{split}
    \label{eq:R6app1}
    \tilde R_{mn} & = \frac{\partial_{(m}\tau\partial_{n)}\bar\tau }{2(\Im\tau)^2}+\frac{\mathrm{e}^{-8A}}{2} \left(\partial_{(m}\mathrm{e}^{4A}\partial_{n)}\mathrm{e}^{4A}-\partial_{(m}\alpha\partial_{n)}\alpha\right)\\
    &\quad - \frac{\mathrm{e}^{4A}}{16\Im\tau} \left( G_{+(m}^{~~~~\widetilde{pq}}\bar G_{-n)pq} +  G_{-(m}^{~~~~\widetilde{pq}}\bar G_{+n)pq}\right) \\
     & \quad  - \frac{\mathrm{e}^{-4A}\tilde g_{mn}}{4} \left(  \tilde\nabla^2\mathrm{e}^{4A} - \frac{\mathrm{e}^{8A} |\tilde G_3|^2}{12 \Im\tau} - \mathrm{e}^{-4A}\left( \left(\tilde\partial\mathrm{e}^{4A}\right)^2 + (\tilde\partial\alpha)^2\right) \right) \,.
\end{split}
\end{equation}
Comparing this with \eqref{eq:R6internalNew} we recognise a difference regarding the $G_3$ terms. Let us check given which assumptions the two equations are the same. 

We start by rewriting the $G_3$ flux term into $G_\pm$ using $G_3 = (G_+-G_-)/2\mathrm{i}$ and $\bar G_3 =- (\bar G_+ -\bar G_-)/2\mathrm{i}$. One finds
\begin{equation}
\label{eq:Gcontributionapp}
    G_{(m}^{~~~\widetilde{pq}}\bar G_{n)pq} = \frac{1}{4} \left( G_{+(m}^{~~\widetilde{pq}}\bar G_{+n)pq} + G_{-(m}^{~~\widetilde{pq}}\bar G_{-n)pq} - G_{(+m}^{~~\widetilde{pq}}\bar G_{-n)pq} - G_{(-m}^{~~\widetilde{pq}}\bar G_{+n)pq} \right)\,.
\end{equation}
By writing the last two terms in complex coordinates, it can be seen that they are only non-zero if the open indices are either both holomorphic or anti-holomorphic. They therefore do not contribute to the trace of $G_{(m}^{~~~\widetilde{pq}}\bar G_{n)pq}$. 

In order to evaluate the first two terms, we make use of the self-duality conditions of $G_\pm$, i.e.,~$\star_6G_\pm =\pm i G_\pm$ and $\star_6\bar G_\pm=\mp i \bar G_\pm$. We calculate  
\begin{equation}
    G_{\pm(m}^{~~~~\widetilde{pq}}\bar G_{\pm n)pq} = (\star_6G_\pm)_{(m}^{~~~~\widetilde{pq}} (\star_6 \bar G_\pm)_{n)pq} = \frac{g_{nm}}{3} |G_\pm|^2 -   G_{\pm(m}^{~~~~\widetilde{pq}}\bar G_{\pm n)pq}\,,
\end{equation}
and therefore find 
\begin{equation}
    G_{\pm(m}^{~~~~\widetilde{pq}}\bar G_{\pm n)pq} = \frac{g_{nm}}{6} |G_\pm|^2 \,.
\end{equation}
All in all, the flux term \eqref{eq:Gcontributionapp} yields
\begin{equation}
    G_{(m}^{~~~\widetilde{pq}}\bar G_{n)pq} = \frac{g_{nm}}{6}|G_3|^2 -\frac{1}{4} \left(  G_{(+m}^{~~\widetilde{pq}}\bar G_{-n)pq} + G_{(-m}^{~~\widetilde{pq}}\bar G_{+n)pq} \right)\,.
    \label{eq:fluxappfinal}
\end{equation}
Plugging in \eqref{eq:fluxappfinal} into \eqref{eq:R6internalNew} we obtain exactly \eqref{eq:R6app1} and therefore find agreement between the literature and the equations of motion derived in this Appendix. 

To summarise, the internal Einstein equations are then given by  
\begin{equation}\label{eq:R6internalNewfull}
\begin{split}
    \tilde R_{mn} & = \frac{\partial_{(m}\tau\partial_{n)}\bar\tau}{2(\Im\tau)^2} + \frac{2\partial_{(m}\Phi_+\partial_{n)}\Phi_-}{(\Phi_++\Phi_-)^2} - \frac{\Phi_++\Phi_-}{32\Im\tau} \left( G_{+(m}^{~~~\widetilde{pq}}\bar G_{-n)pq} + G_{-(m}^{~~~\widetilde{pq}}\bar G_{+n)pq}\right) \\
        & \quad- \frac{\tilde g_{mn}}{2(\Phi_++\Phi_-)} \Biggl( - \frac{(\Phi_++\Phi_-)^2|\tilde G_3|^2}{48\Im\tau } - \frac{2}{\Phi_++\Phi_-}\left((\tilde\partial\alpha)^2 + \left(\tilde\partial\mathrm{e}^{4A}\right)^2 \right)\\
        & \quad+ \tilde\nabla^2\mathrm{e}^{4A}-\frac{\kappa_{10}^2}{2} \mathrm{e}^{2A} (\tilde T^m_m-\tilde T^\mu_\mu )^{\mathrm{loc}}\Biggr) + \kappa_{10}^2 \mathrm{e}^{-2A}\left(\tilde T^{\mathrm{loc}}_{mn} -\frac{\tilde g_{mn}}{4} (\tilde T^p_p)^{\mathrm{loc}} \right) \,.
\end{split}
\end{equation}
All terms in the bracket of the term $\sim \tilde g_{mn}$ are equal to $\tilde {\mathcal{R}}_4$ on-shell when using the Bianchi identity of $\tilde F_5$ and the trace of the 4d Einstein equations.

\section{Integrating out KK-modes and off-shell potentials} \label{app:KK}

In the main text of the paper we consider scalar potentials that are evaluated away from their minima.
This confronts us with the problem that not all higher-dimensional equations of motion can be solved. 
We circumvent this issue by manually subtracting the zero-mode components of these equations that obstruct their integrability.
In this appendix we provide further motivation for the validity of this procedure.

As a simple toy model, we consider the action of a $D$-dimensional scalar field $\phi$, coupled to a source term $\rho$,
\begin{equation}\label{eq:app:Daction}
    S = - \int d^D x \sqrt{-g} \left(\tfrac12 \partial_M \phi \partial^M \phi +  \phi \rho \right) \,.
\end{equation}
Assuming that $\rho$ is independent of $\phi$, the corresponding equation of motion reads
\begin{equation}\label{eq:app:Deom}
    \Box^{(D)} \phi = \rho \,,
\end{equation}
with $\Box^{(D)}$ the usual $D$-dimensional d'Alembert operator.

We consider the compactification of this theory on a compact $d$-dimensional space $X_d$ down to four dimensions, and split our $D$-dimensional coordinates accordingly, $x^M =  \left(x^\mu, y^i\right)$.
For a simple product compactification (e.g., in the absence of warping), this split is also respected by the higher-dimensional d'Alembertian,
\begin{equation}
    \Box^{(D)} = \Box^{(4)} + \Delta^{(d)} \,,
\end{equation}
with $\Delta^{(d)}$ the Laplacian on $X_d$.

In order to preserve all four-dimensional spacetime symmetries, we assume that the source term $\rho$ depends only on the coordinates of the internal space $X_d$,
\begin{equation}
    \rho = \rho(y) \,.
\end{equation}
If $\rho$ has non-zero integral over the internal space, 
we can integrate the equation of motion \eqref{eq:app:Deom} over $X_d$, and use that $\Delta^{(d)} \phi$ is a total derivative,
\begin{equation}
    \int_{X_d} \Box^{(4)} \phi = \int_{X_d} \rho \neq 0 \,,
\end{equation}
to show that any solution for $\phi$ must necessarily have a non-trivial profile along the four-dimensional spacetime directions and cannot be constant.
From an effective, four-dimensional point of view this means that in the compactified theory $\phi$ has a non-trivial potential without a minimum, as otherwise the constant solution would be possible.

We now illustrate how to derive such an effective, off-shell potential for the zero-mode of $\phi$ along $X_d$.
\emph{Off-shell} here simply means that we are not restricting ourselves to the minimum of the potential, and that thus the higher-dimensional equation of motion cannot be solved by $\phi$ being constant in the four-dimensional spacetime directions.
As usual, we proceed by decomposing $\phi$ into eigenmodes of the Laplacian on $X_d$,
\begin{equation}\label{eq:app:phidecomp}
    \phi(x,y) = \sum_I \phi_I(x) Y^I(y) \,,
\end{equation}
that satisfy the eigenvalue equation
\begin{equation}
    \Delta^{(d)} Y^I = - \lambda^I Y^I \,,
\end{equation}
and that we assume to be orthonormal,
\begin{equation}
    \int_{X_d} Y^I Y^J = \delta^{IJ} \,.
\end{equation}
For example, in the case of a simple circle-compactification, we can set  $Y^n \sim \mathrm{e}^{i n y / L}$, and \eqref{eq:app:phidecomp} is just a Fourier decomposition.
We further split the eigenmodes $Y^I$ into zero modes $Y^{I_0}$ and higher KK modes $Y^{\tilde I}$, so that
\begin{equation}
    \lambda^{I_0} = 0 \,,\quad \mathrm{and} \quad \lambda^{\tilde I} \neq 0 \,.
\end{equation}
In the scalar case that we consider here, with $X_d$ connected, there is, of course, only one single zero mode $Y^0 = \mathrm{const}$.
For fields of higher spin, for example, for the deformation modes of the internal metric on a space with non-trivial topology, there can, however, be multiple, different zero modes.

Inserting the decomposition ansatz \eqref{eq:app:phidecomp} into the $D$-dimensional action \eqref{eq:app:Daction} and integrating over $X_d$ gives the four-dimensional action
\begin{equation}\label{eq:app:4daction}
    S_4 = - \int d^4x \sqrt{-g_4} \left(\frac12 \sum_I \partial_\mu \phi_I \partial^\mu \phi_I + V(\phi_I) \right) \,.
\end{equation}
Here, the potential $V(\phi_I)$ is our main object of interest, and is obtained by collecting all terms in the higher-dimensional action without a four-dimensional space-time derivative.
It is given by
\begin{equation}\label{eq:app:4dpotentialfull}
    V(\phi_I) = \sum_I \left( \tfrac12 \lambda^I \phi_I^2 + \rho_I \phi_I \right) \,.
\end{equation}
Equivalently, it is obtained from the action \eqref{eq:app:Deom} by treating $\phi$ as a function of the internal coordinates only, $\phi = \phi(y)$, or equivalently, by treating all expansion coefficients $\phi_I$ in the mode decomposition \eqref{eq:app:phidecomp} as constants,
\begin{equation}
    \int d^4x \sqrt{-g_4} V(\phi_I) = - S \bigr|_{\phi_I = \mathrm{const.}} \,.
\end{equation}

So far, \eqref{eq:app:4daction} is nothing but a rewriting of the original action in internal momentum space.
To obtain an effective, four-dimensional action for the zero mode $\phi_0$, we want to integrate out the higher-order KK modes $\phi_{\tilde I}$ using their equations of motion.
For $\phi = \phi(y)$, the equation of motion \eqref{eq:app:Deom} reads $\Delta^{(d)} \phi = \rho$, and becomes, in terms of the mode decomposition, 
\begin{equation}
    - \lambda^I \phi_I = \rho_I \qquad \mathrm{(no \,\,summation)} \,.
\end{equation}
Evidently, the equation for the zero mode $\phi_0$ only has a solution if $\rho_0 = 0$.
This is the same observation that we have already made above, and warrants the name off-shell potential.

However, since we want to keep the zero mode as a dynamical field in our effective action, we are only interested in the equations for the higher modes $\phi_{\tilde I}$, and can ignore the zero mode equation.
In position space, ignoring the zero mode equation is the same as considering the modified equation
\begin{equation}\label{eq:app:zeromodesubtraction}
    \Delta^{(d)} \phi = \rho - \rho_0 \,, \qquad \text{with} \qquad \rho_0 = \int_{X_d} \rho \,.
\end{equation}
This equation is obtained by subtracting the zero mode contribution from the original equation, and can now be solved as its right hand side integrates to zero over $X_d$.
It can be understood as a projection of the original equation onto the modes orthogonal to the constant zero-mode $Y^0$.
In momentum space its solution is given by
\begin{equation}
    \phi_{\tilde I} = - \frac{\rho_{\tilde I}}{\lambda^{\tilde I}} \,,
\end{equation}
and $\phi_0$ arbitrary.

The same set of equations is equivalently obtained by extremising the potential \eqref{eq:app:4dpotentialfull} with respect to the higher modes $\phi_{\tilde I}$ while keeping the zero mode $\phi_0$ fixed,
\begin{equation}
    \frac{\partial}{\partial \phi_{\tilde I}} V(\phi_0, \phi_{\tilde I}) = 0 \,.
\end{equation}
In either case, inserting the solutions for $\phi_{\tilde I}$ back into \eqref{eq:app:4dpotentialfull} yields an effective potential for $\phi_0$,
\begin{equation}
    V_\mathrm{eff}(\phi_0) =   \phi_0 \rho_0 - \sum_{\tilde I} \frac{\rho_{\tilde I}^2}{2 \lambda_{\tilde I}} \,.
\end{equation}
This potential satisfies
\begin{equation}
    \rho_0 = \frac{\partial V_\mathrm{eff} }{ \partial \phi_0 }  \,,
\end{equation}
and hence nicely reproduces our earlier observation that $V_\mathrm{eff}$ has no minimum unless $\rho_0 = 0$, in which case it is constant.

In the main text of the paper, we face a similar situation as in this toy model, where not all higher-dimensional equations can be solved unless we allow for a non-trivial profile of the fields in the external four-dimensional spacetime directions.
As illustrated here, this situation corresponds to an effective, four-dimensional potential away from its minima or extrema, inducing a rolling (or otherwise non-trivial dynamics) of the four-dimensional fields.

In the main text, we do not perform the full decomposition into eigenmodes of the relevant Laplacian operators explicitly.
However, we still want to separate the equations of motion into a zero mode part, and a part that corresponds to the higher KK modes.
The zero mode part is not solvable unless we are at a critical point of the potential.
The remaining equations, however, can be used to integrate out the KK modes, and to obtain an effective potential for the zero-modes.
In the absence of an explicit mode decomposition, we implement this split by subtracting the zero modes from the equations of motion as in \eqref{eq:app:zeromodesubtraction}, where the additional term has to be chosen so that the right hand side of the equation integrates to zero.

Of course, our toy model is oversimplified in the sense that its equation of motion is linear and does not include any interaction terms.
Therefore, the equations of motion of the KK modes decouple and their solution does not depend on the value of the light field $\phi_0$.
As a consequence, integrating them out has little effect on the effective physics of the zero-mode, and only contributes an additive constant to its potential.
In more realistic setups, such as the one discussed in the main text, these simplifying assumptions are generally not satisfied, and less trivial dynamics may arise from integrating out the higher modes.

\section{Identities from special geometry} \label{app:identities}

In this Appendix, we calculate the contractions of $G_\pm^{(0)}$ used in Section \ref{sec:matching10dand4d} in terms of the harmonic forms $\Omega,\chi_i,\bar\chi_{\bar\imath}$, and $\bar\Omega$ on the Calabi-Yau. 

We start with the definition of $\Omega$ and $\chi_i$ in complex coordinates:
\begin{equation}
    \Omega = \frac{1}{3!} \, \Omega_{\mu\nu\rho} \, \dd z^\mu\wedge \dd z^\nu \wedge \dd z^\rho\,,\qquad \chi_i = \frac{1}{2} (\chi_i)_{\mu\nu\bar\rho} \,\dd z^\mu\wedge \dd z^\nu \wedge \dd \bar z^{\bar\rho}\,,
\end{equation}
where $i=1,\dots,h^{2,1}$. From the definition of the Weil-Petersson metric 
\begin{equation}
\label{eq:WPmetric}
    K_{i\bar\jmath} = - \frac{\int\chi_i\wedge\bar\chi_{\bar\jmath}}{\int\Omega\wedge\bar\Omega} = -\mathrm{i} \mathrm{e}^{K_\mathrm{cs}} \int\chi_i\wedge\bar\chi_{\bar\jmath}\,,
\end{equation}
where
\begin{equation}
    K_\mathrm{cs} = -\ln\left( \mathrm{i} \int \Omega\wedge\bar\Omega \right)\,,
\end{equation}
we can obtain a local version of \eqref{eq:WPmetric} which is given by 
\begin{equation}
    -\chi_i\wedge\bar\chi_{\bar \jmath} = \Omega \wedge \bar\Omega\, K_{i\bar\jmath} + \frac{1}{3!} \star^{(0)} F_{i\bar\jmath}\,.
\end{equation}
Here, $F_{i\bar\jmath}$ is a non-trivial scalar function on $X_6$ such that $\star^{(0)} F_{i\bar\jmath}$ integrates to zero and one recovers \eqref{eq:WPmetric}.
It parametrises the failure of $\chi_i\wedge\bar\chi_{\bar \jmath}$ to be harmonic.
In complex coordinates, this relation can also be written as 
\begin{equation}
    \label{eq:WPlocal}
    (\chi_i)_{\mu\nu\bar\rho} (\bar\chi_{\bar\jmath})^{\mu\nu\bar\rho} = 2 K_{i\bar\jmath} \|\Omega\|^2 +\frac13 F_{i\bar\jmath}\,,
\end{equation}
where we defined $\|\Omega\|^2=\Omega_{\mu\nu\rho}\bar \Omega^{\mu\nu\rho}/3!$.
Together with
\begin{equation}
    G_+^{(0)} = A^i \chi_i + \bar B \bar\Omega \,, \qquad G_-^{(0)} = A\Omega +\bar B^{\bar \imath} \bar\chi_{\bar \imath}\,,
\end{equation}
one then finds 
\begin{equation}\begin{aligned}
    G_-^{(0)} \tilde\cdot\,\bar G_-^{(0)} & = 3! |A|^2\|\Omega\|^2+3\bar B^{\bar \jmath} (\chi_{\bar\jmath})_{\mu\bar\nu\bar\rho}B^i(\chi_i)^{\mu\bar\nu\bar\rho} \\
    &= 3!\|\Omega\|^2\left(|A|^2 + \bar B_i B^i\right) +F_{i\bar\jmath} B^i\bar B ^{\bar\jmath} \,,\\
\end{aligned}\end{equation}
and 
\begin{equation}\begin{aligned}
    G_+^{(0)} \tilde\cdot\, G_-^{(0)} &= 3!A\bar B \|\Omega\|^2 + 3A^i( \chi_i)_{\mu\nu \bar\rho}\bar B^{\bar\jmath}(\bar\chi_{\bar\jmath})^{\mu\nu\bar\rho} \\
    &= 3! \| \Omega \|^2 \left( A \bar B + A^i \bar B_i \right)+F_{i\bar\jmath}A^i \bar B^{\bar\jmath} \,,
\end{aligned}\end{equation}
where we used \eqref{eq:WPlocal}.
Rewriting the flux term in the equation of motion for the metric \eqref{eq:eomg1}, requires an identity with two open spacetime indices. Using the well-known formulas \cite{Candelas:1990pi}
\begin{equation}\label{eq:spgi}
        D_i\Omega =\chi_i\,, \qquad D_i\chi_j= -\mathrm{i\,e}^{K_\mathrm{cs}} \kappa_{ij}^{~~\bar k} (\bar \chi_{\bar k})\,,\qquad \kappa_{ijk} =-\mathrm{i}\int \Omega \wedge D_iD_jD_k \Omega\,,
\end{equation}
we find in complex coordinates
\begin{equation}
    \frac12 \left(D_iD_j\Omega\right)_{\mu\bar\nu\bar\rho} = -\frac{\mathrm{i}}2 \mathrm{e}^{K_\mathrm{cs}} \kappa_{ij}^{~~\bar k} (\bar \chi_{\bar k})_{\mu\bar\nu\bar\rho}\,.
\end{equation}
Contracting this with $\Omega_{\sigma}^{~~\bar\nu\bar\rho}$, and using the symmetric Beltrami differentials
\begin{equation}
    (\mu_i)^\mu_{~\bar\mu} = \frac{1}{2\|\Omega\|^2} \bar\Omega ^{\mu\nu\rho} (\chi_i)_{\nu\rho\bar\mu}\,,
\end{equation}
one finds 
\begin{equation}
\label{eq:identity}
    \frac12 (D_iD_j\Omega)_{\mu\bar\nu\bar\rho} \Omega_{\sigma}^{~~\bar\nu\bar\rho} = -\frac{\mathrm{i}}{2} \mathrm{e}^{K_\mathrm{cs}} \kappa_{ij}^{~~\bar k} (\bar\chi_{\bar k})_{\mu\bar\nu\bar\rho} \Omega_\sigma^{~~\bar\nu\bar\rho} = (\chi_i)_{\mu\nu\bar\rho} (\chi_j)_{\sigma}^{~~\nu\bar\rho}\,.
\end{equation}
With this we obtain 
\begin{equation}
\begin{aligned}
    &(\bar G_+^{(0)})_{(\mu}^{~~\bar\nu\bar \rho}(G_-^{(0)})_{\sigma)\bar\nu\bar \rho} + (\bar G_+^{(0)})_{(\mu}^{~~\nu \rho}(G_-^{(0)})_{\sigma)\nu\rho}+2(G_+^{(0)})_{(\mu}{}^{\bar\nu \rho}(\bar G_-^{(0)})_{\sigma)\bar\nu\rho}+\mathrm{c.c.} \\
    & \qquad =(\bar\chi_{\bar\imath})_{\bar\nu\bar\rho(\mu} \Omega_{\sigma)}{}^{\bar\nu\bar\rho} \left(B \bar B ^{\bar\imath} + A\bar A^{\bar\imath}\right) + 2(\chi_j)_{\nu\bar\rho(\mu}(\chi_k)_{\sigma)}^{~~\nu\bar\rho} A^j B^k +\mathrm{c.c.}\\
    & \qquad = (\bar\chi_{\bar\imath})_{\bar\nu\bar\rho(\mu} \Omega_{\sigma)}{}^{\bar\nu\bar\rho} \left(B \bar B ^{\bar\imath} + A\bar A^{\bar\imath} -\mathrm{i} \mathrm{e}^{K_\mathrm{cs}} A^jB^k \kappa_{jk}{}^{\bar\imath}\right) +\mathrm{c.c.}\,,
\end{aligned}
\end{equation}
where we used \eqref{eq:identity} to obtain the last line.  

\section{Equations of motion at order \texorpdfstring{$1/c^2$}{1/c2}}\label{app:g2mn}

Even though not explicitly needed for estimating the leading order corrections to the scalar potential $V_\mathrm{eff}$ in Section~\ref{sec:10danalysis}, we want to derive the equations of motion at order $1/c^2$ in this Appendix for future reference.
For $\Phi^{(2)}_\pm$ we find
\begin{align}
    \begin{split}
        \tilde\nabla^2\Phi_+^{(2)} & = \frac{G_+^{(1)} \tilde\cdot \,\bar G_+^{(0)}+\mathrm{c.c.} } {24\Im\tau^{(0)}} + \frac{G_+^{(0)} \tilde\cdot \,\bar G_+^{(0)} }{24\Im\tau^{(0)}}\left(\Phi_+^{(1)}+\Phi_-^{(1)}-\frac{\Im\tau^{(1)}}{\Im\tau^{(0)}}\right)\\
        & \qquad + |\tilde\partial\Phi_+^{(1)}|^2 + 4 \kappa_{10}^2 T_3  \,\tilde\rho^{\mathrm{loc}}_3\left(\Phi_+^{(1)}+\Phi_-^{(1)}\right) + \tilde{\mathcal{R}}_4 - \mathcal{C}_{\Phi^{(2)}}\,,
        \label{eq:eomPhi+2}
    \end{split}\\
    \begin{split}
        \tilde\nabla^2\Phi_-^{(2)} & = \frac{G_-^{(1)} \tilde\cdot \,\bar G_-^{(0)}+\mathrm{c.c.} } {24\Im\tau^{(0)}} + \frac{G_-^{(0)} \tilde\cdot \,\bar G_-^{(0)} }{24\Im\tau^{(0)}}\left(\Phi_+^{(1)}+\Phi_-^{(1)}-\frac{\Im\tau^{(1)}}{\Im\tau^{(0)}}\right)\\
        & \qquad + |\tilde\partial\Phi_-^{(1)}|^2  + \tilde{\mathcal{R}}_4 - \mathcal{C}_{\Phi^{(2)}}\,,
        \label{eq:eomPhi-2}
    \end{split}
\end{align}
where it can be shown that, upon using the tadpole cancellation \eqref{eq:bianchi} at order $1/c^2$, zero mode contributions in \eqref{eq:eomPhi+2} and \eqref{eq:eomPhi-2} are the same.
For the dilaton at order $1/c^2$, we obtain (for locally cancelled D7 tadpoles)
\begin{equation}
        \hspace{-.7cm}\tilde\nabla^2\tau^{(2)} = -\frac{\mathrm{i}\,\left(\tilde\partial\tau^{(1)}\right)^2}{\left(\Im\tau^{(0)}\right)^2} -\frac{\mathrm{i}}{24} \left(G_+^{(0)}\tilde\cdot\,G_-^{(1)}+G_+^{(1)}\tilde\cdot\,G_-^{(0)}\right) +\mathrm{i} \frac{\Phi_+^{(1)}+\Phi_-^{(1)}}{48} G_+^{(0)}\tilde\cdot\,G_-^{(0)}-\mathcal{C}_{\tau^{(2)}}\,.
\end{equation}
The equation of motion of $\tilde g^{(2)}_{mn}$ is determined by expanding the internal Einstein equation \eqref{eq:eomR} at order $1/c^2$. Using $\tilde g^{(1)}_{mn}=0$, we obtain 
\begin{equation}
\begin{split}
    \hspace{-.4cm}-\frac12\Delta \tilde g^{(2)}_{mn} &=   \frac{\partial_{(m}\Phi_+^{(1)}\partial_{n)}\Phi_-^{(1)}}{2} - \left(G_{+(m} ^{(0)~~\widetilde{pq}} \bar G_{-n)pq}^{(0)}+\mathrm{c.c.}\right)\left(\frac{\Phi_+^{(1)}+\Phi_-^{(1)}-\frac{2\Im\tau^{(1)}}{\Im\tau^{(0)}}}{32\Im\tau^{(0)}}\right)\\
    & \quad+\frac{\partial_{(m}\tau^{(1)}\partial_{n)}\bar\tau^{(1)}}{2(\Im\tau^{(0)})^2}- \frac{1}{16\Im\tau^{(0)}} \left(G_{+(m} ^{(0)~~\widetilde{pq}} \bar G_{-n)pq}^{(1)} + G_{+(m} ^{(1)~~\widetilde{pq}} \bar G_{-n)pq}^{(0)}+\mathrm{c.c.}\right)\\
    & \quad+ \left(\tilde{\mathcal{R}}_4-\mathcal{C}_{\Phi^{(2)}}\right) \left(\Phi_+^{(1)}+\Phi_-^{(1)}\right)\frac{\tilde g^{(0)}_{mn}}{8} - (\mathcal{C}_{\tilde g^{(2)}})^M(\psi_M)_{mn}\,.
    \end{split}
    \label{eq:eomg2}
\end{equation}
The term $\sim (\tilde{\mathcal{R}}_4-\mathcal{C}_{\Phi^{(2)}})$ comes about as in the derivation of the equation of motion of $\tilde g^{(1)}_{mn}$: The terms in the large bracket in the second and third line in \eqref{eq:eomR} partially cancel upon using the equations of motion for $\Phi_\pm$ at the corresponding order. Due to the triangular form of the set of equations of motion at each order in $1/c$, we can solve the equations for $\Phi_\pm^{(2)}$ without needing to know $\tilde g^{(2)}_{mn}$.
The only leftovers in the large bracket are then the zero mode contributions $(\tilde{\mathcal{R}}_4-\mathcal{C}_{\Phi^{(2)}})$. The zero mode contribution $\mathcal{C}_{\Phi^{(2)}}$ is the sum of the zero mode contributions of the equations of motion of $\Phi_\pm^{(2)}$. 

One can easily convince oneself from \eqref{eq:eomPhi-2} that $(\tilde{\mathcal{R}}_4-\mathcal{C}_{\Phi^{(2)}})$ scales at least linear in $\varepsilon$. Thus, each term in on the right hand side of \eqref{eq:eomg2} is at least linear in $\varepsilon$.

The Bianchi identity and equation of motion for $G_3$ combine to 
\begin{equation}
    \begin{split}
        \dd G_\pm^{(2)}  = & -\frac{1}{2} \left(\dd\left(\Phi_+^{(1)} G_-^{(1)}\right)+\dd\Phi_+^{(2)}\wedge G_-^{(0)}+\dd\left(\Phi_-^{(1)} G_+^{(1)}\right)+\dd\Phi_-^{(2)}\wedge G_+^{(0)}\right) \\
        & -\mathrm{i}\left(\frac{\dd\tau^{(2)}}{\Im\tau^{(0)}}-\frac{\Im(\tau^{(1)})\,\dd\tau^{(1)}}{\left(\Im\tau^{(0)}\right)^2}\right)\wedge \Re G_\pm^{(0)} -\frac{\mathrm{i}\,\dd\tau^{(1)}}{\Im\tau^{(0)}}\wedge \Re G_\pm^{(1)} \\
        & -\frac{\mathrm{i}\,\dd\tau^{(1)}}{2\Im\tau^{(0)}}\wedge\left(\Phi_+^{(1)}\Re G_-^{(0)}+\Phi_-^{(1)}\Re G_+^{(0)}\right)\,.
    \end{split}
\end{equation}

\section{Four-dimensional analysis in the multi-moduli case} \label{app:4d}

\subsection{Generalised no-scale structure} \label{app:many}

Our analysis takes inspiration from \cite{Burgess:2020qsc}. For convenience, we repeat the relevant K\"ahler potential motivated in the main text
\begin{equation}
    K[T^A,\overline{T}^{A},\phi^I,\bar{\phi}^I] = K_k[T^A+\overline{T}^{A}+f^A(\phi^I,\bar{\phi}^I)] + K_{\mathrm{cs}}[\phi^I,\bar{\phi}^I]\, ,
    \label{eq:Kk1}
\end{equation}
where we introduced
\begin{equation}
    K_{\mathrm{cs}}[\phi^I,\bar{\phi}^I] = -\ln\biggl (\mathrm{i}\int \Omega\wedge \ol{\Omega}\biggl )- \ln\left( -\mathrm{i}(\tau-\bar\tau)\right)\,.
\end{equation}
Here $\phi^I=(z^i,\tau)$ collects the complex structure moduli $z^i$ and the axio-dilaton $\tau$.
We then compute
\begin{equation}
    K_A = \dfrac{\partial K}{\partial T^A}\, ,\quad K_I = \dfrac{\partial K}{\partial \phi^I} = (\partial_I f^A) K_A +\partial_I K_{\mathrm{cs}}
\end{equation}
and similarly for $\overline{T}^{A},\bar{\phi}^{I}$. The K\"ahler metric is then given by
\begin{equation}
    K_{M\overline{N}} = \left (\begin{array}{cc}
        K_{A\bar{B}} & K_{A\bar{J}} \\
        K_{I\bar{B}} & K_{I\bar{J}}
    \end{array} \right ) 
\end{equation}
where the mixed components can be written as
\begin{equation}
    K_{A\bar{J}} = (\bar{\partial}_J f^{\bar B}) K_{A\bar{B}}\,.
\end{equation}
Moreover, the corrected metric on complex structure moduli space
\begin{equation}
    K_{I\bar{J}}=(\partial_I\bar{\partial}_J f^A)K_A+(\partial_I f^A)K_{A\bar{B}}(\bar{\partial}_J f^{\bar B})+(K_{\mathrm{cs}})_{I\bar{J}}
\end{equation}
may be expressed in terms of the uncorrected metric
\begin{equation}
    (K_{\mathrm{cs}})_{I\bar{J}} = \dfrac{\partial^2 K_{\mathrm{cs}}}{\partial \phi^I\partial \bar{\phi}^J}\, .
\end{equation}
Note that the functions $f$ can be read as objects with either a plain or with a barred index, $f^A\equiv f^{\bar A}$.

Using the standard formula for inverting a matrix with four blocks
\begin{equation}
    {\displaystyle {\!\!\!\begin{bmatrix}{A}&{B}\\{C}&{D}\end{bmatrix}}^{-1}={\begin{bmatrix}{A}^{-1}+{A}^{-1}{B}\left({D}-{CA}^{-1}{B}\right)^{-1}{CA}^{-1}&\,\,\,-{A}^{-1}{B}\left({D}-{CA}^{-1}{B}\right)^{-1}\\-\left({D}-{CA}^{-1}{B}\right)^{-1}{CA}^{-1}&\left({D}-{CA}^{-1}{B}\right)^{-1}\end{bmatrix}}\, ,}
\end{equation}
we find that the corrected inverse metric on complex structure moduli space is the inverse of
\begin{equation}
    {D}-{CA}^{-1}{B} = K_{I\bar{J}}-(\partial_I f^A)K_{A\bar{B}}(\bar{\partial}_J f^{\bar B}) = (\partial_I\bar{\partial}_J f^A)K_A+(K_{\mathrm{cs}})_{I\bar{J}}\, ,
\end{equation}
that is,
\begin{equation}\label{eq:IKMij4d}
    K^{I\bar{J}} = ((\partial_I\bar{\partial}_J f^A)K_A+(K_{\mathrm{cs}})_{I\bar{J}})^{-1}\, .
\end{equation}
The full inverse metric becomes
\begin{equation}
    K^{M\overline{N}} = \left (\begin{array}{cc}
        K^{A\bar{B}} & K^{A\bar{J}} \\
        K^{I\bar{B}} & K^{I\bar{J}}
    \end{array} \right )
\end{equation}
with
\begin{align}
    K^{A\bar{B}} &= K^{A\bar{B}}_k+K^{A\bar{C}}_k K_{\bar{C}I} K^{I\bar{J}} K_{\bar{J}D}  K^{D\bar{B}}_k\nonumber\\
    &= K^{A\bar{B}}_k+K^{A\bar{C}}_k (\partial_I f^E) K_{E\bar{C}} K^{I\bar{J}} (\bar{\partial}_J f^{\bar F}) K_{D\bar{F}}  K^{D\bar{B}}_k\nonumber\\
    &= K^{A\bar{B}}_k+ (\partial_I f^A)  K^{I\bar{J}} (\bar{\partial}_J f^{\bar B}) \,,
\end{align}
where we used $K_{D\bar{F}} = (K_k)_{D\bar{F}}$.
The mixed terms take the form
\begin{align}
    K^{A\bar{J}} &=-K^{A\bar{C}}_k K_{\bar{C}I} K^{I\bar{J}}=-(\partial_I f^A) K^{I\bar{J}}\, ,
    \label{eq:KAj}
\end{align}
and
\begin{align}
    K^{I\bar{B}} &=-K^{I\bar{J}} K_{\bar{J}D}  K^{D\bar{B}}_k=-(\bar{\partial}_J f^{\bar B}) K^{I\bar{J}}\, .
    \label{eq:KiB}
\end{align}
Assuming a superpotential which depends on complex structure moduli only,
\begin{equation}
    W = W(z^i,\tau)=W(\phi^I)\, ,
\end{equation}
we compute the $F$-term scalar potential
\begin{equation}
    V = \mathrm{e}^K \left( K^{M\bar N} D_M W \, \overline{D_{N} W} -3|W|^2 \right)\, .
\end{equation}
We first note that
\begin{equation}
    D_A W  = K_A W
\end{equation}
and then expand the complex structure $F$-terms as
\begin{equation}
    D_I W  = D_I^{(0)} W+ (\partial_I f^A) K_A W
\end{equation}
in terms of the \emph{uncorrected} $F$-terms
\begin{equation}
     D_I^{(0)} W \equiv \partial_I W+ (\partial_I K_{\mathrm{cs}})W\, .
\end{equation}
Using these expressions, the $F$-term potential becomes
\begin{align}
   \mathrm{e}^{-K}\, V &=  K^{M\bar N} D_M W \, \overline{D_{N} W} -3|W|^2 \nonumber\\[0.3em]
   &= (K^{A\bar B} K_A K_{\bar{B}}-3)|W|^2\nonumber\\[0.3em]
   &\quad +K^{I\bar J} (D_I^{(0)} W+ (\partial_I f^A) K_A W)\, \overline{(D_J^{(0)} W+ (\partial_J f^B) K_B W)} \nonumber\\[0.3em]
   &\quad + K^{I\bar B} (D_I^{(0)} W+ (\partial_I f^A) K_A W)\, K_{\bar{B}}\overline{W} \nonumber\\[0.3em]
   &\quad +K^{A\bar J}  K_A W\, \overline{(D_J^{(0)} W+ (\partial_J f^B) K_B W)}
\end{align}
We start by simplifying the contraction of the K\"ahler moduli part of the inverse metric,
\begin{align}
    K^{A\bar B} K_A K_{\bar{B}} &= K^{A\bar{B}}_k K_A K_{\bar{B}}+ K_A K_{\bar{B}}(\partial_I f^A)  K^{I\bar{J}} (\bar{\partial}_J f^{\bar B}) \nonumber\\
    &= 3+ K_A K_{\bar{B}}(\partial_I f^A)  K^{I\bar{J}} (\bar{\partial}_J f^{\bar B}) \,,
    \label{kabl}
\end{align}
where we used the standard no-scale relation
\begin{equation}
    K^{A\bar{B}}_k (K_k)_A (K_k)_{\bar{B}} = 3\, .
\end{equation}
Next, collecting all terms linear in the uncorrected $F$-terms, we find an exact cancellation
\begin{align}
   \mathrm{e}^{-K}\, V &\supset K^{I\bar J} (D_I^{(0)} W)\, \overline{((\partial_J f^B) K_B W)}  + K^{I\bar B} (D_I^{(0)} W)\, K_{\bar{B}}\overline{W}  + \mathrm{c.c.}\nonumber\\[0.3em]
   &= K^{I\bar J} (D_I^{(0)} W)\, \overline{((\partial_J f^B) K_B W)}  -(\bar{\partial}_J f^{\bar B}) K^{I\bar{J}} (D_I^{(0)} W)\, K_{\bar{B}}\overline{W}  + \mathrm{c.c.}\nonumber\\[0.3em]
   &=0\, .
\end{align}
Here we employed the identity
\begin{equation}
    K^{I\bar J} (\bar{\partial}_J f^{\bar B}) K_{\bar{B}}+K^{I\bar B}K_{\bar{B}} = 0\, .
    \label{kind}
\end{equation}
Finally, collecting all terms which do not involve the uncorrected $F$-terms, we get
\begin{align}
   \dfrac{\mathrm{e}^{-K}\, V}{|W|^2} &\supset (K^{A\bar B} K_A K_{\bar{B}}-3) +K^{I\bar J}  (\partial_I f^A) K_A )\, \overline{( (\partial_J f^B) K_B)} \nonumber\\[0.3em]
   &\quad + K^{I\bar B} ( (\partial_I f^A) K_A )\, K_{\bar{B}} +K^{A\bar J}  K_A \, \overline{( (\partial_J f^B) K_B )}\nonumber\\[0.3em]
   &=2K_A K_{\bar{B}}(\partial_I f^A)  K^{I\bar{J}} (\bar{\partial}_J f^{\bar B})  \nonumber\\[0.3em]
   &\quad -(\bar{\partial}_J f^{\bar B}) K^{I\bar{J}} ( (\partial_I f^A) K_A )\, K_{\bar{B}} -(\partial_I f^A) K^{I\bar{J}}  K_A \, \overline{( (\partial_J f^B) K_B )} \nonumber\\[0.3em]
   &=0\, .
\end{align}
Here we used \eqref{kabl} and the identity \eqref{kind}. All in all, the only term left is the one in \eqref{fullp}.

\subsection{Stability analysis}\label{app:DD:4d:stability}

Let us return to \eqref{fullp} and reinstate the axio-dilaton $\tau$ among the dynamical fields.
In this case, the scalar potential \eqref{fullp} can be written as
\begin{equation}
    V = \mathrm{e}^K K^{I\bar J}\, F_I^{(0)}\, \overline{F}_{\bar{J}}^{(0)}\; ,\quad F_I^{(0)}=D_I^{(0)} W_\mathrm{GVW}\, .
\end{equation}
We emphasise that the quantities $F_I^{(0)} = D_I^{(0)} W_{\mathrm{GVW}}$ are the \emph{uncorrected} $F$-terms, evaluated using the classical Kähler potential. By contrast, $K$ and $K^{I\bar J}$ denote the \emph{full} Kähler potential \eqref{eq:Kk} and the corresponding corrected inverse Kähler metric \eqref{eq:IKMij4d}, which includes the effects of warping.

Our aim is to disentangle the leading-order flux potential from the warping-induced corrections. To this end, we rewrite the scalar potential as
\begin{equation}
    V = \mathrm{e}^{K^{(0)}} \bigl (K_{\mathrm{cs}}^{I\bar J}+\delta K^{I\bar J}+\ldots\bigl )\, F_I^{(0)}\, \overline{F}_{\bar{J}}^{(0)}\, ,
    \label{vkdk}
\end{equation}
where $K_{\mathrm{cs}}^{I\bar J}$ is the leading-order inverse Kähler metric, and $\delta K^{I\bar I}$ is the first sub-leading term in the expansion of the inverse Kähler metric and the Kähler potential $K$ in inverse powers of $c$. This term captures the leading effects of warping in the four-dimensional potential.
Using the explicit form of the full inverse metric given in \eqref{eq:IKMij4d}, we find, using \eqref{eq:Kinverseleadingcorr},
\begin{equation}\label{eq:DD:4d:dKM}
    \delta K^{I\bar J} = -K_{\mathrm{cs}}^{I\bar{L}} (\partial_K\partial_{\bar{L}}f^A)K_A^{(0)}K_{\mathrm{cs}}^{K\bar{J}}+K_{\mathrm{cs}}^{I\bar{J}} f^AK_A^{(0)}\, .
\end{equation}
Here the first term arises from the correction to the inverse metric itself, while the second term originates from expanding the prefactor $\mathrm{e}^K$ to first sub-leading order. 

For notational convenience, we now drop the superscript $(0)$ on the uncorrected quantities and set
\begin{equation}
    F_I \equiv F_I^{(0)}\; , \quad D_I \equiv D_I^{(0)}\, .
\end{equation}
With this notation, the leading-order flux potential is given by
\begin{equation}\label{eq:DD:4d:Vflux}
    V_{\text{flux}} = \mathrm{e}^{K^{(0)}} K_{\mathrm{cs}}^{I\bar J}\, F_I \, \overline{F}_{\bar{J}} \, ,
\end{equation}
while the leading warping correction takes the form
\begin{equation}\label{eq:DD:4d:Vwarp}
    \delta V_{\text{warp}} = \mathrm{e}^{K^{(0)}} \delta K^{I\bar J}\, F_I \, \overline{F}_{\bar{J}} \, .
\end{equation}
Let us emphasise that our analysis proceeds under the assumption that the 4d potential \eqref{eq:DD:4d:Vwarp} correctly encodes the warping contributions contained in the 10d expression \eqref{eq:deltaVeff_summary}. As discussed in the main text (cf.~Section~\ref{sec:K4dgeneral}), this assumption is motivated by an intriguing structural correspondence suggested by the results of \cite{Martucci:2014ska,Martucci:2016pzt}. However, since a complete derivation of the 4d effective potential including all warping effects is presently unavailable, it remains an open question whether this proposal exhausts the full set of warping corrections.

We now compute the corrections to the light directions induced by $\delta V_{\text{warp}}$. In particular, this calculation proceeds in complete analogy to the computation performed in Appendix~\ref{app:DD}. The only essential difference is that the potential \eqref{eq:DD:4d:Vwarp} contains $\delta K^{I\bar J}$ instead of the classical inverse Kähler metric $K_{\mathrm{cs}}^{I\bar J}$ which changes the contractions of indices. Starting from \eqref{eq:DD:4d:Vwarp}, we compute its Kähler-covariant derivatives $D_b\partial_a \delta V_{\text{warp}}$ along a generic direction labelled by $a$. We are interested in the projection of this expression onto the light direction \eqref{eq:light_direction}. Using the relation $\hat{e}_+^a \left(D_a F_b\right) = 0$ which follows from the discussion around Eq.~\eqref{eq:EHatEV}, many of the terms above drop out upon contraction with $\hat{e}_+^a \hat{e}_+^b$. Upon contraction, a non-vanishing contribution takes the form 
\begin{equation}\label{eq:DD:4d:stabTerm}
    \hat{e}_+^a\hat{e}_+^b D_b\partial_a \delta V_{\text{warp}} \supset \mathrm{e}^{K^{(0)}} \dfrac{W_0^2}{||W_0 F||^2}\delta K^{I\bar J} \, U_{IKL}\,\oF^K\,\oF^L  \, \overline{F}_{\bar{J}}
\end{equation}
where $U_{IKL}=D_ID_KD_LW_{\mathrm{GVW}}$ denotes the third covariant derivative of the superpotential. Furthermore, the indices are raised with the leading-order inverse Kähler metric $K_{\mathrm{cs}}^{I\bar J}$.

Finally, using the explicit expression for $\delta K^{I\bar J}$ given in
\eqref{eq:DD:4d:dKM}, we obtain
\begin{equation}
    \delta K^{I\bar J} \, \overline{F}_{\bar{J}} = -K_{\mathrm{cs}}^{I\bar{L}} \overline{F}^K(\partial_K\partial_{\bar{L}}f^A)K_A^{(0)}+ \overline{F}^I f^AK_A^{(0)}\, .
\end{equation}
Plugging this back into \eqref{eq:DD:4d:stabTerm}, we find
\begin{align}
    \label{eq:DD:4d:stabTerm2}
    \hat{e}_+^a\hat{e}_+^b D_b\partial_a \delta V_{\text{warp}} &\supset  \dfrac{\mathrm{e}^{K^{(0)}} W_0^2}{||W_0F||^2}\biggl (- U_{IKL}\,\oF^K\,\oF^L\, K_{\mathrm{cs}}^{I\bar{R}} \overline{F}^P(\partial_P\partial_{\bar{R}}f^A)K_A^{(0)}\nonumber\\[0.4em]
    &\qquad + \,f^AK_A^{(0)} U_{IKL}\, \overline{F}^I\,\oF^K\,\oF^L  \biggl) \, .
\end{align}
While the second term exhibits precisely the contractions required for the additional tuning in \eqref{eq:finetuning} to render its contribution sub-leading (of order $\mathcal{O}(\varepsilon^2)$) the first term involves a slightly different index structure. As a consequence, it is not a priori clear whether, in the case where $F \sim W_0$, the condition \eqref{eq:finetuning} by itself is sufficient to guarantee the positivity of the mass matrix once warping effects are included.
This precisely fits our findings from the ten dimensional perspective: Some terms will be smaller due to the additional tuning but not all of them.

\section{Masses of complex structure moduli}\label{app:DD}

Let us review and generalize the analysis of \cite{Denef:2004cf} for the masses of the moduli from the 4d supergravity point of view, see also \cite{Marsh:2011aa,Marsh:2014nla,Gallego:2017dvd} for similar analyses. Suppose we have a supergravity potential of the form
\begin{equation}
    V = \mathrm{e}^{K}\, (K^{I\bar{J}}D_{I}W\overline{D_JW}-n |W|^2)\, .
\end{equation}
Although our primary interest lies in the no-scale case with $n = 0$, it is nevertheless useful to keep $n$ arbitrary at this stage in order to facilitate a later comparison with \cite{Denef:2004cf}.
To simplify the computation, we introduce
\begin{equation}
    F_I = D_IW\; ,\quad Z_{IJ}=D_ID_JW\; ,\quad U_{IJK}=D_ID_JD_KW
\end{equation}
to write
\begin{equation}
    V = \mathrm{e}^{K}\, (|F|^2-n |W|^2)\, ,
\end{equation}
where
\begin{equation}
    |F|^2 = F_I\, \oF^{I} = K^{I\bar{J}}\, F_I\, \oF_{\bar{J}}\, .
\end{equation}
We may then compute the first and second derivatives of this potential, yielding
\begin{equation}
    \partial_I V = \mathrm{e}^{K}\, (-(n-1) \overline{W}\, F_I +Z_{IJ}\, \oF^J)
\end{equation}
as well as
\begin{align}
    \hspace{-.3cm}D_I \partial_{\bar{J}} V &=  \mathrm{e}^{K}\bigl [-(n-1)K_{I\bar{J}}|W|^2+Z_{IK}\overline{Z}_{\bar{J}}{}^{K}+ R_{I\bar{J}}{}^{K\bar{L}}F_K\oF_{\bar{L}}+K_{I\bar{J}}|F|^2-F_I\oF_{\bar{J}} \bigl ]\, , \label{eq:DD:DbDV}\\[0.5em]
    \hspace{-.3cm}D_I \partial_{J} V &=\mathrm{e}^{K}\bigl [-(n-2)\overline{W}Z_{IJ}+U_{IJK}\oF^{K} \bigl ] \label{eq:DD:DDV}\, .
\end{align}

The extremum condition reads
\begin{equation}\label{eq:DD:dV0}
    \partial_I V = 0 \quad \Leftrightarrow\quad Z_{IJ}\, \oF^J=(n-1) \overline{W}\, F_I\,. 
\end{equation}
From this point onward, it proves convenient to introduce an orthonormal frame or vielbein on moduli space and to work, correspondingly, with frame rather than Einstein indices. For simplicity, we do not introduce new notation but rather declare $I,J$ etc. to be frame indices from now on. This implies in particular that $K_{I\bar{J}}=\delta_{I\bar{J}}$ and hence $F_I=F^{\bar{I}}$ for $I=\bar{I}$. Condition \eqref{eq:DD:dV0} can then be written as
\begin{equation}\label{eq:DD:MinC}
    0 = \begin{pmatrix}
        \partial_IV\\[0.5em]
        \partial_{\bar{I}}V
    \end{pmatrix}  = \begin{pmatrix}
        -(n-1)|W| & \mathrm{e}^{-\I\theta}Z_{IJ}\\[0.5em]
        \mathrm{e}^{\I\theta} \, \bar Z_{\bar I \bar J} & -(n-1)|W|
    \end{pmatrix}\begin{pmatrix}
        \mathrm{e}^{-\I\theta}F^{\bar{J}}\\[0.5em]
        \mathrm{e}^{\I\theta}\oF^{J}
    \end{pmatrix}\, ,
\end{equation}
which amounts to
\begin{equation}
    0 = \bigl [M-(n-1)|W|\bigl ]\cdot\vec{F}
\end{equation}
in terms of
\begin{equation}
    M = \begin{pmatrix}
        0 & \mathrm{e}^{-\I\theta}Z_{IJ}\\[0.5em]
        \mathrm{e}^{\I\theta} \, \bar Z_{\bar I \bar J} & 0
    \end{pmatrix}\; , \quad \vec{F} = \begin{pmatrix}
        \mathrm{e}^{-\I\theta}F^{\bar{I}}\\[0.5em]
        \mathrm{e}^{\I\theta}\oF^{I}
    \end{pmatrix}\, .
\end{equation}
We can then infer that non-supersymmetric vacua correspond to eigenvectors of the matrix $M$ with eigenvalues $(n-1)|W|$.

As explained in \cite{Denef:2004cf}, the matrix $M$ is a Hermitian matrix whose eigenvalues $\lambda_\alpha$ come in pairs $(\lambda_\alpha^+,\lambda_\alpha^-)=(+\lambda_\alpha,-\lambda_\alpha)$ where $\lambda_\alpha\geq 0$.
Following \cite{Denef:2004cf}, we denote the eigenvectors of $M$ with eigenvalues $\lambda_\alpha^{\pm}$ as
\begin{equation}\label{eq:DD:EVsMpm}
    \Psi^+_\alpha = \begin{pmatrix}
        \mathrm{e}^{-\I\theta/2}\,\psi_\alpha \\[0.2em]
        \mathrm{e}^{\I\theta/2}\,\overline{\psi}_\alpha
    \end{pmatrix}\; , \quad\Psi^-_\alpha = \begin{pmatrix}
        \I\mathrm{e}^{-\I\theta/2}\,\psi_\alpha \\[0.2em]
        -\I\mathrm{e}^{\I\theta/2}\,\overline{\psi}_\alpha
    \end{pmatrix}\, ,
\end{equation}
where $\psi_\alpha$ solves
\begin{equation}\label{eq:DD:Zpsi}
    Z\,\overline{\psi}_\alpha = \lambda_\alpha^+\, \psi_\alpha\, .
\end{equation}
The matrix $N=M-(n-1)|W|$ with eigenvector $\vec{F}$ has an eigenvalue equal to zero. Thus, $M$ has an eigenvalue, which we denote as $\lambda_\alpha$, satisfying
\begin{equation}
    \lambda_\alpha = \mathrm{e}^{-\I\theta}\,(n-1)W\,, \qquad \text{with} \qquad  F_I = f\, \mathrm{e}^{\I\theta/2}\, (\psi_\alpha)_I\, .
\end{equation}
The corresponding eigenvector, denoted $\hat e^a_+$ in the main text, is given in Eq.~\eqref{eq:light_direction}. The associated eigenvector $\hat e^a_{-}$ with eigenvalue $\lambda_{\alpha}=-(n-1)|W|$ is given in \eqref{eq:light_direction2}.

Next, following \cite{Denef:2004cf}, let us define the two matrices
\begin{equation}
    V_1^{\prime\prime} = \begin{pmatrix}
        0 & S_1\\[0.2em]
        \overline{S}_1 & 0
    \end{pmatrix}\; , \quad V_2^{\prime\prime} = \begin{pmatrix}
        S_2 & 0\\[0.2em]
        0 & \overline{S}_2
    \end{pmatrix}
\end{equation}
in terms of
\begin{equation}
    S_1 = U_{IJK}\oF^{K}\; ,\quad S_2 = R_{I\bar{J}}{}^{K\bar{L}}F_K\oF_{\bar{L}}+\delta_{I\bar{J}}|F|^2-F_I\oF_{\bar{J}}\, .
\end{equation}
Following \cite{Marsh:2011aa}, we can then write the matrix $\mathcal{H}$ of second derivatives
\begin{equation}\label{eq:DD:CarlH}
    \mathcal{H}= \begin{pmatrix}
    D_I\bar\partial_{\bar J} V & D_{ I} \partial_J V  \\[0.4em]
    \bar D_{\bar I}\bar\partial_{\bar J} V & \bar D_{\bar I}\partial_{J} V 
    \end{pmatrix}
\end{equation}
as (recall Eqs.~\eqref{eq:DD:DbDV}, \eqref{eq:DD:DDV})
\begin{equation}
    \mathcal{H}= (M+|W|)(M-(n-1)|W|) + V_1^{\prime\prime} + V_2^{\prime\prime}
\end{equation}
which generalises Eq.~(2.20) in \cite{Denef:2004cf}.

Assuming that $|F|\ll 1$, the eigenvectors $\Psi_\alpha^{\pm}$ of $M$ provide good approximations to the eigenvectors of the full matrix $\mathcal{H}$ \cite{Denef:2004cf}. We refer the reader, however, to the discussion at the end of this section for important caveats as emphasised in \cite{Marsh:2011aa}. Let us denote the corresponding eigenvalues of $\Psi_\alpha^{\pm}$ as $\lambda_\alpha^{\pm}$. We then have that the values of $H$ in the directions $\Psi_\alpha^{\pm}$ are given by
\begin{equation}
    (m_\alpha^{\pm})^2 = (\lambda_\alpha^{\pm}+|W|)(\lambda_\alpha^{\pm}-(n-1)|W|)\pm \Re(\mathrm{e}^{\I \theta} \bar{\psi}_\alpha\, S_1\, \bar{\psi}_\alpha)+2\bar{\psi}_\alpha\, S_2\, \psi_\alpha
\end{equation}
in terms of $\psi_\alpha$ as defined in Eq.~\eqref{eq:DD:Zpsi}.

Let us denote the direction with the minimal eigenvalue $\lambda_F^{\pm}$ as $\Psi_F^{\pm}$, corresponding to
\begin{equation}
    \lambda_F^{\pm} = \pm(n-1)|W|\, .
\end{equation}
Ignoring terms $\mathcal{O}(F)$ for the moment (see Eqs.~\eqref{eq:DD:m1p} and \eqref{eq:DD:m1m} below), we then have that
\begin{align}
    (m_F^{+})^2 &= (\lambda_F^{+}+|W|)(\lambda_F^{+}-(n-1)|W|)+ \mathcal{O}(F)\nonumber\\[0.5em]
    &=0\cdot|W|^2+\mathcal{O}(F)
\end{align}
independently of the value of $n$ and
\begin{align}
    (m_F^{-})^2 &= (\lambda_F^{-}+|W|)(\lambda_F^{-}-(n-1)|W|)+\mathcal{O}(F)\nonumber\\[0.5em]
    &=(-n+2)(-n+1)2\cdot|W|^2+\mathcal{O}(F) \, ,
\end{align}
where the coefficient on the right hand side for the two cases $n=0,3$ is given by
\begin{equation}
    (-n+2)(-n+1)2 = \begin{cases}
    (-3+2)(-3+1)2 = 4\,, & n=3\, ,\\[0.5em]
    (0+2)(0+1)2 = 4\,, & n=0\, .
    \end{cases}
\end{equation}
Therefore, including the terms $\mathcal{O}(F)$ suppressed above, the two small mass eigenvalues are given by Eqs.~\eqref{eq:DD:m1pmain}, \eqref{eq:DD:m1mmain}, namely \cite{Denef:2004cf}
\begin{equation}\label{eq:DD:m1p}
    (m_F^{+})^2 = \frac{2}{|F|^2} \left(\Re ( \mathrm{e}^{2\I\theta} U_{IJK} \, \oF^I \, \oF^J \, \oF^K ) + R_{I\bar{J}K\bar{L}} \, \oF^I F^{\bar{J}} \oF^K F^{\bar{L}} \right)\,,
\end{equation}
and
\begin{equation}\label{eq:DD:m1m}
    (m_F^{-})^2 = 4|W|^2+\frac{2}{|F|^2} \left(-\Re ( \mathrm{e}^{2\I\theta} U_{IJK} \, \oF^I\, \oF^J \, \oF^K ) + R_{I\bar{J}K\bar{L}} \, \oF^I F^{\bar{J}} \oF^K F^{\bar{L}} \right)\,.
\end{equation}

So far, our analysis applied for a general superpotential. Let us now specialise to the case of Type~IIB flux vacua. In this case, we compute for the second derivatives $Z_{IJ}$ of the GVW superpotential \eqref{eq:GVW} \cite{Denef:2004ze,Denef:2004cf}
\begin{equation}
    Z_{\tau\tau} = 0 \; , \quad Z_{\tau i} \equiv Z_i\; , \quad Z_{ij} = \mathcal{F}_{ijk} \overline{Z}^k\, ,
\end{equation}
where we introduced \cite{Denef:2004ze}
\begin{equation}
    \mathcal{F}_{ijk} = \mathrm{i}\, \mathrm{e}^{K_{\text{cs}}}\int \Omega \wedge D_iD_jD_k \Omega\, ,
\end{equation}
which are related to the $\kappa_{ijk}$ as defined in Eq.~\eqref{eq:kappaijk} by a constant rescaling.
Similarly, the third derivatives can be written in the form
\begin{equation}\label{eq:DD:UGVW}
    U_{\tau \tau i}=0 \; , \quad U_{\tau ij}=\mathcal{F}_{ijk} \oF^k \; , \quad U_{ijk}=D_i \mathcal{F}_{jkl} \overline{Z}^l + \mathcal{F}_{ijk} \oF^\tau\, .
\end{equation}
In general, the presence of the term $D_i \mathcal{F}_{jkl}\,\overline{Z}^i$ in $U_{ijk}$ implies that, whenever $F\sim W$, additional fine tuning is required to ensure that the mass matrix can, at least in principle, be rendered positive definite; see the discussion in Section~\ref{sec:KKLT}. Further details can be found in \cite{Denef:2004cf}.

\bibliographystyle{JHEP}
\bibliography{draft}

\providecommand{\href}[2]{#2}\begingroup\raggedright\begin{thebibliography}{100}

\bibitem{Kachru:2003aw}
S.~Kachru, R.~Kallosh, A.~D. Linde and S.~P. Trivedi, \emph{{De Sitter vacua in
  string theory}},
  \href{http://dx.doi.org/10.1103/PhysRevD.68.046005}{\emph{Phys. Rev. D} {\bf
  68} (2003) 046005}, [\href{http://arxiv.org/abs/hep-th/0301240}{{\tt
  hep-th/0301240}}].

\bibitem{Balasubramanian:2005zx}
V.~Balasubramanian, P.~Berglund, J.~P. Conlon and F.~Quevedo,
  \emph{{Systematics of moduli stabilisation in Calabi-Yau flux
  compactifications}},
  \href{http://dx.doi.org/10.1088/1126-6708/2005/03/007}{\emph{JHEP} {\bf 03}
  (2005) 007}, [\href{http://arxiv.org/abs/hep-th/0502058}{{\tt
  hep-th/0502058}}].

\bibitem{Crino:2020qwk}
C.~Crin{\`o}, F.~Quevedo and R.~Valandro, \emph{{On de Sitter String Vacua from
  Anti-D3-Branes in the Large Volume Scenario}},
  \href{http://dx.doi.org/10.1007/JHEP03(2021)258}{\emph{JHEP} {\bf 03} (2021)
  258}, [\href{http://arxiv.org/abs/2010.15903}{{\tt 2010.15903}}].

\bibitem{McAllister:2024lnt}
L.~McAllister, J.~Moritz, R.~Nally and A.~Schachner, \emph{{Candidate de Sitter
  vacua}}, \href{http://dx.doi.org/10.1103/PhysRevD.111.086015}{\emph{Phys.
  Rev. D} {\bf 111} (2025) 086015},
  [\href{http://arxiv.org/abs/2406.13751}{{\tt 2406.13751}}].

\bibitem{Lust:2022lfc}
S.~L{\"u}st, C.~Vafa, M.~Wiesner and K.~Xu, \emph{{Holography and the KKLT
  scenario}}, \href{http://dx.doi.org/10.1007/JHEP10(2022)188}{\emph{JHEP} {\bf
  10} (2022) 188}, [\href{http://arxiv.org/abs/2204.07171}{{\tt 2204.07171}}].

\bibitem{Bena:2024are}
I.~Bena, Y.~Li and S.~L{\"u}st, \emph{{KKLT Ex Nihilo}},
  \href{http://arxiv.org/abs/2410.22400}{{\tt 2410.22400}}.

\bibitem{Apers:2025pon}
F.~Apers, M.~Montero and I.~Valenzuela, \emph{{Backtracking AdS flux vacua}},
  \href{http://arxiv.org/abs/2506.03314}{{\tt 2506.03314}}.

\bibitem{Bedroya:2025fie}
A.~Bedroya and P.~J. Steinhardt, \emph{{Holographic Constraints on the String
  Landscape}},  \href{http://arxiv.org/abs/2511.15784}{{\tt 2511.15784}}.

\bibitem{Demirtas:2021ote}
M.~Demirtas, M.~Kim, L.~McAllister, J.~Moritz and A.~Rios-Tascon,
  \emph{{Exponentially Small Cosmological Constant in String Theory}},
  \href{http://dx.doi.org/10.1103/PhysRevLett.128.011602}{\emph{Phys. Rev.
  Lett.} {\bf 128} (2022) 011602}, [\href{http://arxiv.org/abs/2107.09065}{{\tt
  2107.09065}}].

\bibitem{Bena:2009xk}
I.~Bena, M.~Grana and N.~Halmagyi, \emph{{On the Existence of Meta-stable Vacua
  in Klebanov-Strassler}},
  \href{http://dx.doi.org/10.1007/JHEP09(2010)087}{\emph{JHEP} {\bf 09} (2010)
  087}, [\href{http://arxiv.org/abs/0912.3519}{{\tt 0912.3519}}].

\bibitem{McGuirk:2009xx}
P.~McGuirk, G.~Shiu and Y.~Sumitomo, \emph{{Non-supersymmetric infrared
  perturbations to the warped deformed conifold}},
  \href{http://dx.doi.org/10.1016/j.nuclphysb.2010.09.008}{\emph{Nucl. Phys. B}
  {\bf 842} (2011) 383--413}, [\href{http://arxiv.org/abs/0910.4581}{{\tt
  0910.4581}}].

\bibitem{Bena:2011hz}
I.~Bena, G.~Giecold, M.~Grana, N.~Halmagyi and S.~Massai, \emph{{On Metastable
  Vacua and the Warped Deformed Conifold: Analytic Results}},
  \href{http://dx.doi.org/10.1088/0264-9381/30/1/015003}{\emph{Class. Quant.
  Grav.} {\bf 30} (2013) 015003}, [\href{http://arxiv.org/abs/1102.2403}{{\tt
  1102.2403}}].

\bibitem{Dymarsky:2011pm}
A.~Dymarsky, \emph{{On gravity dual of a metastable vacuum in
  Klebanov-Strassler theory}},
  \href{http://dx.doi.org/10.1007/JHEP05(2011)053}{\emph{JHEP} {\bf 05} (2011)
  053}, [\href{http://arxiv.org/abs/1102.1734}{{\tt 1102.1734}}].

\bibitem{Bena:2011wh}
I.~Bena, G.~Giecold, M.~Grana, N.~Halmagyi and S.~Massai, \emph{{The
  backreaction of anti-D3 branes on the Klebanov-Strassler geometry}},
  \href{http://dx.doi.org/10.1007/JHEP06(2013)060}{\emph{JHEP} {\bf 06} (2013)
  060}, [\href{http://arxiv.org/abs/1106.6165}{{\tt 1106.6165}}].

\bibitem{Bena:2012bk}
I.~Bena, M.~Grana, S.~Kuperstein and S.~Massai, \emph{{Anti-D3 Branes: Singular
  to the bitter end}},
  \href{http://dx.doi.org/10.1103/PhysRevD.87.106010}{\emph{Phys. Rev. D} {\bf
  87} (2013) 106010}, [\href{http://arxiv.org/abs/1206.6369}{{\tt 1206.6369}}].

\bibitem{Gautason:2013zw}
F.~F. Gautason, D.~Junghans and M.~Zagermann, \emph{{Cosmological Constant,
  Near Brane Behavior and Singularities}},
  \href{http://dx.doi.org/10.1007/JHEP09(2013)123}{\emph{JHEP} {\bf 09} (2013)
  123}, [\href{http://arxiv.org/abs/1301.5647}{{\tt 1301.5647}}].

\bibitem{Dymarsky:2013tna}
A.~Dymarsky and S.~Massai, \emph{{Uplifting the baryonic branch: a test for
  backreacting anti-D3-branes}},
  \href{http://dx.doi.org/10.1007/JHEP11(2014)034}{\emph{JHEP} {\bf 11} (2014)
  034}, [\href{http://arxiv.org/abs/1310.0015}{{\tt 1310.0015}}].

\bibitem{Blaback:2014tfa}
J.~Bl{\r{a}}b{\"a}ck, U.~H. Danielsson, D.~Junghans, T.~Van~Riet and S.~C.
  Vargas, \emph{{Localised anti-branes in non-compact throats at zero and
  finite $T$}}, \href{http://dx.doi.org/10.1007/JHEP02(2015)018}{\emph{JHEP}
  {\bf 02} (2015) 018}, [\href{http://arxiv.org/abs/1409.0534}{{\tt
  1409.0534}}].

\bibitem{Michel:2014lva}
B.~Michel, E.~Mintun, J.~Polchinski, A.~Puhm and P.~Saad, \emph{{Remarks on
  brane and antibrane dynamics}},
  \href{http://dx.doi.org/10.1007/JHEP09(2015)021}{\emph{JHEP} {\bf 09} (2015)
  021}, [\href{http://arxiv.org/abs/1412.5702}{{\tt 1412.5702}}].

\bibitem{Polchinski:2015bea}
J.~Polchinski, \emph{{Brane/antibrane dynamics and KKLT stability}},
  \href{http://arxiv.org/abs/1509.05710}{{\tt 1509.05710}}.

\bibitem{Cohen-Maldonado:2015ssa}
D.~Cohen-Maldonado, J.~Diaz, T.~van Riet and B.~Vercnocke, \emph{{Observations
  on fluxes near anti-branes}},
  \href{http://dx.doi.org/10.1007/JHEP01(2016)126}{\emph{JHEP} {\bf 01} (2016)
  126}, [\href{http://arxiv.org/abs/1507.01022}{{\tt 1507.01022}}].

\bibitem{Bena:2018fqc}
I.~Bena, E.~Dudas, M.~Gra{\~n}a and S.~L{\"u}st, \emph{{Uplifting Runaways}},
  \href{http://dx.doi.org/10.1002/prop.201800100}{\emph{Fortsch. Phys.} {\bf
  67} (2019) 1800100}, [\href{http://arxiv.org/abs/1809.06861}{{\tt
  1809.06861}}].

\bibitem{Armas:2018rsy}
J.~Armas, N.~Nguyen, V.~Niarchos, N.~A. Obers and T.~Van~Riet,
  \emph{{Meta-stable non-extremal anti-branes}},
  \href{http://dx.doi.org/10.1103/PhysRevLett.122.181601}{\emph{Phys. Rev.
  Lett.} {\bf 122} (2019) 181601}, [\href{http://arxiv.org/abs/1812.01067}{{\tt
  1812.01067}}].

\bibitem{Carta:2019rhx}
F.~Carta, J.~Moritz and A.~Westphal, \emph{{Gaugino condensation and small
  uplifts in KKLT}},
  \href{http://dx.doi.org/10.1007/JHEP08(2019)141}{\emph{JHEP} {\bf 08} (2019)
  141}, [\href{http://arxiv.org/abs/1902.01412}{{\tt 1902.01412}}].

\bibitem{Blumenhagen:2019qcg}
R.~Blumenhagen, D.~Kl{\"a}wer and L.~Schlechter, \emph{{Swampland Variations on
  a Theme by KKLT}},
  \href{http://dx.doi.org/10.1007/JHEP05(2019)152}{\emph{JHEP} {\bf 05} (2019)
  152}, [\href{http://arxiv.org/abs/1902.07724}{{\tt 1902.07724}}].

\bibitem{Blaback:2019ucp}
J.~Bl{\r{a}}b{\"a}ck, F.~F. Gautason, A.~Ruip{\'e}rez and T.~Van~Riet,
  \emph{{Anti-brane singularities as red herrings}},
  \href{http://dx.doi.org/10.1007/JHEP12(2019)125}{\emph{JHEP} {\bf 12} (2019)
  125}, [\href{http://arxiv.org/abs/1907.05295}{{\tt 1907.05295}}].

\bibitem{Randall:2019ent}
L.~Randall, \emph{{The Boundaries of KKLT}},
  \href{http://dx.doi.org/10.1002/prop.201900105}{\emph{Fortsch. Phys.} {\bf
  68} (2020) 1900105}, [\href{http://arxiv.org/abs/1912.06693}{{\tt
  1912.06693}}].

\bibitem{Bena:2019sxm}
I.~Bena, A.~Buchel and S.~L{\"u}st, \emph{{Throat destabilization (for profit
  and for fun)}},  \href{http://arxiv.org/abs/1910.08094}{{\tt 1910.08094}}.

\bibitem{Dudas:2019pls}
E.~Dudas and S.~L{\"u}st, \emph{{An update on moduli stabilization with
  antibrane uplift}},
  \href{http://dx.doi.org/10.1007/JHEP03(2021)107}{\emph{JHEP} {\bf 03} (2021)
  107}, [\href{http://arxiv.org/abs/1912.09948}{{\tt 1912.09948}}].

\bibitem{Gao:2020xqh}
X.~Gao, A.~Hebecker and D.~Junghans, \emph{{Control issues of KKLT}},
  \href{http://dx.doi.org/10.1002/prop.202000089}{\emph{Fortsch. Phys.} {\bf
  68} (2020) 2000089}, [\href{http://arxiv.org/abs/2009.03914}{{\tt
  2009.03914}}].

\bibitem{DallAgata:2022abm}
G.~Dall'Agata, M.~Emelin, F.~Farakos and M.~Morittu, \emph{{Anti-brane uplift
  instability from goldstino condensation}},
  \href{http://dx.doi.org/10.1007/JHEP08(2022)005}{\emph{JHEP} {\bf 08} (2022)
  005}, [\href{http://arxiv.org/abs/2203.12636}{{\tt 2203.12636}}].

\bibitem{Junghans:2022exo}
D.~Junghans, \emph{{LVS de Sitter vacua are probably in the swampland}},
  \href{http://dx.doi.org/10.1016/j.nuclphysb.2023.116179}{\emph{Nucl. Phys. B}
  {\bf 990} (2023) 116179}, [\href{http://arxiv.org/abs/2201.03572}{{\tt
  2201.03572}}].

\bibitem{Gao:2022fdi}
X.~Gao, A.~Hebecker, S.~Schreyer and V.~Venken, \emph{{The LVS parametric
  tadpole constraint}},
  \href{http://dx.doi.org/10.1007/JHEP07(2022)056}{\emph{JHEP} {\bf 07} (2022)
  056}, [\href{http://arxiv.org/abs/2202.04087}{{\tt 2202.04087}}].

\bibitem{Junghans:2022kxg}
D.~Junghans, \emph{{Topological constraints in the LARGE-volume scenario}},
  \href{http://dx.doi.org/10.1007/JHEP08(2022)226}{\emph{JHEP} {\bf 08} (2022)
  226}, [\href{http://arxiv.org/abs/2205.02856}{{\tt 2205.02856}}].

\bibitem{Hebecker:2022zme}
A.~Hebecker, S.~Schreyer and V.~Venken, \emph{{Curvature corrections to KPV: do
  we need deep throats?}},
  \href{http://dx.doi.org/10.1007/JHEP10(2022)166}{\emph{JHEP} {\bf 10} (2022)
  166}, [\href{http://arxiv.org/abs/2208.02826}{{\tt 2208.02826}}].

\bibitem{Schreyer:2022len}
S.~Schreyer and V.~Venken, \emph{{{\ensuremath{\alpha}}' corrections to KPV: an
  uplifting story}},
  \href{http://dx.doi.org/10.1007/JHEP07(2023)235}{\emph{JHEP} {\bf 07} (2023)
  235}, [\href{http://arxiv.org/abs/2212.07437}{{\tt 2212.07437}}].

\bibitem{Schreyer:2024pml}
S.~Schreyer, \emph{{Higher order corrections to KPV: The nonabelian brane stack
  perspective}}, \href{http://dx.doi.org/10.1007/JHEP07(2024)075}{\emph{JHEP}
  {\bf 07} (2024) 075}, [\href{http://arxiv.org/abs/2402.13311}{{\tt
  2402.13311}}].

\bibitem{Moritz:2025bsi}
J.~Moritz, \emph{{$G_2$-manifolds from Diophantine equations}},
  \href{http://arxiv.org/abs/2505.15883}{{\tt 2505.15883}}.

\bibitem{VanRiet:2023pnx}
T.~Van~Riet and G.~Zoccarato, \emph{{Beginners lectures on flux
  compactifications and related Swampland topics}},
  \href{http://dx.doi.org/10.1016/j.physrep.2023.11.003}{\emph{Phys. Rept.}
  {\bf 1049} (2024) 1--51}, [\href{http://arxiv.org/abs/2305.01722}{{\tt
  2305.01722}}].

\bibitem{Westphal:2006tn}
A.~Westphal, \emph{{de Sitter string vacua from Kahler uplifting}},
  \href{http://dx.doi.org/10.1088/1126-6708/2007/03/102}{\emph{JHEP} {\bf 03}
  (2007) 102}, [\href{http://arxiv.org/abs/hep-th/0611332}{{\tt
  hep-th/0611332}}].

\bibitem{Cremades:2007ig}
D.~Cremades, M.~P. Garcia~del Moral, F.~Quevedo and K.~Suruliz, \emph{{Moduli
  stabilisation and de Sitter string vacua from magnetised D7 branes}},
  \href{http://dx.doi.org/10.1088/1126-6708/2007/05/100}{\emph{JHEP} {\bf 05}
  (2007) 100}, [\href{http://arxiv.org/abs/hep-th/0701154}{{\tt
  hep-th/0701154}}].

\bibitem{Cicoli:2015ylx}
M.~Cicoli, F.~Quevedo and R.~Valandro, \emph{{De Sitter from T-branes}},
  \href{http://dx.doi.org/10.1007/JHEP03(2016)141}{\emph{JHEP} {\bf 03} (2016)
  141}, [\href{http://arxiv.org/abs/1512.04558}{{\tt 1512.04558}}].

\bibitem{Louis:2012nb}
J.~Louis, M.~Rummel, R.~Valandro and A.~Westphal, \emph{{Building an explicit
  de Sitter}}, \href{http://dx.doi.org/10.1007/JHEP10(2012)163}{\emph{JHEP}
  {\bf 10} (2012) 163}, [\href{http://arxiv.org/abs/1208.3208}{{\tt
  1208.3208}}].

\bibitem{McAllister:2023vgy}
L.~McAllister and F.~Quevedo, \emph{{Moduli Stabilization in String Theory}},
  \href{http://arxiv.org/abs/2310.20559}{{\tt 2310.20559}}.

\bibitem{Saltman:2004sn}
A.~Saltman and E.~Silverstein, \emph{{The Scaling of the no scale potential and
  de Sitter model building}},
  \href{http://dx.doi.org/10.1088/1126-6708/2004/11/066}{\emph{JHEP} {\bf 11}
  (2004) 066}, [\href{http://arxiv.org/abs/hep-th/0402135}{{\tt
  hep-th/0402135}}].

\bibitem{Denef:2004ze}
F.~Denef and M.~R. Douglas, \emph{{Distributions of flux vacua}},
  \href{http://dx.doi.org/10.1088/1126-6708/2004/05/072}{\emph{JHEP} {\bf 05}
  (2004) 072}, [\href{http://arxiv.org/abs/hep-th/0404116}{{\tt
  hep-th/0404116}}].

\bibitem{Denef:2004cf}
F.~Denef and M.~R. Douglas, \emph{{Distributions of nonsupersymmetric flux
  vacua}}, \href{http://dx.doi.org/10.1088/1126-6708/2005/03/061}{\emph{JHEP}
  {\bf 03} (2005) 061}, [\href{http://arxiv.org/abs/hep-th/0411183}{{\tt
  hep-th/0411183}}].

\bibitem{Gallego:2017dvd}
D.~Gallego, M.~C.~D. Marsh, B.~Vercnocke and T.~Wrase, \emph{{A New Class of de
  Sitter Vacua in Type IIB Large Volume Compactifications}},
  \href{http://dx.doi.org/10.1007/JHEP10(2017)193}{\emph{JHEP} {\bf 10} (2017)
  193}, [\href{http://arxiv.org/abs/1707.01095}{{\tt 1707.01095}}].

\bibitem{Honma:2019gzp}
Y.~Honma and H.~Otsuka, \emph{{F-theory Flux Vacua and Attractor Equations}},
  \href{http://dx.doi.org/10.1007/JHEP04(2020)001}{\emph{JHEP} {\bf 04} (2020)
  001}, [\href{http://arxiv.org/abs/1910.10725}{{\tt 1910.10725}}].

\bibitem{Hebecker:2020ejb}
A.~Hebecker and S.~Leonhardt, \emph{{Winding Uplifts and the Challenges of Weak
  and Strong SUSY Breaking in AdS}},
  \href{http://dx.doi.org/10.1007/JHEP03(2021)284}{\emph{JHEP} {\bf 03} (2021)
  284}, [\href{http://arxiv.org/abs/2012.00010}{{\tt 2012.00010}}].

\bibitem{Krippendorf:2023idy}
S.~Krippendorf and A.~Schachner, \emph{{New non-supersymmetric flux vacua in
  string theory}}, \href{http://dx.doi.org/10.1007/JHEP12(2023)145}{\emph{JHEP}
  {\bf 12} (2023) 145}, [\href{http://arxiv.org/abs/2308.15525}{{\tt
  2308.15525}}].

\bibitem{Lanza:2024uis}
S.~Lanza and A.~Westphal, \emph{{Uplifts in the penumbra: features of the
  moduli potential away from infinite-distance boundaries}},
  \href{http://dx.doi.org/10.1007/JHEP05(2025)071}{\emph{JHEP} {\bf 05} (2025)
  071}, [\href{http://arxiv.org/abs/2412.12253}{{\tt 2412.12253}}].

\bibitem{Gukov:1999ya}
S.~Gukov, C.~Vafa and E.~Witten, \emph{{CFT's from Calabi-Yau four folds}},
  \href{http://dx.doi.org/10.1016/S0550-3213(00)00373-4}{\emph{Nucl. Phys. B}
  {\bf 584} (2000) 69--108}, [\href{http://arxiv.org/abs/hep-th/9906070}{{\tt
  hep-th/9906070}}].

\bibitem{Giddings:2001yu}
S.~B. Giddings, S.~Kachru and J.~Polchinski, \emph{{Hierarchies from fluxes in
  string compactifications}},
  \href{http://dx.doi.org/10.1103/PhysRevD.66.106006}{\emph{Phys. Rev. D} {\bf
  66} (2002) 106006}, [\href{http://arxiv.org/abs/hep-th/0105097}{{\tt
  hep-th/0105097}}].

\bibitem{Dasgupta:1999ss}
K.~Dasgupta, G.~Rajesh and S.~Sethi, \emph{{M theory, orientifolds and G -
  flux}}, \href{http://dx.doi.org/10.1088/1126-6708/1999/08/023}{\emph{JHEP}
  {\bf 08} (1999) 023}, [\href{http://arxiv.org/abs/hep-th/9908088}{{\tt
  hep-th/9908088}}].

\bibitem{Grana:2000jj}
M.~Grana and J.~Polchinski, \emph{{Supersymmetric three form flux perturbations
  on AdS(5)}}, \href{http://dx.doi.org/10.1103/PhysRevD.63.026001}{\emph{Phys.
  Rev. D} {\bf 63} (2001) 026001},
  [\href{http://arxiv.org/abs/hep-th/0009211}{{\tt hep-th/0009211}}].

\bibitem{Grana:2001xn}
M.~Grana and J.~Polchinski, \emph{{Gauge / gravity duals with holomorphic
  dilaton}}, \href{http://dx.doi.org/10.1103/PhysRevD.65.126005}{\emph{Phys.
  Rev. D} {\bf 65} (2002) 126005},
  [\href{http://arxiv.org/abs/hep-th/0106014}{{\tt hep-th/0106014}}].

\bibitem{Grimm:2004uq}
T.~W. Grimm and J.~Louis, \emph{{The Effective action of N = 1 Calabi-Yau
  orientifolds}},
  \href{http://dx.doi.org/10.1016/j.nuclphysb.2004.08.005}{\emph{Nucl. Phys. B}
  {\bf 699} (2004) 387--426}, [\href{http://arxiv.org/abs/hep-th/0403067}{{\tt
  hep-th/0403067}}].

\bibitem{DeWolfe:2002nn}
O.~DeWolfe and S.~B. Giddings, \emph{{Scales and hierarchies in warped
  compactifications and brane worlds}},
  \href{http://dx.doi.org/10.1103/PhysRevD.67.066008}{\emph{Phys. Rev. D} {\bf
  67} (2003) 066008}, [\href{http://arxiv.org/abs/hep-th/0208123}{{\tt
  hep-th/0208123}}].

\bibitem{Giddings:2005ff}
S.~B. Giddings and A.~Maharana, \emph{{Dynamics of warped compactifications and
  the shape of the warped landscape}},
  \href{http://dx.doi.org/10.1103/PhysRevD.73.126003}{\emph{Phys. Rev. D} {\bf
  73} (2006) 126003}, [\href{http://arxiv.org/abs/hep-th/0507158}{{\tt
  hep-th/0507158}}].

\bibitem{Burgess:2006mn}
C.~P. Burgess, P.~G. Camara, S.~P. de~Alwis, S.~B. Giddings, A.~Maharana,
  F.~Quevedo et~al., \emph{{Warped Supersymmetry Breaking}},
  \href{http://dx.doi.org/10.1088/1126-6708/2008/04/053}{\emph{JHEP} {\bf 04}
  (2008) 053}, [\href{http://arxiv.org/abs/hep-th/0610255}{{\tt
  hep-th/0610255}}].

\bibitem{Frey:2006wv}
A.~R. Frey and A.~Maharana, \emph{{Warped spectroscopy: Localization of frozen
  bulk modes}},
  \href{http://dx.doi.org/10.1088/1126-6708/2006/08/021}{\emph{JHEP} {\bf 08}
  (2006) 021}, [\href{http://arxiv.org/abs/hep-th/0603233}{{\tt
  hep-th/0603233}}].

\bibitem{Douglas:2007tu}
M.~R. Douglas, J.~Shelton and G.~Torroba, \emph{{Warping and supersymmetry
  breaking}},  \href{http://arxiv.org/abs/0704.4001}{{\tt 0704.4001}}.

\bibitem{Koerber:2007xk}
P.~Koerber and L.~Martucci, \emph{{From ten to four and back again: How to
  generalize the geometry}},
  \href{http://dx.doi.org/10.1088/1126-6708/2007/08/059}{\emph{JHEP} {\bf 08}
  (2007) 059}, [\href{http://arxiv.org/abs/0707.1038}{{\tt 0707.1038}}].

\bibitem{Shiu:2008ry}
G.~Shiu, G.~Torroba, B.~Underwood and M.~R. Douglas, \emph{{Dynamics of Warped
  Flux Compactifications}},
  \href{http://dx.doi.org/10.1088/1126-6708/2008/06/024}{\emph{JHEP} {\bf 06}
  (2008) 024}, [\href{http://arxiv.org/abs/0803.3068}{{\tt 0803.3068}}].

\bibitem{Douglas:2008jx}
M.~R. Douglas and G.~Torroba, \emph{{Kinetic terms in warped
  compactifications}},
  \href{http://dx.doi.org/10.1088/1126-6708/2009/05/013}{\emph{JHEP} {\bf 05}
  (2009) 013}, [\href{http://arxiv.org/abs/0805.3700}{{\tt 0805.3700}}].

\bibitem{Frey:2008xw}
A.~R. Frey, G.~Torroba, B.~Underwood and M.~R. Douglas, \emph{{The Universal
  Kahler Modulus in Warped Compactifications}},
  \href{http://dx.doi.org/10.1088/1126-6708/2009/01/036}{\emph{JHEP} {\bf 01}
  (2009) 036}, [\href{http://arxiv.org/abs/0810.5768}{{\tt 0810.5768}}].

\bibitem{Marchesano:2008rg}
F.~Marchesano, P.~McGuirk and G.~Shiu, \emph{{Open String Wavefunctions in
  Warped Compactifications}},
  \href{http://dx.doi.org/10.1088/1126-6708/2009/04/095}{\emph{JHEP} {\bf 04}
  (2009) 095}, [\href{http://arxiv.org/abs/0812.2247}{{\tt 0812.2247}}].

\bibitem{Martucci:2009sf}
L.~Martucci, \emph{{On moduli and effective theory of N=1 warped flux
  compactifications}},
  \href{http://dx.doi.org/10.1088/1126-6708/2009/05/027}{\emph{JHEP} {\bf 05}
  (2009) 027}, [\href{http://arxiv.org/abs/0902.4031}{{\tt 0902.4031}}].

\bibitem{Douglas:2009zn}
M.~R. Douglas, \emph{{Effective potential and warp factor dynamics}},
  \href{http://dx.doi.org/10.1007/JHEP03(2010)071}{\emph{JHEP} {\bf 03} (2010)
  071}, [\href{http://arxiv.org/abs/0911.3378}{{\tt 0911.3378}}].

\bibitem{Underwood:2010pm}
B.~Underwood, \emph{{A Breathing Mode for Warped Compactifications}},
  \href{http://dx.doi.org/10.1088/0264-9381/28/19/195013}{\emph{Class. Quant.
  Grav.} {\bf 28} (2011) 195013}, [\href{http://arxiv.org/abs/1009.4200}{{\tt
  1009.4200}}].

\bibitem{Marchesano:2010bs}
F.~Marchesano, P.~McGuirk and G.~Shiu, \emph{{Chiral matter wavefunctions in
  warped compactifications}},
  \href{http://dx.doi.org/10.1007/JHEP05(2011)090}{\emph{JHEP} {\bf 05} (2011)
  090}, [\href{http://arxiv.org/abs/1012.2759}{{\tt 1012.2759}}].

\bibitem{Grimm:2012rg}
T.~W. Grimm, D.~Klevers and M.~Poretschkin, \emph{{Fluxes and Warping for Gauge
  Couplings in F-theory}},
  \href{http://dx.doi.org/10.1007/JHEP01(2013)023}{\emph{JHEP} {\bf 01} (2013)
  023}, [\href{http://arxiv.org/abs/1202.0285}{{\tt 1202.0285}}].

\bibitem{Frey:2013bha}
A.~R. Frey and J.~Roberts, \emph{{The Dimensional Reduction and K{\"a}hler
  Metric of Forms In Flux and Warping}},
  \href{http://dx.doi.org/10.1007/JHEP10(2013)021}{\emph{JHEP} {\bf 10} (2013)
  021}, [\href{http://arxiv.org/abs/1308.0323}{{\tt 1308.0323}}].

\bibitem{Martucci:2014ska}
L.~Martucci, \emph{{Warping the K{\"a}hler potential of F-theory/IIB flux
  compactifications}},
  \href{http://dx.doi.org/10.1007/JHEP03(2015)067}{\emph{JHEP} {\bf 03} (2015)
  067}, [\href{http://arxiv.org/abs/1411.2623}{{\tt 1411.2623}}].

\bibitem{Grimm:2014efa}
T.~W. Grimm, T.~G. Pugh and M.~Weissenbacher, \emph{{The effective action of
  warped M-theory reductions with higher derivative terms {\textemdash} part
  I}}, \href{http://dx.doi.org/10.1007/JHEP01(2016)142}{\emph{JHEP} {\bf 01}
  (2016) 142}, [\href{http://arxiv.org/abs/1412.5073}{{\tt 1412.5073}}].

\bibitem{Grimm:2015mua}
T.~W. Grimm, T.~G. Pugh and M.~Weissenbacher, \emph{{The effective action of
  warped M-theory reductions with higher-derivative terms - Part II}},
  \href{http://dx.doi.org/10.1007/JHEP12(2015)117}{\emph{JHEP} {\bf 12} (2015)
  117}, [\href{http://arxiv.org/abs/1507.00343}{{\tt 1507.00343}}].

\bibitem{Martucci:2016pzt}
L.~Martucci, \emph{{Warped K{\"a}hler potentials and fluxes}},
  \href{http://dx.doi.org/10.1007/JHEP01(2017)056}{\emph{JHEP} {\bf 01} (2017)
  056}, [\href{http://arxiv.org/abs/1610.02403}{{\tt 1610.02403}}].

\bibitem{Lust:2022xoq}
S.~L{\"u}st and L.~Randall, \emph{{Effective Theory of Warped Compactifications
  and the Implications for KKLT}},
  \href{http://dx.doi.org/10.1002/prop.202200103}{\emph{Fortsch. Phys.} {\bf
  70} (2022) 2200103}, [\href{http://arxiv.org/abs/2206.04708}{{\tt
  2206.04708}}].

\bibitem{Frey:2025rvf}
A.~R. Frey and R.~Mahanta, \emph{{Dimensional reduction and K{\"a}hler metric
  for metric moduli in imaginary self-dual flux}},
  \href{http://dx.doi.org/10.1007/JHEP07(2025)248}{\emph{JHEP} {\bf 07} (2025)
  248}, [\href{http://arxiv.org/abs/2501.08623}{{\tt 2501.08623}}].

\bibitem{Agarwal:2025rqd}
N.~Agarwal, A.~R. Frey and B.~Underwood, \emph{{Toward an Effective Theory of
  the Volume Modulus}},  \href{http://arxiv.org/abs/2509.18419}{{\tt
  2509.18419}}.

\bibitem{Balasubramanian:2004uy}
V.~Balasubramanian and P.~Berglund, \emph{{Stringy corrections to Kahler
  potentials, SUSY breaking, and the cosmological constant problem}},
  \href{http://dx.doi.org/10.1088/1126-6708/2004/11/085}{\emph{JHEP} {\bf 11}
  (2004) 085}, [\href{http://arxiv.org/abs/hep-th/0408054}{{\tt
  hep-th/0408054}}].

\bibitem{Westphal:2005yz}
A.~Westphal, \emph{{Eternal inflation with alpha-prime-corrections}},
  \href{http://dx.doi.org/10.1088/1475-7516/2005/11/003}{\emph{JCAP} {\bf 11}
  (2005) 003}, [\href{http://arxiv.org/abs/hep-th/0507079}{{\tt
  hep-th/0507079}}].

\bibitem{AbdusSalam:2025twp}
S.~AbdusSalam, C.~Hughes, F.~Quevedo and A.~Schachner, \emph{{Coexisting Flux
  String Vacua from Numerical K{\"a}hler Moduli Stabilisation}},
  \href{http://arxiv.org/abs/2507.00615}{{\tt 2507.00615}}.

\bibitem{Marsh:2011aa}
D.~Marsh, L.~McAllister and T.~Wrase, \emph{{The Wasteland of Random
  Supergravities}},
  \href{http://dx.doi.org/10.1007/JHEP03(2012)102}{\emph{JHEP} {\bf 03} (2012)
  102}, [\href{http://arxiv.org/abs/1112.3034}{{\tt 1112.3034}}].

\bibitem{Marsh:2014nla}
M.~C.~D. Marsh, B.~Vercnocke and T.~Wrase, \emph{{Decoupling and de Sitter
  Vacua in Approximate No-Scale Supergravities}},
  \href{http://dx.doi.org/10.1007/JHEP05(2015)081}{\emph{JHEP} {\bf 05} (2015)
  081}, [\href{http://arxiv.org/abs/1411.6625}{{\tt 1411.6625}}].

\bibitem{Sen:1996vd}
A.~Sen, \emph{{F theory and orientifolds}},
  \href{http://dx.doi.org/10.1016/0550-3213(96)00347-1}{\emph{Nucl. Phys. B}
  {\bf 475} (1996) 562--578}, [\href{http://arxiv.org/abs/hep-th/9605150}{{\tt
  hep-th/9605150}}].

\bibitem{Vafa:1996xn}
C.~Vafa, \emph{{Evidence for F theory}},
  \href{http://dx.doi.org/10.1016/0550-3213(96)00172-1}{\emph{Nucl. Phys. B}
  {\bf 469} (1996) 403--418}, [\href{http://arxiv.org/abs/hep-th/9602022}{{\tt
  hep-th/9602022}}].

\bibitem{Baumann:2008kq}
D.~Baumann, A.~Dymarsky, S.~Kachru, I.~R. Klebanov and L.~McAllister,
  \emph{{Holographic Systematics of D-brane Inflation}},
  \href{http://dx.doi.org/10.1088/1126-6708/2009/03/093}{\emph{JHEP} {\bf 03}
  (2009) 093}, [\href{http://arxiv.org/abs/0808.2811}{{\tt 0808.2811}}].

\bibitem{Baumann:2010sx}
D.~Baumann, A.~Dymarsky, S.~Kachru, I.~R. Klebanov and L.~McAllister,
  \emph{{D3-brane Potentials from Fluxes in AdS/CFT}},
  \href{http://dx.doi.org/10.1007/JHEP06(2010)072}{\emph{JHEP} {\bf 06} (2010)
  072}, [\href{http://arxiv.org/abs/1001.5028}{{\tt 1001.5028}}].

\bibitem{Gandhi:2011id}
S.~Gandhi, L.~McAllister and S.~Sjors, \emph{{A Toolkit for Perturbing Flux
  Compactifications}},
  \href{http://dx.doi.org/10.1007/JHEP12(2011)053}{\emph{JHEP} {\bf 12} (2011)
  053}, [\href{http://arxiv.org/abs/1106.0002}{{\tt 1106.0002}}].

\bibitem{McGuirk:2012sb}
P.~McGuirk, G.~Shiu and F.~Ye, \emph{{Soft branes in supersymmetry-breaking
  backgrounds}}, \href{http://dx.doi.org/10.1007/JHEP07(2012)188}{\emph{JHEP}
  {\bf 07} (2012) 188}, [\href{http://arxiv.org/abs/1206.0754}{{\tt
  1206.0754}}].

\bibitem{DeLuca:2021pej}
G.~B. De~Luca, E.~Silverstein and G.~Torroba, \emph{{Hyperbolic
  compactification of M-theory and de Sitter quantum gravity}},
  \href{http://dx.doi.org/10.21468/SciPostPhys.12.3.083}{\emph{SciPost Phys.}
  {\bf 12} (2022) 083}, [\href{http://arxiv.org/abs/2104.13380}{{\tt
  2104.13380}}].

\bibitem{Lust:2025vyz}
S.~L{\"u}st, M.~Nee and L.~Randall, \emph{{More Effective RS Field Theory}},
  \href{http://arxiv.org/abs/2510.11771}{{\tt 2510.11771}}.

\bibitem{Burgess:2005jx}
C.~P. Burgess, C.~Escoda and F.~Quevedo, \emph{{Nonrenormalization of flux
  superpotentials in string theory}},
  \href{http://dx.doi.org/10.1088/1126-6708/2006/06/044}{\emph{JHEP} {\bf 06}
  (2006) 044}, [\href{http://arxiv.org/abs/hep-th/0510213}{{\tt
  hep-th/0510213}}].

\bibitem{Burgess:2020qsc}
C.~P. Burgess, M.~Cicoli, D.~Ciupke, S.~Krippendorf and F.~Quevedo, \emph{{UV
  Shadows in EFTs: Accidental Symmetries, Robustness and No-Scale
  Supergravity}},
  \href{http://dx.doi.org/10.1002/prop.202000076}{\emph{Fortsch. Phys.} {\bf
  68} (2020) 2000076}, [\href{http://arxiv.org/abs/2006.06694}{{\tt
  2006.06694}}].

\bibitem{Gao:2022uop}
X.~Gao, A.~Hebecker, S.~Schreyer and V.~Venken, \emph{{Loops, local corrections
  and warping in the LVS and other type IIB models}},
  \href{http://dx.doi.org/10.1007/JHEP09(2022)091}{\emph{JHEP} {\bf 09} (2022)
  091}, [\href{http://arxiv.org/abs/2204.06009}{{\tt 2204.06009}}].

\bibitem{Cicoli:2007xp}
M.~Cicoli, J.~P. Conlon and F.~Quevedo, \emph{{Systematics of String Loop
  Corrections in Type IIB Calabi-Yau Flux Compactifications}},
  \href{http://dx.doi.org/10.1088/1126-6708/2008/01/052}{\emph{JHEP} {\bf 01}
  (2008) 052}, [\href{http://arxiv.org/abs/0708.1873}{{\tt 0708.1873}}].

\bibitem{vonGersdorff:2005bf}
G.~von Gersdorff and A.~Hebecker, \emph{{Kahler corrections for the volume
  modulus of flux compactifications}},
  \href{http://dx.doi.org/10.1016/j.physletb.2005.08.024}{\emph{Phys. Lett. B}
  {\bf 624} (2005) 270--274}, [\href{http://arxiv.org/abs/hep-th/0507131}{{\tt
  hep-th/0507131}}].

\bibitem{Berg:2005ja}
M.~Berg, M.~Haack and B.~Kors, \emph{{String loop corrections to Kahler
  potentials in orientifolds}},
  \href{http://dx.doi.org/10.1088/1126-6708/2005/11/030}{\emph{JHEP} {\bf 11}
  (2005) 030}, [\href{http://arxiv.org/abs/hep-th/0508043}{{\tt
  hep-th/0508043}}].

\bibitem{Berg:2005yu}
M.~Berg, M.~Haack and B.~Kors, \emph{{On volume stabilization by quantum
  corrections}},
  \href{http://dx.doi.org/10.1103/PhysRevLett.96.021601}{\emph{Phys. Rev.
  Lett.} {\bf 96} (2006) 021601},
  [\href{http://arxiv.org/abs/hep-th/0508171}{{\tt hep-th/0508171}}].

\bibitem{Berg:2007wt}
M.~Berg, M.~Haack and E.~Pajer, \emph{{Jumping Through Loops: On Soft Terms
  from Large Volume Compactifications}},
  \href{http://dx.doi.org/10.1088/1126-6708/2007/09/031}{\emph{JHEP} {\bf 09}
  (2007) 031}, [\href{http://arxiv.org/abs/0704.0737}{{\tt 0704.0737}}].

\bibitem{Cicoli:2021rub}
M.~Cicoli, F.~Quevedo, R.~Savelli, A.~Schachner and R.~Valandro,
  \emph{{Systematics of the {\ensuremath{\alpha}}' expansion in F-theory}},
  \href{http://dx.doi.org/10.1007/JHEP08(2021)099}{\emph{JHEP} {\bf 08} (2021)
  099}, [\href{http://arxiv.org/abs/2106.04592}{{\tt 2106.04592}}].

\bibitem{Epple:2004ra}
F.~T.~J. Epple, \emph{{Induced gravity on intersecting branes}},
  \href{http://dx.doi.org/10.1088/1126-6708/2004/09/021}{\emph{JHEP} {\bf 09}
  (2004) 021}, [\href{http://arxiv.org/abs/hep-th/0408105}{{\tt
  hep-th/0408105}}].

\bibitem{Haack:2015pbv}
M.~Haack and J.~U. Kang, \emph{{One-loop Einstein-Hilbert term in minimally
  supersymmetric type IIB orientifolds}},
  \href{http://dx.doi.org/10.1007/JHEP02(2016)160}{\emph{JHEP} {\bf 02} (2016)
  160}, [\href{http://arxiv.org/abs/1511.03957}{{\tt 1511.03957}}].

\bibitem{Becker:2002nn}
K.~Becker, M.~Becker, M.~Haack and J.~Louis, \emph{{Supersymmetry breaking and
  alpha-prime corrections to flux induced potentials}},
  \href{http://dx.doi.org/10.1088/1126-6708/2002/06/060}{\emph{JHEP} {\bf 06}
  (2002) 060}, [\href{http://arxiv.org/abs/hep-th/0204254}{{\tt
  hep-th/0204254}}].

\bibitem{Kallosh:2014oja}
R.~Kallosh, A.~Linde, B.~Vercnocke and T.~Wrase, \emph{{Analytic Classes of
  Metastable de Sitter Vacua}},
  \href{http://dx.doi.org/10.1007/JHEP10(2014)011}{\emph{JHEP} {\bf 10} (2014)
  011}, [\href{http://arxiv.org/abs/1406.4866}{{\tt 1406.4866}}].

\bibitem{Anderson:2020hux}
L.~B. Anderson, M.~Gerdes, J.~Gray, S.~Krippendorf, N.~Raghuram and F.~Ruehle,
  \emph{{Moduli-dependent Calabi-Yau and SU(3)-structure metrics from Machine
  Learning}}, \href{http://dx.doi.org/10.1007/JHEP05(2021)013}{\emph{JHEP} {\bf
  05} (2021) 013}, [\href{http://arxiv.org/abs/2012.04656}{{\tt 2012.04656}}].

\bibitem{Douglas:2020hpv}
M.~R. Douglas, S.~Lakshminarasimhan and Y.~Qi, \emph{{Numerical Calabi-Yau
  metrics from holomorphic networks}},
  \href{http://arxiv.org/abs/2012.04797}{{\tt 2012.04797}}.

\bibitem{Jejjala:2020wcc}
V.~Jejjala, D.~K. Mayorga~Pena and C.~Mishra, \emph{{Neural network
  approximations for Calabi-Yau metrics}},
  \href{http://dx.doi.org/10.1007/JHEP08(2022)105}{\emph{JHEP} {\bf 08} (2022)
  105}, [\href{http://arxiv.org/abs/2012.15821}{{\tt 2012.15821}}].

\bibitem{Larfors:2022nep}
M.~Larfors, A.~Lukas, F.~Ruehle and R.~Schneider, \emph{{Numerical metrics for
  complete intersection and Kreuzer\textendash{}Skarke Calabi\textendash{}Yau
  manifolds}}, \href{http://dx.doi.org/10.1088/2632-2153/ac8e4e}{\emph{Mach.
  Learn. Sci. Tech.} {\bf 3} (2022) 035014},
  [\href{http://arxiv.org/abs/2205.13408}{{\tt 2205.13408}}].

\bibitem{SeverinFabianSimon}
S.~Lüst, F.~Ruehle and S.~Schreyer, \emph{{Warped numerical Calabi-Yau
  metrics}}, {\emph{\text{work in progress}} }.

\bibitem{Minasian:2015bxa}
R.~Minasian, T.~G. Pugh and R.~Savelli, \emph{{F-theory at order $\alpha'^3$}},
  \href{http://dx.doi.org/10.1007/JHEP10(2015)050}{\emph{JHEP} {\bf 10} (2015)
  050}, [\href{http://arxiv.org/abs/1506.06756}{{\tt 1506.06756}}].

\bibitem{Klaewer:2020lfg}
D.~Klaewer, S.-J. Lee, T.~Weigand and M.~Wiesner, \emph{{Quantum corrections in
  4d $N$ = 1 infinite distance limits and the weak gravity conjecture}},
  \href{http://dx.doi.org/10.1007/JHEP03(2021)252}{\emph{JHEP} {\bf 03} (2021)
  252}, [\href{http://arxiv.org/abs/2011.00024}{{\tt 2011.00024}}].

\bibitem{Kim:2023sfs}
M.~Kim, \emph{{On string one-loop correction to the Einstein-Hilbert term and
  its implications on the K{\"a}hler potential}},
  \href{http://dx.doi.org/10.1007/JHEP07(2023)044}{\emph{JHEP} {\bf 07} (2023)
  044}, [\href{http://arxiv.org/abs/2302.12117}{{\tt 2302.12117}}].

\bibitem{Kim:2023eut}
M.~Kim, \emph{{On one-loop corrected dilaton action in string theory}},
  \href{http://dx.doi.org/10.4310/ATMP.2023.v27.n7.a2}{\emph{Adv. Theor. Math.
  Phys.} {\bf 27} (2023) 1965--2044},
  [\href{http://arxiv.org/abs/2305.08263}{{\tt 2305.08263}}].

\bibitem{Cho:2023mhw}
M.~Cho and M.~Kim, \emph{{A worldsheet description of flux compactifications}},
  \href{http://dx.doi.org/10.1007/JHEP05(2024)247}{\emph{JHEP} {\bf 05} (2024)
  247}, [\href{http://arxiv.org/abs/2311.04959}{{\tt 2311.04959}}].

\bibitem{Cvetic:2024wsj}
M.~Cveti{\v{c}} and M.~Wiesner, \emph{{Nonperturbative resolution of strong
  coupling singularities in 4D N=1 heterotic M-theory}},
  \href{http://dx.doi.org/10.1103/PhysRevD.110.106008}{\emph{Phys. Rev. D} {\bf
  110} (2024) 106008}, [\href{http://arxiv.org/abs/2408.12458}{{\tt
  2408.12458}}].

\bibitem{Sethi:2017phn}
S.~Sethi, \emph{{Supersymmetry Breaking by Fluxes}},
  \href{http://dx.doi.org/10.1007/JHEP10(2018)022}{\emph{JHEP} {\bf 10} (2018)
  022}, [\href{http://arxiv.org/abs/1709.03554}{{\tt 1709.03554}}].

\bibitem{Carta:2021sms}
F.~Carta, A.~Mininno, N.~Righi and A.~Westphal, \emph{{Gopakumar-Vafa
  hierarchies in winding inflation and uplifts}},
  \href{http://dx.doi.org/10.1007/JHEP05(2021)271}{\emph{JHEP} {\bf 05} (2021)
  271}, [\href{http://arxiv.org/abs/2101.07272}{{\tt 2101.07272}}].

\bibitem{Hebecker:2017lxm}
A.~Hebecker, P.~Henkenjohann and L.~T. Witkowski, \emph{{Flat Monodromies and a
  Moduli Space Size Conjecture}},
  \href{http://dx.doi.org/10.1007/JHEP12(2017)033}{\emph{JHEP} {\bf 12} (2017)
  033}, [\href{http://arxiv.org/abs/1708.06761}{{\tt 1708.06761}}].

\bibitem{Hebecker:2018fln}
A.~Hebecker, D.~Junghans and A.~Schachner, \emph{{Large Field Ranges from
  Aligned and Misaligned Winding}},
  \href{http://dx.doi.org/10.1007/JHEP03(2019)192}{\emph{JHEP} {\bf 03} (2019)
  192}, [\href{http://arxiv.org/abs/1812.05626}{{\tt 1812.05626}}].

\bibitem{Cole:2019enn}
A.~Cole, A.~Schachner and G.~Shiu, \emph{{Searching the Landscape of Flux Vacua
  with Genetic Algorithms}},
  \href{http://dx.doi.org/10.1007/JHEP11(2019)045}{\emph{JHEP} {\bf 11} (2019)
  045}, [\href{http://arxiv.org/abs/1907.10072}{{\tt 1907.10072}}].

\bibitem{Dubey:2023dvu}
A.~Dubey, S.~Krippendorf and A.~Schachner, \emph{{JAXVacua {\textemdash} a
  framework for sampling string vacua}},
  \href{http://dx.doi.org/10.1007/JHEP12(2023)146}{\emph{JHEP} {\bf 12} (2023)
  146}, [\href{http://arxiv.org/abs/2306.06160}{{\tt 2306.06160}}].

\bibitem{Chauhan:2025rdj}
A.~Chauhan, M.~Cicoli, S.~Krippendorf, A.~Maharana, P.~Piantadosi and
  A.~Schachner, \emph{{Deep observations of the Type IIB flux landscape}},
  \href{http://dx.doi.org/10.1007/JHEP07(2025)271}{\emph{JHEP} {\bf 07} (2025)
  271}, [\href{http://arxiv.org/abs/2501.03984}{{\tt 2501.03984}}].

\bibitem{Bena:2020xrh}
I.~Bena, J.~Bl{\r{a}}b{\"a}ck, M.~Gra{\~n}a and S.~L{\"u}st, \emph{{The tadpole
  problem}}, \href{http://dx.doi.org/10.1007/JHEP11(2021)223}{\emph{JHEP} {\bf
  11} (2021) 223}, [\href{http://arxiv.org/abs/2010.10519}{{\tt 2010.10519}}].

\bibitem{Carroll:2004st}
S.~M. Carroll, \emph{{Spacetime and Geometry}: {An Introduction to General
  Relativity}}.
\newblock Cambridge University Press, 7, 2019,
  \href{http://dx.doi.org/10.1017/9781108770385}{10.1017/9781108770385}.

\bibitem{Candelas:1990pi}
P.~Candelas and X.~de~la Ossa, \emph{{Moduli Space of {Calabi-Yau} Manifolds}},
  \href{http://dx.doi.org/10.1016/0550-3213(91)90122-E}{\emph{Nucl. Phys. B}
  {\bf 355} (1991) 455--481}.

\end{thebibliography}\endgroup

\end{document}